\documentclass[twocolumn]{emulateapj}
\usepackage{amsmath}
\usepackage{nicefrac}
\usepackage[none]{hyphenat}

\def\Rsun{R$_{\odot}$}
\def\Msun{M$_{\odot}$}
\def\Zsun{Z$_{\odot}$}

\begin{document}

\title{The Close Binary Fraction of Solar-type Stars is \\ Strongly Anti-correlated with Metallicity}

\author{Maxwell Moe\altaffilmark{1,2}, Kaitlin M.  Kratter\altaffilmark{1}, and Carles Badenes$^{3}$}

\altaffiltext{1}{Steward Observatory, University of Arizona, 933~N.~Cherry~Ave.,~Tucson,~AZ 85721,~USA}

\altaffiltext{2}{Einstein Fellow; moem@email.arizona.edu}

\altaffiltext{3}{Department of Physics and Astronomy and Pittsburgh Particle~Physics, Astrophysics and Cosmology Center~(PITT~PACC), University~of~Pittsburgh, 3941~O'Hara~Street, Pittsburgh, PA 15260, USA}

\begin{abstract}
 There is now strong evidence that the close binary fraction ($P$~$<$~10$^4$~days; $a$~$<$~10~AU) of solar-type stars ($M_1$~$\approx$~0.6\,-\,1.5\,\Msun) decreases significantly with metallicity.  Although early surveys showed that the {\it observed} spectroscopic binary (SB) fractions in the galactic disk and halo are similar (e.g., Carney-Latham sample), these studies did not correct for incompleteness. In this study, we examine five different surveys and thoroughly account for their underlying selection biases to measure the intrinsic occurrence rate of close solar-type binaries. We re-analyze: (1)~a volume-limited sample of solar-type stars (Raghavan et al. 2010), (2)~the SB survey of high-proper-motion stars (Latham et al. 2002), (3)~various SB samples of metal-poor giants (Carney et al. 2003; Hansen et al. 2015,\,2016), (4)~the APOGEE survey of radial velocity (RV) variables (Badenes et al. 2018), and (5)~eclipsing binaries (EBs) discovered by {\it Kepler} (Kirk et al. 2016).  The observed APOGEE RV variability fraction and {\it Kepler} EB fraction both decrease by a factor of $\approx$\,4 across $-$1.0~$<$~[Fe/H]~$<$~0.5 at the 22$\sigma$ and 9$\sigma$ confidence levels, respectively. After correcting for incompleteness, all five samples / methods exhibit a quantitatively consistent anti-correlation between the intrinsic close binary fraction ($a$~$<$~10~AU) and metallicity: $F_{\rm close}$ = 53\%\,$\pm$\,12\%, 40\%\,$\pm$\,6\%, 24\%\,$\pm$\,4\%, and 10\%\,$\pm$\,3\% at [Fe/H]~=~$-$3.0, $-$1.0, $-$0.2 (mean field metallicity), and +0.5, respectively.  We present simple fragmentation models that explain why the close binary fraction of solar-type stars strongly decreases with metallicity while the wide binary fraction, close binary fraction of OB~stars, and initial mass function are all relatively constant across $-$1.5~$\lesssim$~[Fe/H]~$<$~0.5. The majority of solar-type stars with [Fe/H]~$\lesssim$~$-$1.0 will interact with a stellar companion, which has profound implications for binary evolution in old and metal-poor environments such as the galactic halo, bulge, thick disk, globular clusters, dwarf galaxies, and high-redshift universe.
\end{abstract}


\section{Introduction}
\label{Introduction}

Variations in the close binary fraction ($a$~$\lesssim$~10~AU) with respect to metallicity have been continuously debated over the years \citep[][additional references below]{Carney1983,Latham2002,Carney2005,Machida2009,Raghavan2010,Rastegaev2010,Moe2013,Bate2014,Badenes2018}.  Some observations indicate no dependence on metallicity \citep{Latham2002,Carney2005,Moe2013}, others find the close binary fraction and metallicity are positively correlated \citep{Carney1983,Abt1987,Hettinger2015}, while yet others have found that the close binary fraction decreases with metallicity \citep{Grether2007,Raghavan2010,Gao2014,Badenes2018}.  Studying how the close binary fraction varies with primary mass, metallicity, age, and environment provides significant insight into the processes of protobinary fragmenation, accretion, and orbital migration \citep{Kratter2008,Kratter2010,Duchene2013,Moe2017,Moe2018}.  The close binary fraction is also a crucial input parameter in population synthesis studies of blue stragglers, chemically peculiar stars, cataclysmic variables, Type Ia and Ib/c supernovae, X-ray binaries, mergers of compact objects, short gamma-ray bursts, and sources of gravitational waves \citep{Hurley2002,Eggleton2006,Belczynski2008,Sana2012,DeMarco2017}  A substantial change in the close binary fraction with respect to metallicity would have dramatic consequences for the predicted rates and properties of various channels of binary evolution. The apparent discrepancies in the inferred close binary fraction as a function of metallicity must be reconciled in order to more fully understand binary star formation and to make reliable predictions for binary evolution.

The primary goal of this study is to reconcile the conflicting results reported in the literature in order to accurately measure the bias-corrected close binary fraction of solar-type stars as a function of metallicity.  In \S\ref{Overview}, we overview the methods, results, and potential caveats associated with previous results. In \S\ref{Spectroscopic}, we correct for incompleteness within the Carney-Latham sample and other spectroscopic binary surveys to determine if a large change in the close binary fraction with respect to metallicity is apparent in these earlier datasets.  In \S\ref{APOGEE}, we analyze the \citet{Badenes2018} sample of APOGEE stars to measure precisely how the radial velocity variability fraction and bias-corrected close binary fraction change as a function of metallicity.  We next measure the eclipsing binary fraction of solar-type dwarfs in the {\it Kepler} sample, providing a new and independent method for determining how the close binary fraction varies with metallicity (\S\ref{Kepler}).  We combine and summarize the observational constraints in \S\ref{Summary}, where we show all five samples / methods investigated in this study exhibit a remarkably consistent anti-correlation between metallicity and close binary fraction.  We also discuss the overall binary fraction and period distribution as a function of mass and metallicity, and highlight the resulting implications for binary evolution.  In \S\ref{Models}, we investigate fragmentation models to explain why the close binary fraction of solar-type stars strongly decreases with metallicity while the wide binary fraction, close binary fraction of massive stars, and initial mass function are relatively constant. We conclude in \S\ref{Conclusions}.

\section{Overview of Previous Observations}
\label{Overview}

\noindent {\it Carney-Latham Sample}. For solar-type (FGK) dwarfs, early observations indicated the spectroscopic binary (SB) fraction of metal-poor halo stars was slightly lower than that of metal-rich stars in the galactic disk \citep{Carney1983,Abt1987}.  Subsequent surveys instead found the SB fraction was relatively independent of metallicity \citep{Stryker1985,Ryan1992,Latham2002,Carney2005}.  In particular, \citet{Latham2002} and \citet{Carney2005} investigated a large sample of 1,464 FGK stars with high proper motion in the disk and halo.  They identified SBs as stars that exhibited larger radial velocity (RV) variations compared to their RV measurement uncertainties.  They obtained a median of $N_{\rm RV}$ = 12 RV measurements per star, and so they were able to fit robust orbital parameters for the majority of their SBs. \citet{Latham2002} measured the halo and disk SB fractions to be 14.5\%\,$\pm$\,1.8\% and 15.6\%\,$\pm$\,1.5\%, respectively, which are consistent with each other within the uncertainties.  They also showed the observed SB period distributions in the disk and halo are similar (see their Fig.~8).  \citet{Carney2005} refined the sample by excluding stars with too few RV measurements or large uncertainties in the RVs or metallicities, leaving 994 systems.   \citet{Carney2005} measured a slightly larger SB fraction of 24\%\,$\pm$\,2\% for their refined sample, but still found the SB fraction was nearly constant across $-$2.5~$<$~[m/H]~$<$~0.0 (see their Fig.~2). 

However, \citet{Latham2002} and \citet{Carney2005} did not correct for incompleteness. Although the {\it observed} SB fraction appears to be independent of metallicity, the true bias-corrected close binary fraction could be substantially different.  In fact, to explain the small deficit in the halo SB fraction (14.5\%) compared to the disk SB fraction (15.6\%), \citet{Latham2002} hinted at the likelihood that their halo measurement was more incomplete.  They stated, ``This might be the result of an observational bias, because halo binaries have lower metallicity and therefore weaker lines, with a corresponding poorer velocity precision and higher threshold for the detection of binaries." This effect likely explains why the earlier observations by \citet{Carney1983} and \citet{Abt1987} found a smaller SB fraction for metal-poor stars. In \S\ref{Latham}, we demonstrate that this selection bias reverses the inferred trends in the Carney-Latham SB samples, and therefore the intrinsic close binary fraction of metal-poor halo stars is actually larger than that of metal-rich disk stars. 

\vspace*{0.2cm}

\noindent {\it Volume-limited Samples}. \citet{Grether2007} and \citet{Raghavan2010} provided the earliest statistically significant evidence that the binary fraction of solar-type stars is anti-correlated with metallicity.  \citet{Raghavan2010} utilized spectroscopic RV observations, long-baseline and speckle interferometry, adaptive optics, and common proper motion to investigate the multiplicity statistics of 454 FGK dwarfs within 25 pc.  In their sample,  411 stars have reliable metallicity measurements across $-$0.9~$<$~[Fe/H]~$<$~0.4.  As shown in their Fig.~19, \citet{Raghavan2010} found the overall binary fraction decreases from 66\%\,$\pm$\,7\% across $-$0.9~$<$~[Fe/H]~$<$~$-$0.4 ($N$~=~44 systems) to 39\%\,$\pm$\,3\% across $-$0.3~$<$~[Fe/H]~$<$~0.4 ($N$~=~343; uncertainties derive from binomial statistics). The overall binary fraction decreases with metallicity by a factor of 1.7\,$\pm$\,0.2, statistically significant at the 3.8$\sigma$ level.  Although the \citet{Raghavan2010} survey is slightly incomplete \citep{Chini2014,Moe2017}, it is difficult to explain how selection biases alone could cause the observed anti-correlation between binary fraction and metallicity.  

\vspace*{0.2cm}

\noindent {\it Close versus Wide Solar-type Binaries}. The anti-correlation between metallicity and binary fraction appears to be limited to shorter orbital separations.  Of the 44 systems in the  \citet{Raghavan2010} sample with $-$0.9~$<$~[Fe/H]~$<$~$-$0.4, 22 (50\%\,$\pm$\,8\%) have companions with log\,$P$\,(days)~$<$~6 ($a$~$\lesssim$~200~AU) and 7 (16\%\,$\pm$\,5\%) are wide binaries with log\,$P$\,(days)~$>$~6 ($a$~$\gtrsim$~200~AU).  Meanwhile, of the 343 systems with $-$0.3~$<$~[Fe/H]~$<$~0.4, 87 (25\%\,$\pm$\,2\%) and 47 (14\%\,$\pm$\,2\%) have companions below and above $a$~$\approx$~200~AU, respectively.  Hence, the very wide binary fraction ($a$~$\gtrsim$~200~AU) remains constant within the uncertainties.  Common proper motion and CCD imaging surveys also demonstrate the wide binary fraction of solar-type stars is independent of metallicity \citep{Chaname2004,Zapatero2004}. Meanwhile,  the  binary fraction below $a$~$\lesssim$~200~AU in the \citet{Raghavan2010} sample decreases by a factor of 2.0\,$\pm$\,0.3 between [Fe/H]~$\approx$~$-$0.6 and 0.0, statistically significant at the 3.2$\sigma$ level.

\citet{Rastegaev2010} combined spectroscopy, speckle interferometry, and visual observations to measure the full multiplicity properties of metal-poor FGK stars ([m/H]~$<$~$-$1).  After correcting for incompleteness, they measured an overall binary fraction of $\approx$\,40\%, which is consistent with the binary fraction of 46\%\,$\pm$\,2\% measured by \citet{Raghavan2010} for solar-type stars within 25 pc. Compared to metal-rich systems, however, \citet{Rastegaev2010} showed metal-poor binaries are significantly skewed toward close to intermediate separations, exhibiting a factor of $\approx$\,2\,-\,3 excess across log\,$P$\,(days)~=~1\,-\,4 ($a$ $\approx$ 0.1\,-\,10 AU; see their Fig.~10). Their combined spectroscopic and speckle interferometric survey is relatively complete across this parameter space, and so the factor of $\approx$\,2\,-\,3 excess observed across $a$~$\approx$~0.1\,-\,10 AU for metal-poor FGK binaries is likely a real effect.

\vspace*{0.2cm}

\noindent {\it Wide Companions to KM Subdwarfs}. Speckle, {\it HST}, and adaptive optics imaging of metal-poor KM subdwarfs all indicate a lower wide binary fraction compared to their solar-metallicity counterparts \citep{Riaz2008,Jao2009,Lodieu2009,Ziegler2015}. However, these surveys specifically targeted metal-poor stars based on their photometric colors and absolute magnitudes, i.e., KM subdwarfs in the HR diagram that lie well below the main-sequence relation of solar-metallicity dwarfs.  A metal-poor subdwarf with an equally bright companion would appear photometrically as a normal metal-rich dwarf, and so would not have been included in their samples.  Late-K and M-type binaries are weighted toward equal-mass companions \citep{Janson2012,Dieterich2012,Duchene2013}.  A bias against equally bright companions would dramatically reduce the inferred binary fraction of metal-poor KM subdwarfs. In their adaptive optics survey of metal-poor KM subdwarfs, \citet{Ziegler2015} specifically noted a substantial shortage of low-contrast companions with $\Delta$i~$<$~2~mag compared to metal-rich KM dwarfs (see their Fig.~10).  A deficit of binaries with nearly equal brightnesses is naturally explained by their subdwarf photometric selection criteria.  These surveys are heavily influenced by this selection bias and we conclude there is little or no change in the wide binary fraction of KM stars as a function of metallicity. 

\vspace*{0.2cm}

\noindent {\it Recent Wide-field Surveys}. Over the past few years, there have been several wide-field spectroscopic surveys that measured the chemical abundances and radial velocities of hundreds of thousands of stars.  Some of these spectroscopic surveys obtained multiple epochs of individual stars, allowing for a statistical measurement of the RV variability fraction as a function of metallicity. Utilizing multi-epoch SDSS spectra of F-type dwarfs (resolution R~$\approx$~2,000), \citet{Hettinger2015} measured the RV variability fraction increases by $\approx$\,30\% between [Fe/H]~=~$-$1.7 and $-$0.5 (see their Fig.~5).  Based on SEGUE and LAMOST spectra of FGK dwarfs (R~$\approx$~2,000), \citet{Gao2014}, \citet{Gao2017}, and \citet{Tian2018} instead found the RV variability fraction decreases by a factor of $\approx$\,2 between their metal-poor ([Fe/H] $<$ $-$1.1) and metal-rich ([Fe/H] $>$ $-$0.6) samples.  They also determined the RV variability fraction increases by a factor of $\approx$\,2 between K-type and F-type dwarfs, consistent with other studies that show the close binary fraction strongly increases above $M_1$~$\gtrsim$~1\Msun\ \citep{Abt1990,Raghavan2010,Sana2012,Duchene2013,Moe2017,Murphy2018}. Utilizing SEGUE spectra of extremely metal-poor stars with [Fe/H]~$\lesssim$~$-$3.0, \citet{Aoki2015} estimated the binary fraction below $P$~$<$~1,000~days is $\approx$\,20\%, nearly double that of their metal-rich counterparts. 

Most recently, \citet{Badenes2018} analyzed multi-epoch APOGEE spectra of $\approx$\,90,000 FGK stars, which had superior spectral resolution R~$\approx$~22,500 and higher signal-to-noise ratios S/N~$>$~40.  They searched for RV variables that exhibited large enough amplitudes $\Delta$RV$_{\rm max}$~$>$~10~km~s$^{-1}$ between epochs to be nearly 100\% certain they were real binary stars.  \citet{Badenes2018}  demonstrated the RV variability fraction decreases by a factor of $\approx$\,2\,-\,3 between their low-metallicity tercile ([Fe/H] $\lesssim$ $-$0.3) and high-metallicity tercile ([Fe/H]~$\gtrsim$~0.0).  They observed this factor of $\approx$\,2\,-\,3 metallicity effect for stars of varying surface gravities 0.0~$\lesssim$~log\,$g$\,(cm\,s$^{-2}$)~$\lesssim$~5.0 (see their Fig.~13). This suggests the anti-correlation between binary fraction and metallicity occurs for both close companions orbiting small main-sequence stars and for slightly wider companions orbiting large giants.  We investigate a subset of the APOGEE data in \S\ref{APOGEE} to quantify more precisely how the RV variability fraction and close binary fraction change as a continuous function of metallicity.

Other observational methods corroborate that the binary fraction of FGK stars decreases with metallicity, but to a lesser extent than the factor of $\approx$\,2\,-\,3 effect determined by \citet{Badenes2018}.  For example, \citet{Yuan2015} analyzed the properties of binaries discovered through the stellar locus outlier method. These are unresolved binaries in which the companions are bright enough to sufficiently shift the combined photometric colors to be inconsistent with single stars. They found the unresolved binary fraction decreases by a factor of $\approx$\,1.4 between [Fe/H]~$\approx$~$-$1.7 and $-$0.3. Similarly, \citet{ElBadry2018} identified double-lined spectroscopic binaries (SB2s) with luminous secondaries in the APOGEE dataset.  For SB2s that exhibited significant orbital motion between epochs, i.e., $\Delta$RV$_{\rm max}$~$>$~10~km\,s$^{-1}$ as adopted in \citet{Badenes2018}, \citet{ElBadry2018} confirmed the close binary fraction decreases by a factor of $\approx$\,1.6 between their low-metallicity tercile ([Fe/H] $<$ $-$0.2) and high-metallicity tercile ([Fe/H] $>$ 0.0).  However, for their larger population of wider SB2s that did not show RV variability, \citet{ElBadry2018} found the binary fraction was consistent with being constant with respect to metallicity. Taken as a whole, these recent observations suggest the close binary fraction of solar-type stars is strongly anti-correlated with metallicity while the wide binary fraction is independent of metallicity.  Photometric binaries \citep{Yuan2015} and SB2s \citep{ElBadry2018}, which include both close and wide binaries, exhibit a weaker trend with metallicity compared to close binaries exclusively.  

\vspace*{0.2cm}

\noindent {\it Close Massive Binaries}. Meanwhile, the close binary fraction of massive stars does not vary significantly with metallicity \citep{Moe2013,Dunstall2015,Almeida2017}. \citet{Moe2013} measured the eclipsing binary (EB) fraction of early-B stars ($M_1$~$\approx$~6\,-\,16\,\Msun) based on OGLE observations of the Small ([Fe/H]~$\approx$~$-$0.7) and Large ([Fe/H]~$\approx$~$-$0.4) Magellanic Clouds (SMC/LMC) and {\it Hipparcos} observations of nearby systems in the Milky Way (MW; [Fe/H]~$\approx$~0.0).  They found the fraction of early-B stars that have eclipsing companions across orbital periods $P$~=~2\,-\,20~days and eclipse depths $\Delta$m~=~0.25\,-\,0.65~mag is 0.70\%\,$\pm$\,0.06\%, 0.69\%\,$\pm$\,0.03\%, and 1.00\%\,$\pm$\,0.25\% for the SMC, LMC, and MW, respectively (see their Table~1).  Although EB observations are less complete due to geometrical selection effects, they are not affected by the spectroscopic selection bias discussed above and are therefore more robust in detecting variations in the close binary fraction with respect to metallicity.  Nevertheless, after correcting for incompleteness in their spectroscopic RV observations, the close binary fraction of O stars \citep{Almeida2017} and early-B stars \citep{Dunstall2015} in the LMC is consistent with their solar-metallicity counterparts in the MW.  For massive stars ($M_1$~$\gtrsim$~6\,\Msun), the close binary fraction is relatively independent of metallicity, at least within the $\delta F_{\rm close}$/$F_{\rm close}$ $\approx$ 30\% measurement uncertainties and across the range of metallicities $-$0.7~$\lesssim$~[Fe/H]~$\lesssim$~0.1 probed by the observations. 

\vspace*{0.2cm}

\noindent {\it Initial Mass Function}. Similarly, the initially mass function (IMF) is fairly universal across two orders of magnitude in metallicity $-$1.5~$\lesssim$~[Fe/H]~$\lesssim$~0.5 \citep[][references therein]{Bastian2010,Kroupa2013}.  Young metal-poor associations and clusters in the LMC ([Fe/H]~$\approx$~$-$0.4; \citealt{DaRio2009}), in the SMC ([Fe/H]~$\approx$~$-$0.7; \citealt{Sirianni2002, Schmalzl2008}), and in the outer regions of the MW ([Fe/H]~$\approx$~$-$0.8; \citealt{Yasui2016a,Yasui2016b}) all have IMFs consistent with the canonical IMF. The low-mass end of the IMF ($M_1$~$\approx$~0.1\,-\,0.9\Msun)  is invariant across galactic open clusters and globular clusters that span a wide range of metallicities $-$2.3~$\lesssim$~[Fe/H]~$\lesssim$~0.3 \citep{vonHippel1996,DeMarchi2010,Bastian2010}.  Although some observations indicate the IMF becomes top-heavy toward lower metallicities \citep{Marks2012,Geha2013,Kroupa2013}, this trend is not statistically significant until the metallicity falls below at least [Fe/H]~$\lesssim$~$-$1.5. 

\section{Spectroscopic Versus Intrinsic \\ Close Binary Fraction}
\label{Spectroscopic}

\subsection{Carney-Latham Sample}
\label{Latham}

\subsubsection{Description of Observations}

Of the 1,464 stars with high proper motion in the Carney-Latham sample, \citet{Latham2002} cataloged detailed information for 1,359 single-lined stars.  They listed the stellar properties, e.g., metallicity [m/H], effective temperature $T_{\rm eff}$, and rotational velocity $v_{\rm rot}$, of the template spectrum that most closely matched the observed spectra.  The full temperature range is $T_{\rm eff}$~$\approx$~3,800\,-7,700\,K, but 1,301 of the systems (96\%) have $T_{\rm eff}$~$\approx$~4,500\,-\,6,300\,K, corresponding approximately to F7\,-\,K4 spectral types. The template spectra are in large metallicity increments of $\Delta$[m/H]~=~0.5, but 1,349 of their 1,359 single-lined stars span a large range of $-$3.0~$\le$~[m/H]~$\le$~0.5 to provide sufficient leverage for investigating metallicity effects. \citet{Latham2002} derived robust orbital solutions for 156 SB1s (all with $P$~$<$~7,000 days) and presented preliminary orbits for an additional 15 SB1s (mostly with $P$~=~5,000\,-\,10,000 days).  They also cataloged 17 large-amplitude RV variables that likely have wide stellar companions but lack the necessary phase coverage to measure orbital elements (see their Fig.~6). In a separate study, \citet{Goldberg2002}  measured stellar parameters and orbital solutions for 34 SB2s from the Carney-Latham sample, all of which have $P$~$<$~5,000~days and $-$2.5~$\le$~[m/H]~$\le$~0.0.  Neither \citet{Latham2002} nor \citet{Goldberg2002} fitted the surface gravities log\,$g$ directly, but instead adopted log\,$g$~=~4.5 for cooler stars ($T_{\rm eff}$ $\lesssim$ 6,000K) and log\,$g$~=~4.0 for hotter stars ($T_{\rm eff}$ $\gtrsim$ 6,000K). About 10\% of the high-proper-motion stars in the Carney-Latham sample are likely subgiants or giants \citep{Laird1988,Carney1994}, and the fraction is probably larger for systematically older halo stars.

\citet{Latham2002} listed the Julian dates, RVs, and RV uncertainties $\sigma_{\rm RV}$ for each of the $N_{\rm RV}$ observations of each single-lined star. We compile their data and compute the mean RV uncertainty $\langle\sigma_{RV}\rangle$ for each system.  In Fig.~\ref{Latham_RV}, we show the average of and 1$\sigma$ spread in $\langle\sigma_{RV}\rangle$ as a function of metallicity [m/H].  As indicated in \citet{Latham2002}, the metal-poor stars in their sample have systematically larger RV uncertainties due to their weaker absorption lines.  The mean uncertainties double from $\langle \sigma_{RV}\rangle$~=~0.5~km\,s$^{-1}$ for solar-metallicity to $\langle \sigma_{RV}\rangle$~=~1.0~km\,s$^{-1}$ for metal-poor stars with [m/H]~$\le$~$-$2.0.  

\citet{Latham2002} observed their single-lined stars with varying cadence (see their Fig.~3). For their full sample, the median number of RV measurements is $N_{\rm RV}$ = 12, and the 10\,-\,90 percentile range spans $N_{\rm RV}$ = 8\,-\,39.  Similarly, the median timespan is $\Delta t$~=~9~yr between first and final visits, and the 10\,-\,90 percentile interval is $\Delta t$~=~8\,-\,14~yr. There is no trend in the number or timespan of RV measurements as a function of metallicity.  The median number of RV observations is $N_{\rm RV}$~=~13 for the 544 metal-poor single-lined stars with $-$3.0~$\le$~[m/H]~$\le$~$-$0.8 and $N_{\rm RV}$~=~11 for the 805 metal-rich stars with $-$0.8~$<$~[m/H]~$\le$~0.5.  The median timespan, which is most important parameter for estimating completeness rates (see below and \S\ref{APOGEE}), is $\Delta t$~=~9~yr for both the metal-poor and metal-rich subsamples.

\subsubsection{Corrections for Incompleteness}
\label{LathamCorrections}

We next perform Monte Carlo simulations to determine the probability of detecting SBs as a function of $\langle\sigma_{RV}\rangle$. In our simulations, we fix the mass of the primary to be $M_1$~=~1.0\,\Msun\, and draw  period, mass-ratio, and eccentricity distributions consistent with solar-type binaries in the field \citep{Duquennoy1991,Raghavan2010,Tokovinin2014,Moe2017}.  Specifically, we adopt a log-normal period distribution with a peak at log\,$P$\,(days)~=~4.9 and dispersion of $\sigma_{\rm logP}$ = 2.3, but only select binaries from the short-period tail across the interval 0.0~$<$~log\,$P$\,(days)~$<$~4.0 ($a$~$\lesssim$~10~AU) we are investigating.  We assume a uniform mass-ratio distribution across $q$~=~$M_2$/$M_1$~=~0.1\,-\,1.0 and that very close binaries with $P$~$<$~$P_{\rm circ}$~=~10~days are tidally circularized.  Toward longer periods $P$~$>$~$P_{\rm circ}$, we adopt a uniform eccentricity distribution across the interval 0.0~$<$~$e$~$<$~$e_{\rm max}$($P$), where the upper envelope of the eccentricity versus period distribution derives from conservation of orbital angular momentum during tidal evolution \citep{Badenes2018}:

\begin{equation}
e_{\rm max} = \bigg(1 - \Big(\frac{P}{P_{\rm circ}}\Big)^{\nicefrac{-2}{3}}\bigg)^{\nicefrac{1}{2}}.
\label{tide}
\end{equation}

\noindent We assume random orientations, which have an inclination probability distribution of $p$ = sin\,$i$ and a uniform distribution for arguments of periastron. Reasonable variations in the period, mass-ratio, or eccentricity distributions yield only minor changes in the simulated detection efficiencies.

\begin{figure}[t!]
\centerline{
\includegraphics[trim=0.3cm 0.3cm 0.3cm 0.5cm, clip=true, width=3.4in]{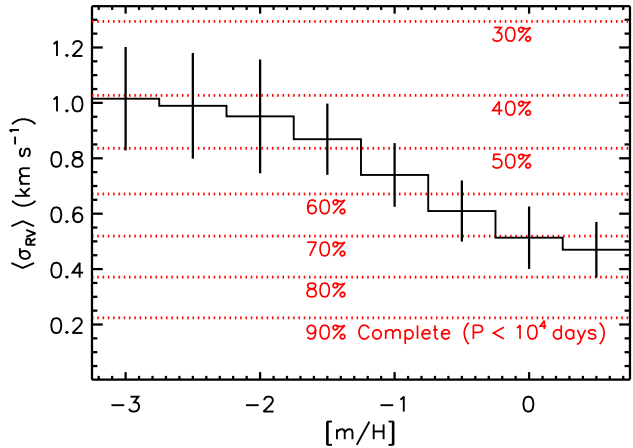}}
\caption{The mean RV uncertainty $\langle \sigma_{RV}\rangle$ as a function of metallicity [m/H] in the \citet{Latham2002} sample (black).  We simulate the completeness rates for a population of binaries with $P$~$<$~10$^{4}$ days (dotted red).  As the RV uncertainties decrease from $\langle \sigma_{RV}\rangle$ = 1.0 km\,s$^{-1}$ to 0.5 km\,s$^{-1}$ between metal-poor and metal-rich stars, the completeness fractions increase from $\approx$\,40\% to $\approx$\,70\%.}
\label{Latham_RV}
\end{figure}

For each binary, we generate RVs at $N_{\rm RV}$~=~12 epochs randomly distributed across a timespan of $\Delta t$~=~9~yr, matching the median cadence and median baseline of the \citet{Latham2002} observations.  For each RV measurement, we add Gaussian random noise according to $\langle\sigma_{RV}\rangle$.  A large-amplitude RV variable will exhibit a larger variance of RVs compared to the variance implied by its measurement uncertainties.  We therefore use an F-variance test to measure the probability $p$ that each generated system has a constant RV.  In the \citet{Latham2002} catalog, the majority of constant RV stars have $p$~$>$~5$\times$10$^{-7}$ while nearly all systems with $p$~$<$~5$\times$10$^{-7}$ are cataloged as SBs, the majority of which have measured orbital parameters.  We adopt the criterion that $p$~$<$~5$\times$10$^{-7}$ for a simulated binary to be considered an RV variable, corresponding to a 5.0$\sigma$ level of significance.

We show the results of our Monte Carlo simulations in Fig.~\ref{Latham_RV}.  Given a small RV uncertainty $\langle\sigma_{RV}\rangle$~=~0.2~km\,s$^{-1}$, $\approx$\,90\% of the binaries with $P$~$<$~10$^4$~days would appear as spectroscopic RV variables with $p$~$<$~5$\times$10$^{-7}$.  The remaining $\approx$\,10\% of the binaries are generally in wide orbits ($P$~$\approx$~5,000\,-\,10,000~days) with low-mass companions ($q$ $\approx$ 0.1\,-\,0.3).  Meanwhile, given a mean uncertainty of $\langle\sigma_{RV}\rangle$~=~1.3~km\,s$^{-1}$ and 12 random epochs across 9 years, only $\approx$\,30\% of binaries with $P$~$<$~10$^4$ days would appear as RV variables.  Across the interval of interest, the completeness rate increases from $\approx$\,40\% for metal-poor halo stars ([m/H]~$\le$~$-$2.0; $\langle\sigma_{RV}\rangle$~$\approx$~1.0~km~s$^{-1}$) to $\approx$\,70\% for metal-rich disk stars ([m/H]~$\ge$~0.0; $\langle\sigma_{RV}\rangle$~$\approx$~0.5~km~s$^{-1}$).  The \citet{Latham2002} spectroscopic survey is $\approx$\,1.8 times more complete in detecting close binary companions to metal-rich disk stars compared to metal-poor halo stars. 

\subsubsection{Binary Mass Functions}

\begin{figure}[t!]
\centerline{
\includegraphics[trim=0.3cm 0.2cm 0.3cm 0.2cm, clip=true, width=3.4in]{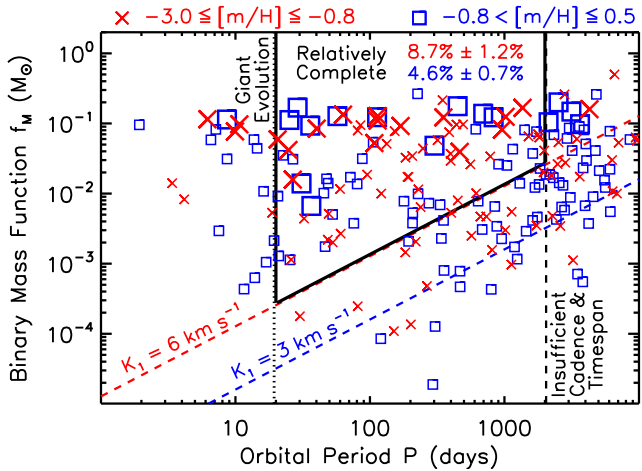}}
\caption{The measured binary mass functions and orbital periods for the 169 SB1s with $P$~$<$~10$^4$ days from \citet[][smaller symbols]{Latham2002} and 34 SB2s from \citet[][larger symbols]{Goldberg2002} divided into metal-poor ($-$3.0~$\le$~[m/H]~$\le$~$-$0.8; red crosses) and metal-rich ($-$0.8~$<$~[m/H]~$\le$~0.5; blue squares) subsets. The samples are biased against very close binaries with $P$~$<$~20~days (left of dotted black line) due to contamination by subgiants and giants while wide binaries beyond $P$~$>$~2,000~days (right of dashed black line) are incomplete given the median number $N_{\rm RV}$~=~12 and timespan $\Delta t$~=~9~yr of the RV observations. SBs with small velocity semi-amplitudes $K_1$~$<$~6$\langle\sigma_{\rm RV}\rangle$ are also incomplete, corresponding to $K_1$~$<$~3~km~s$^{-1}$ for metal-rich systems (dashed blue) and $K_1$~$<$ 6~km~s$^{-1}$ for metal-poor systems (dashed red).  Within the relatively complete and unbiased parameter space (solid black lines), the SB fraction decreases by a factor of 1.9\,$\pm$\,0.4 from 8.7\%\,$\pm$\,1.2\% for the metal-poor subsample to 4.6\%\,$\pm$\,0.7\% for the metal-rich subsample at the 3.0$\sigma$ significance level.}
\label{Latham_fM}
\end{figure}

The observed distribution of binary mass functions $f_{\rm M}$ = ($M_2$\,sin\,$i$)$^3$/($M_1$\,+\,$M_2$)$^2$ = $P$\,$K_1^3$\,(1\,$-$\,$e^2$)$^{\nicefrac{3}{2}}$/(2$\pi$G) also demonstrates that metal-poor SBs are less complete.  In Fig.~\ref{Latham_fM}, we show the measured binary mass functions versus orbital periods for the 169 SB1s with $P$~$<$~10$^4$~days in the \citet{Latham2002} sample.  We also display with slightly larger symbols the 34 SB2s from \citet{Goldberg2002}, which concentrate toward larger binary mass functions $f_{\rm M}$ = 0.007\,-\,0.2\,\Msun\ as expected. We divide the sample into a metal-poor subset with $-$3.0~$\le$~[m/H]~$\le$~$-$0.8 (red crosses; $N_{\rm SB}$~=~91~SBs with measured orbital elements; $N$~=~562~stars) and a metal-rich subset with $-$0.8~$<$~[m/H]~$\le$~0.5 (blue squares; $N_{\rm SB}$~=~114, $N$~=~821). Both subsamples are measurably incomplete toward wide separations and small ratios.  However, the metal-rich SB1s, which have systematically smaller RV uncertainties, extend toward smaller binary mass functions and longer orbital periods.  A KS test demonstrates that the observed 71 SBs with $P$~$>$~100~days in our metal-rich subset are weighted toward smaller velocity semi-amplitudes compared to the 57 metal-poor SBs with $P$~$>$~100~days at the 2.7$\sigma$ confidence level ($p_{\rm KS}$ = 0.004).  For reference, we also show $f_{\rm M}$ as a function of $P$ for a fixed eccentricity of $e$~=~0.5 and a velocity semi-amplitude of $K_1$~=~6$\langle\sigma_{RV}\rangle$, corresponding to $K_1$~=~3~km~s$^{-1}$ for metal-rich stars (dashed blue line in Fig.~\ref{Latham_fM}) and $K_1$~=~6~km~s$^{-1}$ for metal-poor stars (dashed red). The \citet{Latham2002} SB1 sample is measurably incomplete below these relations.

The samples of SB1s and SB2s with measured orbital solutions are relatively complete across $P$~=~20\,-\,2,000~days and above binary mass functions $f_{\rm M}$ corresponding to $K_1$~=~6~km~s$^{-1}$ and $e$~=~0.5.  We display this relatively complete parameter space by solid black lines in Fig.~\ref{Latham_fM}.  Enclosed within this area, the SB fraction is 49/554 = 8.7\%\,$\pm$\,1.2\% for our metal-poor subsample ($-$3.0~$\le$~[m/H]~$\le$~$-$0.8).  Meanwhile, the SB fraction within the same region of $P$ and $f_{\rm M}$ is only 38/821 = 4.6\%\,$\pm$\,0.7\% for our metal-rich subsample ($-$0.8~$<$~[m/H]~$\le$~0.5). By focusing on this relatively complete parameter space, we demonstrate that the close binary fraction decreases by a factor of 1.9\,$\pm$\,0.4 at the 3.0$\sigma$ significance level between our metal-poor and metal-rich subsamples.

The sample of SBs with measured orbital solutions is incomplete beyond $P$~$>$~2,000~days (right of black dashed line in Fig.~\ref{Latham_fM}). The handful of systems in this part of the parameter space required substantially more RV measurements and longer timespans to fit the orbits.  For example, the median number and timespan of RV measurements for the 15 long-period SB1s with preliminary orbits are $N_{\rm RV}$~=~57 and $\Delta t$~=~18~yr, respectively, which are considerably larger than the median values of $N_{\rm RV}$ = 12 and $\Delta t$~=~9~yr for the \citet{Latham2002} sample as a whole. In addition, the 17 SB1s without orbital solutions in the \citet{Latham2002} catalog likely have $P$~$>$~2,000~days, but simply lack the number of observations and/or timespan to fit the RVs (see their Fig.~6).

The Carney-Latham SB sample is also slightly biased against very close binaries with $P$~$<$~20~days due to contamination by subgiants and giants.   As stars in very close binaries expand beyond the main-sequence (MS), they undergo Roche-lobe overflow, thereby preventing evolution toward the giant stage.    \citet{Badenes2018} thoroughly discussed this effect of giant evolution truncating the short-period tail of the binary period distribution as a function of giant surface gravity, an indicator of radius. In volume-limited samples of solar-type dwarfs, the very close binary fraction below $P$~$<$~20~days is 4\%\,$\pm$\,1\% \citep{Duquennoy1991,Raghavan2010,Tokovinin2014,Moe2017}.  In our metal-rich subsample with $-$0.8~$<$~[m/H]~$\le$~0.5, however, the observed very close binary fraction is only 13/821 = 1.6\%\,$\pm$\,0.4\% (see systems left of dotted black line in Fig.~\ref{Latham_fM}).  The very close binary fraction in our metal-poor subsample with $-$3.0~$\le$~[m/H]~$\le$~$-$0.8 is lower still at 6/562 = 1.1\%\,$\pm$\,0.4\%, likely due to a larger contamination by giants for systematically older halo stars.  We estimate that the close binary fraction should increase by 1\% and 2\% for our metal-rich and metal-poor subsamples, respectively, in order to correct for this selection bias.  

\subsubsection{Intrinsic Close Binary Fraction}

\begin{figure}[t!]
\centerline{
\includegraphics[trim=0.1cm 0.3cm 0.3cm 0.5cm, clip=true, width=3.5in]{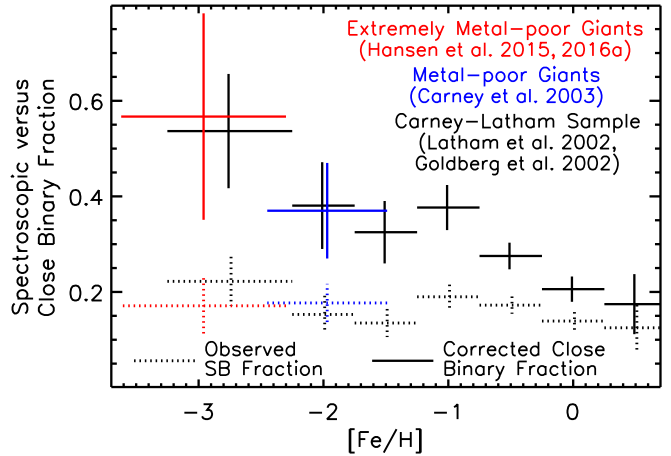}}
\caption{As a function of metallicity, the observed SB fraction (dotted) versus the intrinsic close binary fraction ($P$~$<$~10$^4$~days; $a$~$\lesssim$~10~AU) after correcting for incompleteness and the removal of very close binaries due to giant evolution (solid).  We compare the samples of extremely metal-poor giants \citep[red;][]{Hansen2015,Hansen2016a}, metal-poor giants \citep[blue;][]{Carney2003}, and solar-type stars (mostly dwarfs) with high proper motion in the Carney-Latham survey \citep[black;][]{Latham2002,Goldberg2002}.  Although the observed SB fraction of $\approx$\,15\%\,-\,20\% is relatively independent of metallicity, the true bias-corrected close binary fraction decreases from $F_{\rm close}$ $\approx$ 35\%\,-\,55\% across $-$3.5~$<$~[Fe/H]~$<$~$-$1.0 to $F_{\rm close}$~$\approx$~20\% at [Fe/H]~=~0.0. }
\label{Latham_binfrac}
\end{figure}

In Fig.~\ref{Latham_binfrac}, we show the observed SB fraction as a function of metallicity for the combined \citet{Latham2002} and \citet{Goldberg2002} samples (dotted black data points). The observations are consistent with a constant $\approx$\,15\%\,-\,20\% SB fraction across the full metallicity range $-$3.0~$\le$~[m/H]~$\le$~0.5 as reported in \citet{Latham2002} and \citet{Carney2005}. We correct the observed distribution according to our simulated completeness rates displayed in Fig.~\ref{Latham_RV}.  For example, the observed SB fraction for [m/H]~=~0.0 is 14\%\,$\pm$\,2\%.  For this metallicity, we estimate $\approx$\,70\% of binaries with $P$~$<$~10$^4$~days are detectable as SBs (Fig.~\ref{Latham_RV}), implying a corrected close binary fraction of (0.14\,$\pm$\,0.02)/0.70 = 20\%\,$\pm$\,3\%. We add the 1\% of very close metal-rich binaries ($P$~$<$~20~days) that were excluded due to contamination by subgiants and giants, resulting in our final value of $F_{\rm close}$~=~21\%\,$\pm$\,3\% for [m/H]~=~0.0.  We repeat this procedure for each of the metallicity intervals, but add 2\% to the close binary fraction of metal-poor stars ([m/H]~$\le$~$-$1) to account for the increased contamination by evolved giants in the older metal-poor populations.

 We display in Fig.~\ref{Latham_binfrac} our bias-corrected close binary fraction as a function of metallicity based on the Carney-Latham sample (solid black).   The corrected close binary fraction decreases by a factor of 3.2$_{-0.9}^{+1.9}$ from $F_{\rm close}$ = 54\%\,$\pm$\,12\% at [m/H]~$\approx$~$-$2.7 to  $F_{\rm close}$ = 17\%\,$\pm$\,6\% at [m/H]~$\approx$~0.5. Attempting to fit a constant close binary fraction to the seven black data points in Fig.~\ref{Latham_binfrac} results in a reduced $\chi^2$/$\nu$~=~3.5 with $\nu$~=~6 degrees of freedom.  The probability to exceed this value is $p$~=~0.0016, i.e., the bias-corrected close binary fraction decreases with metallicity at the 3.0$\sigma$ significance level.  This is identical to the level of significance determined by comparing the metal-poor and metal-rich SB fractions across the parameter space in  Fig.~\ref{Latham_fM} that was relatively complete.  

Focusing on a narrower metallicity interval, the close binary fraction decreases by a factor of 2.2$_{-0.6}^{+1.2}$ between [m/H] = $-$1.0 and +0.5 in Fig.~\ref{Latham_binfrac}.  A factor of $\approx$\,2\,-\,4 decrease in the close binary fraction across this metallicity interval, as indicated in \citet{Badenes2018} and measured by us in \S\ref{APOGEE}, is fully consistent with the Carney-Latham observations. We conclude that once corrections for incompleteness and selection biases are considered, the Carney-Latham sample is not only consistent with a large anti-correlation between metallicity and the close binary fraction, but actually supports such a trend at the 3.0$\sigma$ significance level.  

\subsection{Metal-poor Giants}
\label{Giants}

The SB fractions of metal-poor giants \citep{Carney2003} and extremely metal-poor giants enriched with r-process elements or carbon \citep{Hansen2015,Hansen2016a} are $\approx$\,15\%\,-\,20\%. These values are consistent with the observed SB fractions of metal-poor dwarfs in the halo \citep{Latham2002,Carney2005}.  We re-emphasize that the observed SB fractions are lower limits to the true close binary fractions, especially for metal-poor stars that have weaker absorption lines. In the following, we account for incompleteness within these additional samples of metal-poor stars in order to compute their intrinsic close binary fractions.

\subsubsection{Carney et al. (2003) Sample}

\citet{Carney2003} obtained a median of $N_{\rm RV}$~=~13 RV measurements of 91 metal-poor field giants with an average precision of $\langle \sigma_{\rm RV} \rangle$~=~0.65~km~s$^{-1}$ and a median timespan of $\Delta t$~=~13.8~yr. This is similar in frequency but with improved sensitivity and duration compared to the \citet{Latham2002} survey of metal-poor dwarfs in the halo.   The metallicities of the giants span $-$4.0~$<$~[Fe/H]~$<$~$-$0.9, resulting in a mean and 1$\sigma$ spread of [Fe/H] = $-$2.0\,$\pm$\,0.5.  These metal-poor giants are some of the oldest stars in the galaxy, and therefore have masses $M_1$ $\approx$ 0.8\,-\,1.1\,\Msun\ corresponding to MS-turnoff ages of $\tau$~$\approx$~7\,-\,13 Gyr. \citet{Carney2003} identified 16 SB1s in their sample and measured robust orbital periods spanning $P$~$\approx$~40\,-\,5,200~days for 14 of them.  As shown in Fig.~\ref{Latham_binfrac}, the observed SB fraction is 16/91 = 18\%\,$\pm$\,4\%. 

The most luminous giants in the \citet{Carney2003} sample exhibit significant RV jitter due to radial pulsations, convective instabilities in the tenuous upper layers, or intermittent starspots modulated by rotation.  They found $\approx$\,40\% of giants with absolute magnitudes $M_{\rm V}$~$<$~$-$1.4 display detectable RV jitter $\sigma_{\rm RV,jitter}$~$\gtrsim$~1~km~s$^{-1}$. \citet{Hekker2008} later showed that non-periodic RV jitter occurs in smaller, less luminous giants, but simply the magnitude increases from $\sigma_{\rm RV,jitter}$~=~0.03~km~s$^{-1}$ at log\,$g$~$\approx$~3.0 to $\sigma_{\rm RV,jitter}$~=~0.3~km~s$^{-1}$ at log\,$g$~$\approx$~1.5.  Stochastic variations in the RVs due to intrinsic fluctuations in the atmospheres inhibit the detection of SBs with small velocity semi-amplitudes.  We therefore remove the nine giants in the \citet{Carney2003} sample that exhibit significant RJ jitter (dark systems in their Fig.~8).  One of these objects, HD\,218732, is also an SB in which the velocity semi-amplitude $K_1$~=~2.9~km~s$^{-1}$ induced by the companion is larger than the RV jitter $\sigma_{\rm RV,jitter}$~$\approx$~1~km~s$^{-1}$. The observed SB fraction for our refined subsample remains unchanged at 15/82 = 18\%\,$\pm$\,4\%.

The metal-poor giants in the \citet{Carney2003} sample also span a broad range of radii $R_1$~=~4.3\,-\,112\,\Rsun, providing a mean of $\langle R_1 \rangle$ = 23\,\Rsun.  Adopting typical parameters $M_1$~$\approx$ 1.0\,\Msun\ and $q$~=~0.5, then very close binaries with $P$~$\lesssim$~35~days would have already filled their Roche lobes by the time the primaries evolved to $R_1$ = 23\,\Rsun\ \citep{Eggleton1983}.  The \citet{Carney2003} sample is therefore significantly biased against very close binaries with $P$~$\lesssim$~35~days. Their closet binary, i.e., BD~+13$^{\circ}$3683 with $P$~$\approx$~40~days, happens to contain the smallest giant ($R_1$~=~4.3\Rsun) in their sample.  We correct for incompleteness and this selection bias using two different methods described below.

First, we perform a Monte Caro simulation as done in \S\ref{LathamCorrections} to measure the completeness rate, but adopt $N_{\rm RV}$~=~13, $\Delta t$~=~13.8~yr, and $\langle \sigma_{\rm RV} \rangle$~=~0.65~km~s$^{-1}$ to match the median cadence and sensitivity of the \citet{Carney2003} observations. We increase the circularization period to  $P_{\rm circ}$~=~100~days in Eqn.~\ref{tide} to account for the larger tidal radius of the giants.  We also generate close binaries across the 
interval $P$~=~35\,-\,10$^4$~days because very close binaries with $P$~$<$~35~days have effectively been removed from the \citet{Carney2003} sample of giants. Of all the metal-poor giants with companions across $P$~=~35\,-\,10$^4$~days, we calculate 55\% would have been detected as SBs by \citet{Carney2003} at the $>$5$\sigma$ significance level.  This is slightly lower than the completeness rate of 62\% for $\langle \sigma_{\rm RV} \rangle$~=~0.65~km~s$^{-1}$ inferred from Fig.~\ref{Latham_RV}.  Despite the increased timespan of the \citet{Carney2003} observations, the removal of very close binaries with $P$~$<$~35~days, which are easier to detect, causes the overall completeness rate to decrease. The details of tidal circularization during the giant phase have a negligible effect on our corrections for incompleteness; we repeat our Monte Carlo simulation with $P_{\rm circ}$~=~20 and 500 days, and calculate completeness rates of 54\% and 56\%, respectively.  The corrected binary fraction of metal-poor giants in the \citet{Carney2003} sample is (0.18\,$\pm$\,0.04)/0.55 = 33\%\,$\pm$\,7\% across $P$~=~35\,-\,10$^4$~days.  According to our adopted log-normal period distribution for solar-type binaries, 17\% of close binaries with log\,$P$\,(days)~=~0\,-\,4 have very short periods $P$~=~1\,-\,35 days.  The close binary fraction (log\,$P$~=~0\,-\,4; $a$ $\lesssim$~10\,AU) of metal-poor {\it dwarfs} is therefore $F_{\rm close}$ = (0.33\,$\pm$\,0.07)/0.83 = 40\%\,$\pm$\,8\% after accounting for the bias against very close binaries in giant systems.

\begin{figure}[t!]
\centerline{
\includegraphics[trim=0.3cm 0.1cm 0.3cm 0.2cm, clip=true, width=3.5in]{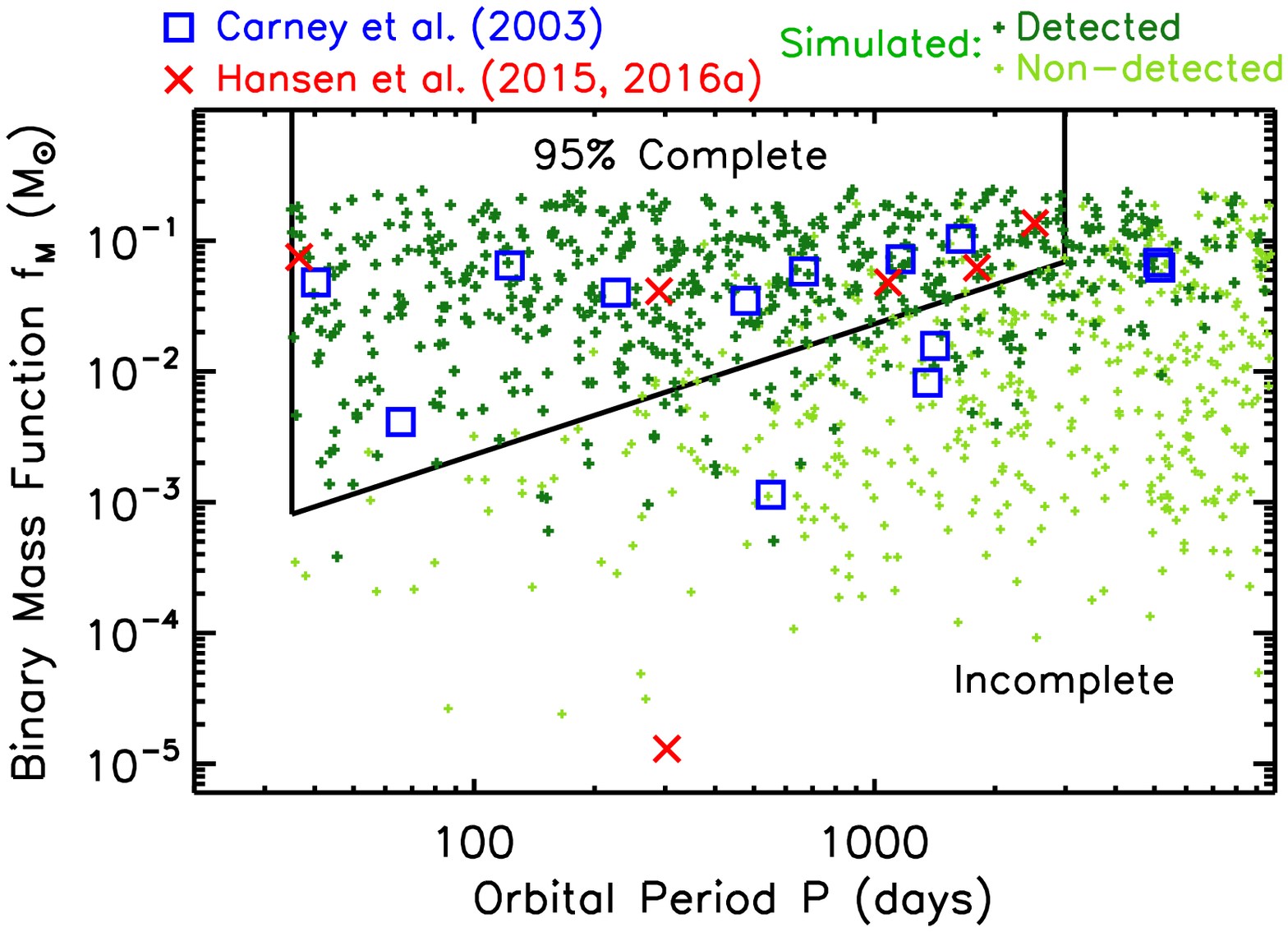}}
\caption{Similar to Fig.~\ref{Latham_fM}, but for the 13 SBs with orbital solutions and no significant RV jitter in the \citet{Carney2003} sample of metal-poor giants (blue squares) and 6 SBs with orbital solutions in the \citet{Hansen2015,Hansen2016a} samples of extremely metal-poor giants chemically enriched with r-process elements or carbon (red crosses).  We also display a random subset of 1,000 binaries from our Monte Carlo simulations (green pluses) that match the cadence and sensitivity of the \citet{Carney2003} observations.  The simulated binaries that exhibit RV variability above a $>$5$\sigma$ significance level are indicated with darker, larger symbols. The observations are $\approx$\,95\% complete across $P$~=~35\,-\,3,000 days and above binary mass functions $f_{\rm M}$ corresponding to $K_1$~=~7~km~s$^{-1}$ and $e$~=~0.5 (black lines).  }
\label{Carney_fM}
\end{figure}

Second, we examine in Fig.~\ref{Carney_fM} the binary mass functions and periods of the 13 SBs with measured orbital elements and no significant RV jitter in \citet{Carney2003}, similar to our analysis of the the Carney-Latham SBs (see Fig.~\ref{Latham_fM}). We also show in Fig.~\ref{Carney_fM} a random subset of 1,000 binaries spanning $P$~=~35\,-\,10$^4$ days from our Monte Carlo simulation with $P_{\rm circ}$~=~100~days, indicating those that were detectable above the $>$5$\sigma$ level with darker, thicker symbols.  The observed density of SBs in the $P$ versus $f_M$ parameter space follow our simulated detections quite well.  Our analysis confirms that the \citet{Carney2003} SB survey is incomplete toward long periods and small binary mass functions. In our Monte Carlo model, 37\% of binaries have $P$~=~35\,-\,3,000~days and binary mass functions $f_M$ greater than that corresponding to $K_1$~=~7~km~s$^{-1}$ and $e$~=~0.5.  We indicate this parameter space, which is $\approx$\,95\% complete, in Fig.~\ref{Carney_fM}.  We find eight of the SBs from the \citet{Carney2003} sample are located within this relatively complete region, indicating a corrected binary fraction of 8/82/0.37/0.95 = 28\%\,$\pm$\,10\%.  After accounting for the bias against very close binaries with $P$~$<$~35~days, the close binary fraction of metal-poor dwarfs is $F_{\rm close}$ =(0.28\,$\pm$\,0.10)/0.83 = 34\%\,$\pm$\,12\%.

The bias-corrected close binary fraction determined from our forward-modeling method ($F_{\rm close}$~=~40\%\,$\pm$\,8\%) is consistent with our inversion technique ($F_{\rm close}$~=~34\%\,$\pm$\,12\%).  We adopt an average of $F_{\rm close}$~=~37\%\,$\pm$\,10\%, and present the result in Fig.~\ref{Latham_binfrac}.  The bias-corrected close binary fraction measured for the \citet{Carney2003} sample of metal-poor giants matches the  close binary fraction determined for metal-poor halo stars with high proper motion in the Carney-Latham sample.

\subsubsection{Hansen et al. (2015, 2016a) Samples}

We next combine the samples of extremely metal-poor giants enriched with r-process elements \citep{Hansen2015} and with carbon \citep{Hansen2016a}.  We do not include extremely metal-poor giants enriched with s-process elements, e.g., barium, which exhibit a very large SB fraction of $\approx$80\% and are clearly the result of post-MS binary mass transfer \citep{Jorissen1998,Lucatello2005,Hansen2016b}.  \citet{Hansen2015,Hansen2016a} concluded the abundances of extremely metal-poor giants enriched with r-process elements and carbon are primordial, i.e., the  enhanced elements were imprinted on their natal molecular clouds.  Our combined sample contains 41 extremely metal-poor giants that span $-$5.8~$<$~[Fe/H]~$<$~$-$1.6, providing a mean and 1$\sigma$ spread of [Fe/H] = $-$3.0\,$\pm$\,0.7. Within this sample, \citet{Hansen2015,Hansen2016a} found seven SBs, six of which have orbital solutions.  We display the observed SB fraction of 7/41 = 17\,$\pm$\,6\% in Fig.~\ref{Latham_binfrac}.

\citet{Hansen2015,Hansen2016a} observed their 41 targets with varying cadence.  In particular, 11 of their extremely metal-poor giants were observed only $N_{\rm RV}$ = 2\,-\,7 times.  For comparison, both \citet{Latham2002} and \citet{Carney2003} obtained at least $N_{\rm RV}$ $\ge$ 7 measurements for each of their targets, $\approx$\,90\% of which were observed $N_{\rm RV}$ $\ge$ 9 times.  A small number $N_{\rm RV}$ = 2\,-\,7 of RV measurements reduces the probability of detecting RV variability, and makes it nearly impossible to fit robust orbital solutions.  We therefore remove the 11 objects with $N_{\rm RV}$ = 2\,-\,7, none of which were identified as SBs, leaving 30 extremely metal-poor giants in our culled sample.

The mean RV precision of the extremely metal-poor giants in the \citet{Hansen2015,Hansen2016a} samples ranged significantly from $\langle \sigma_{\rm RV} \rangle$ = 0.012~km~s$^{-1}$ to 2.5~km~s$^{-1}$.  With such a large variance in $\langle \sigma_{\rm RV} \rangle$, a Monte Carlo simulation with a single value of $\langle \sigma_{\rm RV} \rangle$ is no longer valid.  We instead rely on the measured binary mass functions $f_M$ and periods $P$ of the 6 SBs with orbital solutions, which are displayed in Fig.~\ref{Carney_fM}.  One of the SBs,  HE\,1523−0901, has an extremely small binary mass function of $f_{\rm M}$ = 1.3$\times$10$^{-5}$\,\Msun\ \citep{Hansen2015}.  This object was observed with superior precision $\langle \sigma_{\rm RV} \rangle$ = 0.016~km~s$^{-1}$ and more times ($N_{\rm RV}$ = 34) than any other targets in the \citet{Hansen2015,Hansen2016a} samples. If the other targets were SBs with such small binary mass functions, they would not be detected.  

Meanwhile, the other five SBs with orbital solutions in the \citet{Hansen2015,Hansen2016a} survey extend across the upper middle region in Fig.~\ref{Carney_fM}, spanning $f_{\rm M}$ $\approx$ 0.04\,-\,0.14\,\Msun\ and $P$ $\approx$ 37\,-\,2,500 days.  The fact that 5 of the 30 extremely metal-poor giants with $N_{\rm RV}$~$\ge$~8 are SBs with such large binary mass functions strongly suggests the intrinsic close binary fraction is particularly large.  These five SBs occupy the same parameter space that is $\approx$\,95\% complete according to our Monte Carlo model that simulates the cadence and sensitivity of the \citet{Carney2003} observations.  Although the \citet{Hansen2015,Hansen2016a} surveys had variable precision, we also expect this parameter space to be $\approx$\,95\% complete.  We therefore use the same inversion technique to correct for incompleteness, resulting in an intrinsic binary fraction of 5/30/.37/.95 = 47\%\,$\pm$\,19\% across $P$~=~35\,-\,10$^4$~days.  After accounting for the bias against very close binaries with $P$~$<$~35~days, the primordial close binary fraction of extremely metal-poor dwarfs is $F_{\rm close}$ = (0.47\,$\pm$\,0.19)/0.83 = 57\%\,$\pm$\,22\%. We display our result in Fig.~\ref{Latham_binfrac}, which is consistent with our measurement of  $F_{\rm close}$ = 54\%\,$\pm$\,12\% for extremely metal-poor stars with [m/H] = $-$2.7\,$\pm$\,0.7 selected from the Carney-Latham sample.  The close binary fraction of metal-poor dwarfs, metal-poor giants, and extremely metal-poor giants are all $F_{\rm close}$~$\approx$~35\%\,-\,55\%, substantially larger than the close binary fraction $F_{\rm close}$~$\approx$~20\% of solar-metallicity FGK dwarfs in the disk.

\section{APOGEE RV Variables}
\label{APOGEE}

\subsection{Sample Selection and Description}

\begin{figure}[t!]
\centerline{
\includegraphics[trim=0.2cm 0.2cm 0.3cm 0.3cm, clip=true, width=3.4in]{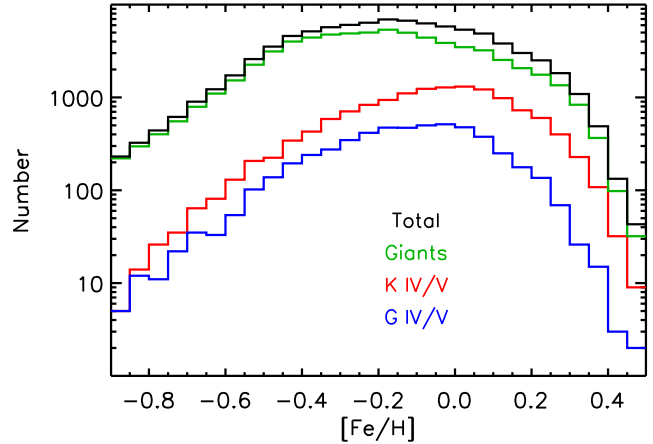}}
\caption{Metallicity distribution of APOGEE stars in our selected total sample (black) and giant (green), K\,IV/V (red), and G\,IV/V (blue) subsamples.}
\label{Zdist}
\end{figure}

The SDSS-IV/APOGEE near-infrared spectroscopic survey (data release 13) measured the effective temperatures, surface gravities, metallicities, and RVs of $\approx$\,164,000 stars in various environments including the galactic disk, bulge, and halo \citep{Zasowski2013,Holtzman2015,Nidever2015,Albareti2017}.  After calibrating their observations to both synthetic spectra and empirical relations, APOGEE measured the stellar parameters to high precision, e.g., $\delta T_{\rm eff}$~$\approx$~90\,K, $\delta$log\,$g$~$\approx$~0.11\,dex, and $\delta$[Fe/H]~$\approx$~0.15\,dex \citep{Holtzman2015}. In their study, \citet{Badenes2018} removed targets in open clusters and stars with effective temperatures or surface gravities that were inadequately measured, leaving 122,141 objects. They then examined the spectra and RV measurements for each star, keeping only the individual visits with spectral S/N~$>$~40. A total of 91,246 stars with $N_{\rm RV}$~$\ge$~2 high-quality RV measurements (78\% which have $N_{\rm RV}$~$\ge$~3~epochs) were included in the \citet{Badenes2018} analysis. We further remove the 2,893 stars (mostly giants) with [Fe/H]~$<$~$-$0.9 and 7 systems with [Fe/H]~$>$~0.5, leaving 88,346 stars across the interval $-$0.9~$\le$~[Fe/H]~$\le$~0.5 in our final sample.  The metallicity distribution is adequately modeled by a Gaussian with mean of $\langle$[Fe/H]$\rangle$~=~$-$0.16 and dispersion of $\sigma_{\rm [Fe/H]}$~=~0.26~dex (see Fig.~\ref{Zdist}).  

We divide our sample according to the measured surface gravities and effective temperatures.  Of the 88,346 stars in our full sample, 20,649 are MS dwarfs or Hertzsprung gap (HG) subgiants with 3.2~$\le$~log\,$g$~$<$~5.0 while the remaining 67,697 are giants with 0.1~$<$~log\,$g$~$<$~3.2.  The giants mostly have primary masses $M_1$~$\approx$~1.1\,-\,2.0\,\Msun\ with an average of $M_1$~$\approx$~1.5\,\Msun\  (see Fig.~2 and Fig.~4 in \citealt{Badenes2018}).  Our giant subsample includes both normal and red clump giants.  APOGEE red clump giants were targeted differently \citep{Zasowski2013}, and as a result are slightly biased against close binaries \citep{Badenes2018}.  Fortunately, only $\approx$\,20\% of the APOGEE giants occupy the red clump \citep{Badenes2018}, and so the bias in the RV variability fraction can at most be 20\% for our overall giant subsample.  For our MS/HG stars, a majority (13,864 objects; 67\%) have effective temperatures $T_{\rm eff}$~=~4,000\,-\,5,000\,K, corresponding roughly to K\,IV/V stars with primary masses $M_1$~$\approx$~0.6\,-\,1.1\,\Msun.  Another 5,375 MS/HG stars (26\%) have $T_{\rm eff}$~=~5,000\,-\,6,000\,K, corresponding approximately to G\,IV/V stars with $M_1$~$\approx$~0.9\,-\,1.4\,\Msun.  The remaining 1,410 MS/HG stars (7\%) are either cool early-M dwarfs ($T_{\rm eff}$~=~3,500\,-\,4,000\,K) or hot late-F stars ($T_{\rm eff}$~=~6,000\,-\,6,500\,K). In the following, we separately analyze our three main subsamples: giants ($N$~=~67,697), K\,IV/V stars ($N$~=~13,864), and G\,IV/V stars ($N$~=~5,375). As shown in Fig.~\ref{Zdist}, giants dominate the total sample and peak at [Fe/H]~$\approx$~$-$0.2.  Meanwhile, K\,IV/V and G\,IV/V stars are systematically younger and peak at slightly larger metallicities [Fe/H]~$\approx$~0.0.

The resolution R~$\approx$~22,500 (13~km~s$^{-1}$) of the APOGEE spectra is similar to the \citet{Latham2002} and \citet{Carney2005} observations (R~$\approx$~35,000; 9~km~s$^{-1}$).  However, our selected subsample of high-quality APOGEE spectra has an average $\langle$S/N$\rangle$~$\approx$~110, which is a factor of six times larger than the mean $\langle$S/N$\rangle$~$\approx$~15\,-\,20  of the Carney-Latham observations.  The average RV measurement uncertainties are $\langle\sigma_{\rm RV,meas}\rangle$~=~0.02~km~s$^{-1}$, 0.04 km~s$^{-1}$, and 0.05 km~s$^{-1}$ for our giant, K\,IV/V, and G\,IV/V subsamples, respectively. For our K\,IV/V subsample,  the 1\,-\,99 percentile range in the RV measurement uncertainties is $\sigma_{\rm RV,meas}$~=~0.006\,-\,0.152~km~s$^{-1}$. The APOGEE RVs are substantially more precise than the mean RV uncertainties $\langle\sigma_{\rm RV}\rangle$~=~0.5\,-\,1.0~km~s$^{-1}$ in the \citet{Latham2002} sample (see Fig.~\ref{Latham_RV}).

The number and timespan of the APOGEE RV observations are comparatively smaller, but fortunately they do not vary significantly with metallicity.  For metal-poor ($-$0.9~$<$~[Fe/H]~$<$~$-$0.7) and metal-rich (0.3~$<$~[Fe/H]~$<$~0.5) K\,IV/V stars, the mean numbers of RV measurements are $\langle N_{\rm RV} \rangle$ = 2.93 and 3.04, respectively, and the median timespans are $\Delta t$ = 33~days and 37~days, respectively. We find similar results for the giant and G\,IV/V subsamples. The APOGEE sample is incomplete toward SBs with longer periods due to the limited timespan, but the superior RV precision helps significantly to offset this effect.  The timespans of the APOGEE observations vary substantially from system to system.  For K\,IV/V stars, the 15\,-\,85 percentile range in the timespan is $\Delta t$~=~23\,-\,305 days.  When correcting for incompleteness (see below), we assume the cadence is independent of metallicity but account for the small number of observations and wide distribution in the timespans.

The RV uncertainties in our APOGEE sample decrease with metallicity, similar to the trend in the Carney-Latham sample. In particular, the mean RV measurement uncertainty for K\,IV/V stars decreases by a factor of $\approx$\,2.9 from $\langle\sigma_{\rm RV,meas}\rangle$~=~0.08 km~s$^{-1}$ across $-$0.9~$<$~[Fe/H]~$<$~$-$0.7 to $\langle\sigma_{\rm RV,meas}\rangle$~=~0.03 km~s$^{-1}$ across 0.3~$<$~[Fe/H]~$<$~0.5. It is therefore crucial that we do not follow \citet{Latham2002} and \citet{Carney2005} by defining the SB fraction according to those systems that exhibit RV variability above some statistical significance.

\begin{figure}[t!]
\centerline{
\includegraphics[trim=0.3cm 0.2cm 0.3cm 0.2cm, clip=true, width=3.5in]{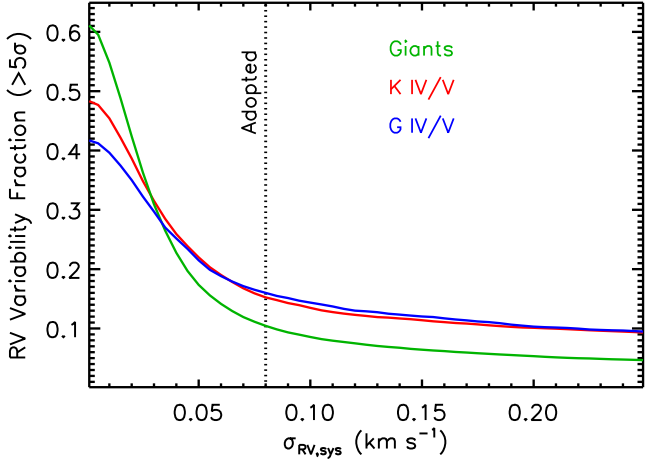}}
\caption{The fraction of APOGEE stars that exhibit RV variability above the 5$\sigma$ significance level as a function of an assumed value of systematic uncertainty $\sigma_{\rm RV,sys}$ for our giant (green), K\,IV/V (red), and G\,IV/V (blue) subsamples.
Assuming no systematic uncertainty, a significant fraction of APOGEE stars are spurious RV variables due to either RV jitter and/or the APOGEE pipeline underestimating the true RV uncertainties. The curves begin to significantly flatten beyond $\sigma_{\rm RV,sys}$~$\gtrsim$~0.08~km s$^{-1}$, and so we add a systematic uncertainty of $\sigma_{\rm RV,sys}$~$=$~0.08~km s$^{-1}$ (dotted) in quadrature with all the measurement uncertainties.}
\label{RVsys}
\end{figure}

Another reason to avoid this definition is because a substantial fraction of our giants are RV variables due to RV jitter. The mean surface gravity of giants in our sample is log\,$g$~=~2.4, which exhibit an average RV jitter of $\sigma_{\rm RV,jitter}$=~0.07~km~s$^{-1}$ according to Fig.~3 in \citet{Hekker2008}. In addition, we find the APOGEE pipeline underestimates the true RV uncertainties for systems with very small measurement uncertainties $\sigma_{\rm RV,meas}$~$\lesssim$~0.1~km~s$^{-1}$. Many RV variables with very small amplitudes are actually spurious.  To account for both RV jitter and systematic effects in the APOGEE pipeline, we add a systematic uncertainty of $\sigma_{\rm RV,sys}$ in quadrature with each of the measurement uncertainties $\sigma_{\rm RV,meas}$.  As shown in Fig.~\ref{RVsys}, the fraction of systems that exhibit RV variability above the 5$\sigma$ significance level decreases as the assumed value for $\sigma_{\rm RV,sys}$ increases.  The curves in Fig.~\ref{RVsys} rapidly decline and then begin to flatten beyond $\sigma_{\rm RV,sys}$~$\gtrsim$~0.08~km~s$^{-1}$.  We therefore adopt a systematic uncertainty of $\sigma_{\rm RV,sys}$~=~0.08~km~s$^{-1}$ for all three subsamples. Systems that exhibit statistically significant RV variability well above the total RV uncertainty $\sigma_{\rm RV,tot}$ = ($\sigma_{\rm RV,meas}^2$\,+\,$\sigma_{\rm RV,sys}^2$)$^{\nicefrac{1}{2}}$ are real SBs.

\vspace*{0.6cm}

\subsection{RV Variability Fractions}

\begin{figure}[t!]
\centerline{
\includegraphics[trim=0.3cm 0.2cm 0.3cm 0.2cm, clip=true, width=3.4in]{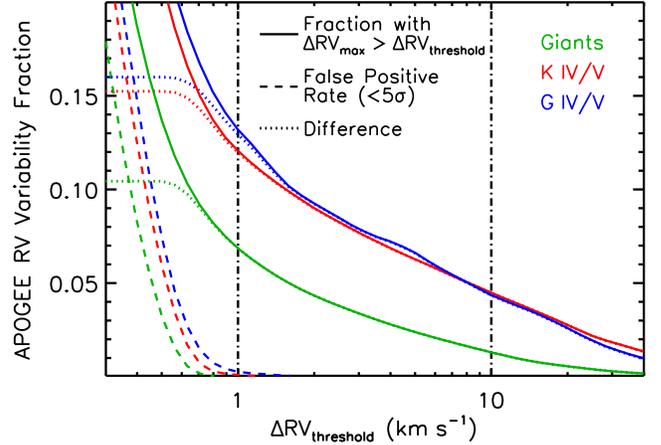}}
\caption{The fraction of APOGEE stars that exhibit RV variability above $\Delta$RV$_{\rm max}$~$>$~$\Delta$RV$_{\rm threshold}$ (solid), the fraction of systems that have $\Delta$RV$_{\rm max}$~$>$~$\Delta$RV$_{\rm threshold}$ but are consistent with constant RV within the 5$\sigma$ tolerance level (dashed), and the difference between these two distributions (dotted) for our giant (green), K\,IV/V (red) and G\,IV/V (blue) subsamples.  \citet{Badenes2018} adopted a conservative threshold of $\Delta$RV$_{\rm threshold}$~=~10~km~s$^{-1}$ (right dash-dotted line) to be 100\% certain all RV variables were real SBs. We adopt a threshold of $\Delta$RV$_{\rm threshold}$~=~1~km~s$^{-1}$ (left dash-dotted line) in order to retain a significant majority of the real SBs while simultaneously keeping the false positive rate below $<$\,1\% for all three subsamples and across all metallicities.}
\label{RVthresh}
\end{figure}

As advocated in \citet{Badenes2018}, we instead measure the RV variability fraction according to the fraction of stars that exhibit a maximum difference in radial velocities $\Delta$RV$_{\rm max}$ between any two epochs above a certain threshold $\Delta$RV$_{\rm threshold}$.  Based on this definition, the close binary fraction is directly proportional to the observed RV variability fraction, i.e.,  corrections for incompleteness are independent of metallicity.   In Fig.~\ref{RVthresh}, we show the RV variability fraction as a function of $\Delta$RV$_{\rm threshold}$ for our giant, K\,IV/V, and G\,IV/V subsamples.  For the K\,IV/V and G\,IV/V subsamples, the RV variability fraction increases from $\approx$\,4.4\% for $\Delta$RV$_{\rm max}$~$>$~10 km\,s$^{-1}$ to $\approx$\,12\%\,-\,13\% for $\Delta$RV$_{\rm max}$~$>$~1 km\,s$^{-1}$.  The similarity in their RV variability distributions, both in terms of functional form and normalization, suggests K\,IV/V stars and G\,IV/V stars have the same close binary fraction and period distribution.  The relative change in the close binary fraction between these two subsamples can at most be $\Delta F_{\rm close}$/$F_{\rm close}$ $<$ 20\%  (2$\sigma$ confidence level).  This is consistent with previous studies that show the close binary fraction changes only slightly between early-M dwarfs and G-dwarfs \citep{Fischer1992,Raghavan2010,Clark2012,Duchene2013,Murphy2018}. The RV variability fraction for our giant subsample is considerably lower, increasing from $\approx$\,1.3\% for $\Delta$RV$_{\rm max}$~$>$~10~km\,s$^{-1}$ to $\approx$\,6.9\% for $\Delta$RV$_{\rm max}$~$>$~1~km\,s$^{-1}$.   As discussed in \S\ref{Spectroscopic} and \citet{Badenes2018}, giant evolution truncates the short-period tail of the binary period distribution, thereby removing SBs with large RV amplitudes.

We display the false positive rate in Fig.~\ref{RVthresh}, i.e., the fraction of systems that have both $\Delta$RV$_{\rm max}$~$>$~$\Delta$RV$_{\rm threshold}$ and a difference in RVs that are discrepant with each other by {\it less} than 5$\sigma$. We also display the difference between the RV variability fraction and false positive rate, which provides the real SB fraction.  \citet{Badenes2018} chose a very conservative threshold of $\Delta$RV$_{\rm threshold}$~$=$~10~km\,s$^{-1}$ in order to be certain all of their RV variables were real binaries (see their Fig.~9).  Indeed, we find 100\% of RV variables with $\Delta$RV$_{\rm max}$~$>$~10~km\,s$^{-1}$ are real, i.e., the false positive rate is zero for all three subsamples (see Fig.~\ref{RVthresh}).  The false positive rate remains zero down to $\Delta$RV$_{\rm threshold}$~$=$~2~km~s$^{-1}$ and then steadily increases below $\Delta$RV$_{\rm threshold}$~$\lesssim$~1~km~s$^{-1}$.  Systems with $\Delta$RV$_{\rm max}$~$\lesssim$~0.4~km\,s$^{-1}$ are consistent with constant RV or exhibit RV jitter. 
 
We adopt a threshold of $\Delta$RV$_{\rm threshold}$~=~1~km~s$^{-1}$ (Fig.~\ref{RVthresh}), but we also keep track of large-amplitude RV variables with $\Delta$RV$_{\rm max}$~$>$~3~km~s$^{-1}$ and $\Delta$RV$_{\rm max}$~$>$~10~km~s$^{-1}$ to perform consistency checks (see below).  A significant majority ($\approx$\,70\%\,-\,80\%) of the real SBs have $\Delta$RV$_{\rm max}$~$>$~1~km~s$^{-1}$. The false positive rate is also negligible above $\Delta$RV$_{\rm max}$~$>$~1~km~s$^{-1}$, e.g., 0.0\%, 0.1\% and 0.3\% for our giant, K\,IV/V, and G\,IV/V subsamples, respectively.  Our threshold of $\Delta$RV$_{\rm threshold}$~=~1~km~s$^{-1}$ is well above the systematic uncertainty $\sigma_{\rm RV,sys}$~$\approx$~0.08~km~s$^{-1}$. The few false positives with $\Delta$RV$_{\rm max}$~$\approx$~1.0\,-\,1.5~km\,s$^{-1}$ simply have larger measurement uncertainties $\sigma_{\rm RV,meas}$~$\approx$~0.2~km~s$^{-1}$ compared to average.  The false positive rate increases slightly toward lower metallicities for our adopted threshold. Nonetheless, the false positive rate is extremely small across all metallicities, especially compared to the RV variability fraction.  For instance, the false positive rate for metal-poor K\,IV/V stars with $-$0.9~$<$~[Fe/H]~$<$~$-$0.5 is 0.8\% above $\Delta$RV$_{\rm max}$~$>$~1~km\,s$^{-1}$.  For this same metal-poor subset, the ratio of the false positive rate to RV variability fraction is only 4.3\%.  In other words, $\approx$\,96\% of metal-poor K\,IV/V RV variables with $\Delta$RV$_{\rm max}$~$>$~1~km\,s$^{-1}$ are real SBs. A systematic uncertainty of $\delta F_{\rm close}$/$F_{\rm close}$~$\approx$~4\% in the inferred close binary fraction due to spurious RV variables is much smaller than the measurement uncertainties and other sources of systematic error (see below).

\subsection{Variations with Metallicity}
\label{APOGEEMetallicity}

\begin{figure}[t!]
\centerline{
\includegraphics[trim=0.3cm 0.2cm 0.3cm 0.2cm, clip=true, width=3.4in]{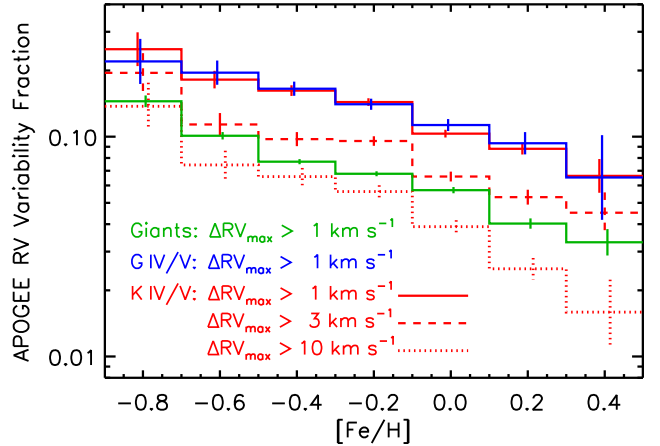}}
\caption{The fraction of APOGEE stars that exhibit RV variability above $\Delta$RV$_{\rm max}$~$>$~1~km~s$^{-1}$ (solid) for our giant (green), G\,IV/V (blue), and K\,IV/V (red) subsamples.  We also display the fraction of K\,IV/V stars with $\Delta$RV$_{\rm max}$~$>$~3~km~s$^{-1}$ (dashed red) and $\Delta$RV$_{\rm max}$~$>$~10~km~s$^{-1}$ (dotted red). The RV variability fraction decreases with metallicity at a similar rate for all three subsamples and RV thresholds.  Combining the giant, G\,IV/V, and K\,IV/V subsamples, the RV variability fraction decreases by a factor of 4.0\,$\pm$\,0.5 across $-$0.9~$<$~[Fe/H]~$<$~0.5 at the 22$\sigma$ confidence level.}
\label{DeltaRV}
\end{figure}

As displayed in Fig.~\ref{DeltaRV}, the fraction of APOGEE stars that exhibit RV variability above $\Delta$RV$_{\rm max}$~$>$~1~km~s$^{-1}$ decreases dramatically with metallicity for all three subsamples. For K\,IV/V stars, the RV variability fraction decreases by a factor of 3.8$_{-0.9}^{+1.2}$ from 25\%\,$\pm$\,5\% across $-$0.9~$<$~[Fe/H]~$<$~$-$0.7 to 6.6\%\,$\pm$\,1.3\% across 0.3~$<$~[Fe/H]~$<$~0.5.  Attempting to fit a uniform RV variability fraction for K\,IV/V stars across the seven metallicity bins in Fig.~\ref{DeltaRV} results in a reduced $\chi^2$/$\nu$~=~19.7 with $\nu$~=~6 degrees of freedom.  The probability to exceed this value is $p$ = 4$\times$10$^{-23}$, i.e., the RV variability fraction of K\,IV/V stars decreases with metallicity at the 9.9$\sigma$ significance level.  The G\,IV/V histogram in Fig.~\ref{DeltaRV} is consistent with the K\,IV/V histogram, but has larger uncertainties due to the smaller sample size.  The RV variability fraction of giants is measurably smaller due to the effective removal of very close binaries, but nonetheless exhibits the same metallicity trend.  The giant RV variability fraction decreases by a factor of 4.4$_{-0.6}^{+0.8}$ from 14.5\%\,$\pm$\,0.9\% at [Fe/H] $\approx$ $-$0.8 to 3.3\%\,$\pm$\,0.5\% at [Fe/H]~$\approx$~0.4.  A model of a uniform RV variability fraction for giants results in an even larger reduced $\chi^2$/$\nu$~=~62.1 that can be rejected at the 18.6$\sigma$ confidence level ($p$~=~2$\times$10$^{-77}$). By combining the results from our three independent subsamples, the RV variability fraction decreases by a factor of 4.0\,$\pm$\,0.5 across $-$0.9~$<$~[Fe/H]~$<$~0.5 at the 21.9$\sigma$ significance level.  

The relative decrease in the RV variability fraction as a function of metallicity is consistent among our K\,IV/V, G\,IV/V, and giant subsamples.  This indicates the slope of the anti-correlation between the close binary fraction and metallicity is similar across primary masses $M_1$~$\approx$~0.6\,-\,1.5\,\Msun. The consistency also suggests the binary fraction decreases with metallicity at a similar rate for both very close companions orbiting small MS/HG stars and for slightly wider companions orbiting larger giants.

We also display in Fig.~\ref{DeltaRV} the fraction of K\,IV/V stars with $\Delta$RV$_{\rm max}$~$>$~3~km~s$^{-1}$ and $\Delta$RV$_{\rm max}$~$>$~10~km~s$^{-1}$, which both exhibit the same metallicity trend as K\,IV/V binaries with smaller RV amplitudes. Utilizing the K\,IV/V histogram with $\Delta$RV$_{\rm max}$~$>$~1~km~s$^{-1}$ as a template, we multiply this distribution by a reduction factor $R$ to fit the other K\,IV/V histograms.  We measure $R_{\rm 3\,to\,1}$ = $N$($\Delta$RV$_{\rm max}$~$>$~3~km~s$^{-1}$)/$N$($\Delta$RV$_{\rm max}$~$>$~1~km~s$^{-1}$) = 0.65\,$\pm$\,0.03 with goodness-of-fit parameter $\chi^2$/$\nu$ = 0.43 ($p$~=~0.86).  Similarly, we fit $R_{\rm 10\,to\,1}$ = $N$($\Delta$RV$_{\rm max}$~$>$~10~km~s$^{-1}$)/$N$($\Delta$RV$_{\rm max}$~$>$~1~km~s$^{-1}$) = 0.38\,$\pm$\,0.02 with $\chi^2$/$\nu$~=~1.9 ($p$~=~0.08).  If spurious RV variables with $\Delta$RV$_{\rm max}$~=~1\,-\,3~km~s$^{-1}$ had significantly contaminated metal-poor systems, we would have expected the $\Delta$RV$_{\rm max}$~$>$~1~km~s$^{-1}$ distribution to be steeper than the $\Delta$RV$_{\rm max}$~$>$~3~km~s$^{-1}$ distribution. Instead, all three K\,IV/V histograms in Fig.~\ref{DeltaRV} have the same slope, which further demonstrates false positives negligibly affect the distribution with $\Delta$RV$_{\rm max}$~$>$~1~km~s$^{-1}$.  The consistency also suggests the frequency of very close binaries, which dominate the large-amplitude RV tail with $\Delta$RV$_{\rm max}$~$>$~10~km~s$^{-1}$, decreases with metallicity at a similar rate as slightly wider binaries.

As discussed in \citet{Badenes2018}, systematic uncertainties can potentially bias the measured relation between the RV variability fraction and metallicity, but to a substantially smaller degree than the observed anti-correlation.  For example, metal-poor stars are systematically older and therefore have a larger fraction of close white dwarf (WD) companions.  In the field, $\approx$20\% of close companions to solar-type stars are WDs \citep{Moe2017,Murphy2018}. The close binary fraction therefore increases by $\Delta F_{\rm close}$/$F_{\rm close}$ $\approx$ 5\%\,-\,10\% between metal-rich field stars and slightly older metal-poor field stars due to the larger frequency of close WDs. Similarly, older metal-poor binaries have had more time for tidal friction and magnetic braking to harden the orbit, thereby boosting the RV variability fraction.  However, only $\approx$2\% of solar-type stars in volume-limited samples have $P$~$<$~10~days \citep{Duquennoy1991,Raghavan2010,Tokovinin2014,Moe2017}, and so tidal friction and magnetic braking alone cannot explain the observed RV variability fraction of 25\%\,$\pm$\,5\% for metal-poor K\,IV/V stars.  Finally, we selected our giant, K\,IV/V, and G\,IV/V subsamples according to fixed intervals of surface gravity and temperature, {\it not} mass. By interpolating the Dartmouth stellar evolutionary tracks \citep{Dotter2008}, we find a $M_1$~=~0.9\,\Msun\ star with [Fe/H]~=~0.4 and age $\tau$~=~5~Gyr has log\,$g$~$\approx$~4.53 and $T_{\rm eff}$~$\approx$~5,100~K. Meanwhile, a metal-poor star with [Fe/H]~=~$-$0.8 of the same mass and age is substantially smaller (log\,$g$~=~4.43) and hotter ($T_{\rm eff}$~=~6,300~K) due to the decreased opacities.  To extend down to $T_{\rm eff}$~$\approx$~5,100~K, a metal-poor dwarf with [Fe/H]~=~$-$0.8 must be $M_1$~$\approx$~0.67\,\Msun.  Given the same cuts in log\,$g$ and $T_{\rm eff}$, the metal-poor stars in our APOGEE subsamples are $\Delta M_1$~$\approx$~0.2\,\Msun\ less massive than their metal-rich counterparts.  The close binary fraction increases slightly with primary mass across $M_1$~$\approx$~0.5\,-\,1.5\,\Msun. Our selection criteria therefore leads to a $\approx$\,10\% bias in the metallicity versus binary relation in the {\it positive} direction.  This effect is opposite the observed anti-correlation, i.e., consideration of this particular selection bias strengthens our overall conclusion. In any case, the systematic uncertainty $\delta F_{\rm close}$/$F_{\rm close}$ $\approx$ 10\% in the inferred close binary fraction is insignificant compared to the observed factor of 4.0\,$\pm$\,0.5 decrease across $-$0.9~$<$~[Fe/H]~$<$~0.5.  We confirm the conclusion of \citet{Badenes2018} that the RV variability fraction and thus the intrinsic close binary fraction strongly decreases with metallicity.

\subsection{Cumulative Metallicity Distributions}
\label{cumAPOGEE}

\begin{figure}[t!]
\centerline{
\includegraphics[trim=4.7cm 0.2cm 5.5cm 0.1cm, clip=true, width=3.2in]{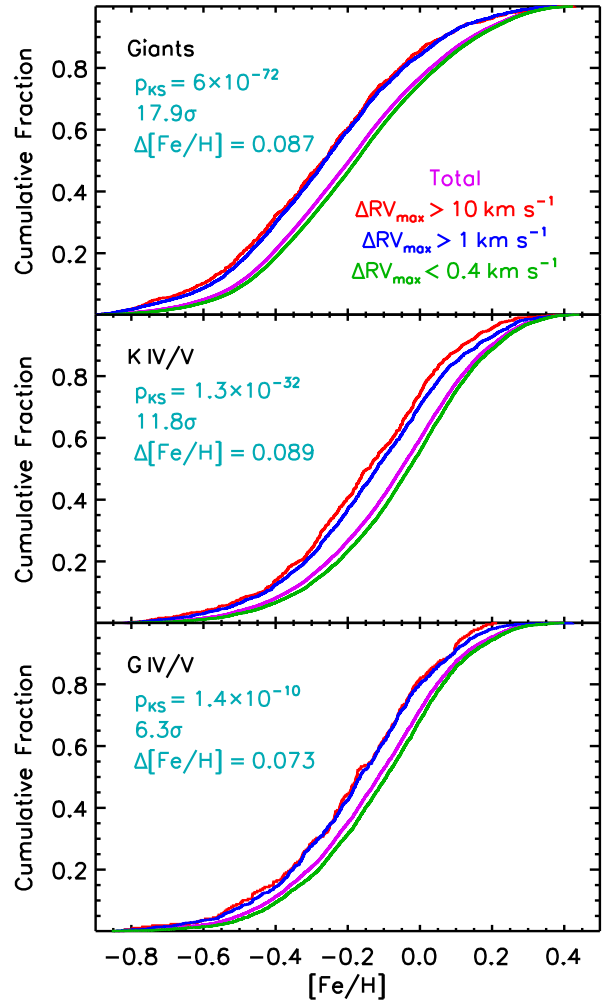}}
\caption{Cumulative metallicity distributions of giants (top), K\,IV/V stars (middle) and G\,IV/V stars (bottom) for the total populations (magenta), RV variables with $\Delta$RV$_{\rm max}$~$>$~10~km~s$^{-1}$ (red), RV variables with $\Delta$RV$_{\rm max}$~$>$~1~km~s$^{-1}$ (blue), and constant RV stars with $\Delta$RV$_{\rm max}$~$<$~0.4~km~s$^{-1}$ (green). We indicate in cyan the probability the blue and green distributions are drawn from the same parent distribution according to a KS test, the corresponding level of significance, and the difference in their median metallicities.   After correcting for incompleteness, the median metallicities of close binaries are $\Delta$[Fe/H] = 0.13\,$\pm$\,0.03 smaller than single stars.}
\label{cumRV}
\end{figure}

In Fig.~\ref{cumRV}, we display the cumulative distribution of metallicities for our giant, K\,IV/V, and G\,IV subsamples.  For each subsample, we show the metallicity distributions for large-amplitude RV variables with $\Delta$RV$_{\rm max}$~$>$~10~km~s$^{-1}$, small-amplitude RV variable with $\Delta$RV$_{\rm max}$~$>$~1~km~s$^{-1}$, and constant RV stars with $\Delta$RV$_{\rm max}$~$<$~0.4~km~s$^{-1}$.  The distributions of small-amplitude and large-amplitude RV variables are consistent with each other.  For K\,IV/V stars, a KS test shows the probability the $\Delta$RV$_{\rm max}$~$>$~1~km~s$^{-1}$ and $\Delta$RV$_{\rm max}$~$>$~10~km~s$^{-1}$ histograms are drawn from the same parent distribution is $p_{\rm KS}$~=~0.20.  For giants and G\,IV/V stars, the RV variability distributions are even closer, resulting in $p_{\rm KS}$~=~0.71 and $p_{\rm KS}$~=~0.99, respectively. This further demonstrates that false positives negligibly affect RV variables with $\Delta$RV$_{\rm max}$~=~1.0\,-\,2.0~km~s$^{-1}$ and that very close binaries that produce large-amplitude RV variations follow the same metallicity trend as slightly wider binaries.

Meanwhile, RV variables are noticeably shifted toward smaller metallicities compared to both the total population and especially the constant RV stars.   KS tests demonstrate the $\Delta$RV$_{\rm max}$~$>$~1~km~s$^{-1}$ and $\Delta$RV$_{\rm max}$~$<$~0.4~km~s$^{-1}$ distributions are discrepant with each other at the 17.9$\sigma$~($p_{\rm KS}$~=~6$\times$10$^{-72}$), 11.8$\sigma$~($p_{\rm KS}$~=~1.3$\times$10$^{-32}$), and 6.3$\sigma$~($p_{\rm KS}$~=~1.4$\times$10$^{-10}$) confidence levels for the giant, K\,IV/V, and G\,IV/V subsamples, respectively.  These levels of statistical significance are similar to those found above, but are based on the discrete metallicity distributions instead of the binned RV variability fractions. Both the $\chi^2$ and KS tests confirm the close binary fraction decreases with metallicity at the $\approx$\,20$\sigma$ confidence level. 

Close binaries have systematically smaller metallicities compared to single stars and wide binaries. We measure the differences between the median metallicities of the $\Delta$RV$_{\rm max}$~$>$~1~km~s$^{-1}$ and total populations to be $\Delta$[Fe/H]~=~0.068, 0.067, and 0.051 for the giant, K\,IV/V and G\,IV/V subsamples, respectively.  The metallicity differences between the $\Delta$RV$_{\rm max}$~$>$~1~km~s$^{-1}$ and $\Delta$RV$_{\rm max}$~$<$~0.4~km~s$^{-1}$ distributions are slightly larger at $\Delta$[Fe/H]~=~0.087, 0.089, and 0.073. Constant RV stars mainly consist of single stars and wide binaries, but also include close binaries that have small velocity amplitudes or were observed with unfavorable cadence to detect RV variations. As we calculate in \S\ref{RVcorr}, the fraction of close binaries ($P$~$<$~10$^4$; $a$~$\lesssim$~10~AU) that are detectable as APOGEE RV variables with $\Delta$RV$_{\rm max}$~$>$~1~km~s$^{-1}$ is $\approx$\,60\%.  The median metallicities of close binaries are therefore $\Delta$[Fe/H]~=~0.11\,$\pm$\,0.02 smaller than single stars and wide binaries with $a$~$\gtrsim$~10~AU.  Very wide binaries with $a$~$\gtrsim$~200~AU do not depend significantly on metallicity, while solar-type binaries with intermediate separations $a$~$\approx$~10\,-\,200~AU likely exhibit a weak metallicity anti-correlation (see \S\ref{Overview} and \S\ref{Summary}). We estimate the median metallicities of close binaries are $\Delta$[Fe/H]~=~0.13\,$\pm$\,0.03 smaller than single stars and very wide binaries with $a$~$\gtrsim$~200~AU. This difference may seem relatively small compared to the broad metallicity distribution of solar-type stars.  However, the mean metallicities of large stellar populations, such as the APOGEE sample, are measured to extremely high precision $\delta \langle$[Fe/H]$\rangle$~$\approx$~0.02~dex. A metallicity difference of $\Delta$[Fe/H]~=~0.13\,$\pm$\,0.03 between close binaries and single stars therefore represents a relatively substantial offset.

\subsection{Corrections for Incompleteness}
\label{RVcorr}

We next correct for incompleteness to recover the intrinsic close binary fraction from the observed RV variability fraction.   Accounting for the distribution of giant surface gravities, how close binaries evolve during giant expansion, the larger RV jitter associated with very luminous giants, and the differences in target selection of red clump versus normal giants is beyond the scope of this paper (see \citealt{Badenes2018}). A more detailed analysis of RV variability in APOGEE giants utilizing the more recent data release~14 is the subject of a future study (Mazzola et al., in prep.). In the present study, we combine our K\,IV/V and G\,IV/V subsamples, and we account only for incompleteness to measure the close binary fraction.

We modify our Monte Carlo model in \S\ref{LathamCorrections} to compute the completeness fraction $C$ of close binaries with $P$~=~1\,-\,10$^4$~days that are detectable as APOGEE RV variables.  We adopt a primary mass of $M_1$~=~0.9\,\Msun\ appropriate for the combined GK\,IV/V subsample.  We calculate the probability to detect RV variations as a continuous function of timespan $\Delta t$.  We generate RVs at $N_{\rm RV}$ = 2, 3 (average) and 4 epochs.  For $N_{\rm RV}$~=~2, the two epochs span $\Delta t$, while for $N_{\rm RV}$~=~3 and 4 the additional epochs are randomly distributed across $\Delta t$.  We do not add noise to the simulated RVs because the RV uncertainties are below our adopted RV thresholds. We simply calculate the fraction of close binaries that have $\Delta$RV$_{\rm max}$~$>$~$\Delta$RV$_{\rm threshold}$ for $\Delta$RV$_{\rm threshold}$ =  1, 3, and 10~km~s$^{-1}$.

\begin{figure}[t!]
\centerline{
\includegraphics[trim=0.5cm 0.2cm 0.3cm 0.2cm, clip=true, width=3.4in]{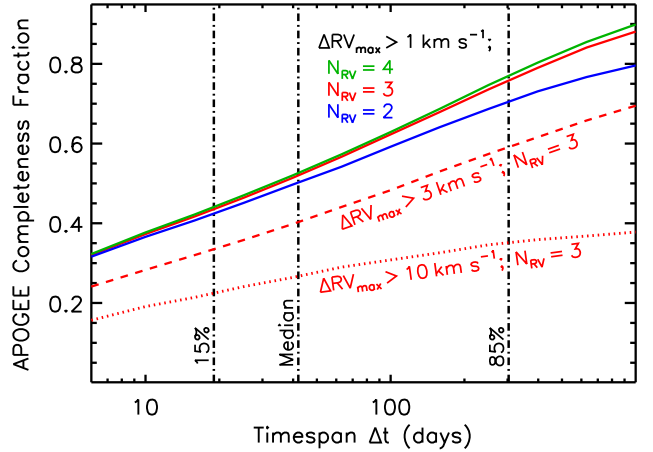}}
\caption{The simulated fraction of close binaries below $P$~$<$~10$^4$~days ($a$~$\lesssim$~10~AU) that exhibit RV variability above $\Delta$RV$_{\rm max}$~$>$~1~km~s$^{-1}$ (solid), 3~km~s$^{-1}$ (dashed), and 10~km~s$^{-1}$ (dotted) given $N_{\rm RV}$ = 2 (blue), 3 (red), and 4 (green) RV measurements as a function of timespan $\Delta t$ between first and final visits. APOGEE observed the 19,239 GK\,IV/V stars in our sample with varying cadence, and we indicate the 15$^{\rm th}$-percentile, median, and 85$^{\rm th}$-percentile in  timespans with vertical dash-dotted lines. }
\label{RVcomp}
\end{figure}

We display in Fig.~\ref{RVcomp} the simulated completeness fractions $C$ as a function of $\Delta t$ for the different values of $N_{\rm RV}$ and $\Delta$RV$_{\rm threshold}$. The fraction of close binaries that are detectable as RV variables increases nearly linearly with respect to log\,$\Delta t$.  Given $N_{\rm RV}$~=~3, the fraction of close binaries that have $\Delta$RV$_{\rm max}$~$>$~1~km~s$^{-1}$ increases from $C$~=~37\% for $\Delta t$~=~10~days to $C$~=~88\% for $\Delta t$~=~1,000~days. The number $N_{\rm RV}$ of RV observations only slightly affects the detection rates. In particular, a fourth RV measurement negligibly increases the completeness fraction unless it also extends the timespan between first and final visits. The completeness curves for $\Delta$RV$_{\rm max}$~$>$~3~km~s$^{-1}$ and $\Delta$RV$_{\rm max}$~$>$~10~km~s$^{-1}$ are substantially smaller, and the latter is also flatter with respect to $\Delta t$.  Even with an infinite number and timespan of RV observations, only $C$~$\approx$~45\% of close binaries with $P$~=~1\,-\,10$^4$~days produce large-amplitude RV variations above $\Delta$RV$_{\rm max}$~$>$~10~km~s$^{-1}$.  

For our combined GK\,IV/V subsample, the 15$^{\rm th}$-percentile, median, and 85$^{\rm th}$-percentile in timespans are $\Delta t$ = 19, 42, and 303 days, respectively, which we indicate in Fig.~\ref{RVcomp}. Given the wide spread in timespans, we do not adopt the median but instead weight our Monte Carlo models according to the actual cadence of the APOGEE observations. We calculate weighted completeness fractions of $C$ = 0.57, 0.40, and 0.24 for $\Delta$RV$_{\rm max}$~$>$~1, 3, and 10~km~s$^{-1}$, respectively.  

Our Monte Carlo model, which incorporates the short-period tail of a log-normal period distribution (see \S\ref{LathamCorrections}), accurately reproduces the observed distribution of RV amplitudes. For example, the modeled ratio $R_{\rm 3\,to\,1}$ = $C$($\Delta$RV$_{\rm max}$~$>$~3~km~s$^{-1}$)/$C$($\Delta$RV$_{\rm max}$~$>$~1~km~s$^{-1}$) = 0.40/0.57 = 0.70 between the completeness fractions is consistent with the observed ratio $R_{\rm 3\,to\,1}$ = 0.65\,$\pm$\,0.03 between the corresponding number of RV variables (see \S\ref{APOGEEMetallicity} and Fig.~\ref{DeltaRV}).  Similarly, the simulated ratio  $R_{\rm 10\,to\,1}$ = 0.24/0.57 = 0.42 is slightly larger than but still consistent with the observed ratio $R_{\rm 10\,to\,1}$ = 0.38\,$\pm$\,0.02 between the number of large-amplitude and small-amplitude RV variables. If we instead adopt a uniform distribution in log\,$P$, i.e., Opik's law, then we simulate larger completeness fractions of $C$ = 0.75, 0.63, and 0.47 for $\Delta$RV$_{\rm max}$~$>$~1, 3, and 10~km~s$^{-1}$, respectively, because more of the close binaries are weighted toward shorter periods.  However, Opik's law predicts ratios $R_{\rm 3\,to\,1}$ = 0.63/0.75 = 0.84 and $R_{\rm 10\,to\,1}$ = 0.47/0.75 = 0.63 that are clearly discrepant with the observed ratios 0.65\,$\pm$\,0.03 and 0.38\,$\pm$\,0.02, respectively.  Both metal-poor and metal-rich solar-type binaries therefore follow the same short-period tail of a log-normal period distribution. Metal-poor solar-type stars simply have a larger close binary fraction.

\begin{figure}[t!]
\centerline{
\includegraphics[trim=0.1cm 0.2cm 0.3cm 0.2cm, clip=true, width=3.4in]{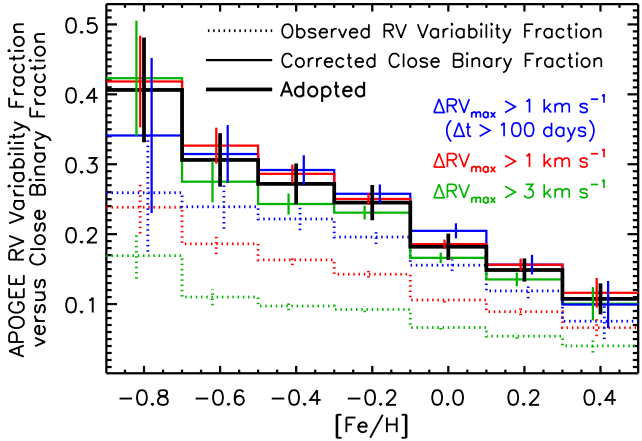}}
\caption{As a function of metallicity, the observed fraction of all GK\,IV/V APOGEE stars in our sample that exhibit RV variability above $\Delta$RV$_{\rm max}$ $>$ 1~km~s$^{-1}$ (dotted red) and 3~km~s$^{-1}$ (dotted green), and the observed fraction of GK\,IV/V APOGEE stars monitored during a timespan of at least $\Delta t$~$>$~100~days that exhibit RV variability above $\Delta$RV$_{\rm max}$ $>$ 1~km~s$^{-1}$ (dotted blue). We divide these three histograms by their respective completeness fractions of $C$ = 0.40, 0.57, and 0.76, resulting in the bias-corrected close binary fractions (thin colored lines). We adopt a weighted average and a systematic uncertainty of $\delta F_{\rm close}$/$F_{\rm close}$~=~10\%, providing an intrinsic close binary fraction that decreases from  $F_{\rm close}$~=~41\%\,$\pm$\,7\% at [Fe/H]~=~$-$0.8 to $F_{\rm close}$~=~11\%\,$\pm$\,2\% at [Fe/H]~=~+0.4 (thick solid black). }
\label{RV_binfrac}
\end{figure}

Similar to Fig.~\ref{DeltaRV}, we display in Fig.~\ref{RV_binfrac} the fraction of GK\,IV/V stars with $\Delta$RV$_{\rm max}$~$>$~1~km~s$^{-1}$ and $\Delta$RV$_{\rm max}$~$>$~3~km~s$^{-1}$ as a function of metallicity.  Of the 19,239 GK\,IV/V stars in our combined sample, 5,394 (28\%) were observed by APOGEE during a timespan of at least $\Delta t$~$>$~100~days.  As shown in Fig.~\ref{RV_binfrac}, this subset exhibits a noticeably higher fraction of RV variables with $\Delta$RV$_{\rm max}$~$>$~1~km~s$^{-1}$ compared to the total GK\,IV/V sample. By fitting across all metallicities, we find the RV variability fraction of GK\,IV/V stars observed with longer timespans is $R_{\rm long/total}$ = 1.37\,$\pm$\,0.05 times larger than the total GK\,IV/V population ($\chi^2$/$\nu$~=~0.49, $p$~=~0.82).  With increased timespans, the APOGEE observations become more complete toward detecting SBs with longer periods (see Fig.~\ref{RVcomp}).  We weight our Monte Carlo model according to the cadence of RV observations for the 5,394 GK\,IV/V stars with $\Delta t$~$>$~100~days.  The resulting completeness fraction of $C$ = 0.76 is $R_{\rm long/total}$ = 0.76/0.57 = 1.33 times larger than the completeness fraction for the total GK\,IV/V population. The simulated ratio nearly matches the observed ratio, providing another confirmation our Monte Carlo model accurately describes close solar-type binaries.

In Fig.~\ref{RV_binfrac}, we divide the observed RV variability fractions by their corresponding completeness fractions.  The three resulting completeness-corrected close binary fractions are all consistent with each other.  We adopt a weighted average of the three histograms and the measurement uncertainties from the distribution based on all GK\,IV/V RV variables with $\Delta$RV$_{\rm max}$ $>$ 1~km~s$^{-1}$. For each metallicity bin, we add a systematic uncertainty of $\delta F_{\rm close}$/$F_{\rm close}$~=~10\% in quadrature with the measurement uncertainties to account for the small selection biases discussed in \S\ref{APOGEEMetallicity}.  We present our final completeness-corrected close binary fraction of GK\,IV/V stars as the thick black histogram in Fig.~\ref{RV_binfrac}.  The intrinsic close binary fraction ($P$~$<$~10$^4$ days; $a$~$\lesssim$~10~AU) decreases from $F_{\rm close}$ = 41\%\,$\pm$\,7\% at [Fe/H] = $-$0.8 to $F_{\rm close}$ = 11\%\,$\pm$\,2\% at [Fe/H] = +0.4.  The metallicity-dependent close binary fraction inferred from the APOGEE RV variables and the Carney-Latham SB samples (see Fig.~\ref{Latham_binfrac}) are consistent with each other.  Our APOGEE RV sample of 19,239 GK\,IV/V stars is a factor of 14 times larger than the \citet{Latham2002} sample.  Moreover, APOGEE measured the RVs and metallicities of their targets to substantially higher precision.  The anti-correlation between the close binary fraction and metallicity is therefore even more pronounced and measured to much higher statistical significance with the APOGEE dataset.

\section{Kepler Eclipsing Binaries}
\label{Kepler}

\subsection{Sample Selection and Description}

The primary {\it Kepler} mission monitored nearly 200,000 solar-type stars for four years with exquisite photometric precision.  Designed to discover transiting exoplanets, {\it Kepler} also identified and characterized 2,878 EBs and non-eclipsing binary ellipsoidal variables \citep{Prsa2011,Kirk2016}. About a third of the systems in the {\it Kepler} EB catalog have very short periods $P$~$<$~1~day, the majority of which are evolved contact or ellipsoidal binaries.  Most of the {\it Kepler} EBs with longer periods are in pre-mass-transfer detached configurations.  A few EBs have especially long periods $P$~=~1,000\,-\,1,100 days, but geometrical selection effects and the four-year lifetime of the main {\it Kepler} mission severely limited the discovery of such wide binaries. We initially select the 1,924 EBs with $P$~=~1\,-\,1,000~days in the third revision of the {\it Kepler} EB catalog \citep{Kirk2016}.  

\subsubsection{Sample with Photometric Metallicities}
\label{photmetal}

\citet{Brown2011} utilized photometry, stellar isochrones, and a Bayesian model of the galactic stellar population to estimate $T_{\rm eff}$, log\,$g$, and [Fe/H] for all stars in the {\it Kepler} input catalog.  Specifically, they measured the spectral energy distribution (SED) of each {\it Kepler} star based on broadband optical photometry (griz), 2MASS near-infrared photometry (JHK), and an intermediate-band filter (D51) centered on the Fraunhofer b absorption lines near 515\,nm that are associated with Mg and Fe.  \citet{Brown2011} then fitted the measured SEDs to synthetic colors from ATLAS9 model atmospheres \citep{Castelli2004} assuming the dust extinction varied as a simple function of distance and galactic latitude.  They also incorporated Bayesian priors in $T_{\rm eff}$, log\,$g$, and [Fe/H] according to the observed distributions in the solar neighborhood.  \citet{Huber2014} revised and significantly improved the measured parameters of 196,468 {\it Kepler} stars.  They updated the photometry with recent observations, calibrated $T_{\rm eff}$ according to empirical relations, incorporated more accurate stellar isochrones from the Dartmouth evolutionary tracks \citep{Dotter2008}, and treated dust extinction in a more realistic manner.  \citet{Huber2014} adopted Bayesian priors in log\,$g$ and [Fe/H] similar to those in \citet{Brown2011}, but developed a slightly more sophisticated method for sampling the distributions. 

\citet{Brown2011} and \citet{Huber2014} stressed the measured surface gravities and metallicities in their catalogs are highly uncertain and should not be used on a star-by-star basis.  Nevertheless, they argued the distributions of surface gravities and metallicities are statistically accurate and can therefore be utilized to study broad trends across these parameters.  \citet{Brown2011} and \citet{Huber2014} also identified regions in the HR diagram where the photometric solutions for log\,$g$ and [Fe/H] are highly degenerate and most uncertain, notably for subgiants and cool late-K and M-type dwarfs.  We therefore select the $N_{\rm phot}$ = 142,951 solar-type dwarfs in the \citet{Huber2014} catalog with photometric parameters $T_{\rm eff}$ = 4,800\,-\,6,800\,K, log\,$g$ = 4.0\,-\,5.0, and $-$1.7~$<$~[Fe/H]~$<$~0.5, corresponding approximately to F3V\,-\,K3V stars. 

\citet{Berger2018} recently utilized {\it Gaia} parallactic distances to measure the stellar radii of {\it Kepler} stars, and found $\approx$\,65\%, 23\%, and 12\% are MS stars, subgiants, and giants, respectively.  They concluded contamination by subgiants in the {\it Kepler} sample is smaller than previously thought. Moreover, a non-negligible fraction of the \citet{Berger2018} subgiants, which were identified because they lie slightly above the MS relation in the HR diagram, are actually twin binaries with MS components of comparable luminosity. Thus a significant majority of the solar-type dwarfs in our photometric sample are truly MS stars. 

The metallicity distribution of our photometric sample of {\it Kepler} solar-type dwarfs follows a Gaussian with mean of $\langle$[Fe/H]$\rangle$~=~$-$0.17 and dispersion of $\sigma_{\rm [Fe/H]}$~=~0.26~dex.  \citet{Huber2014} estimated the uncertainties in the photometric metallicities of {\it Kepler} stars is $\delta$[Fe/H]~$\approx$~0.3~dex.  We can therefore examine metallicity trends across the much broader interval $-$1.7~$<$~[Fe/H]~$<$~0.5. Of the 1,924 {\it Kepler} EBs with $P$~=~1\,-\,1,000~days, $N_{\rm EB,phot}$ = 1,292 systems satisfy our selection criteria of $T_{\rm eff}$ = 4,800\,-\,6,800\,K, log\,$g$ = 4.0\,-\,5.0, and $-$1.7~$<$~[Fe/H]~$<$~0.5 according to the \citet{Huber2014} photometric catalog.  The observed EB fraction in our photometric sample of {\it Kepler} solar-type dwarfs is $F_{\rm EB,phot}$ = 1,291/142,951 = 0.90\%\,$\pm$\,0.03\%.  

The presence of a binary companion can potentially bias the metallicities inferred from fitting single-star isochrones to the measured photometry.  The photometric metallicities of EBs in particular may be substantially inaccurate if the observations in the different filters correspond to different orbital phases, e.g., during versus outside of eclipse.  In addition, the majority of very close binaries with $P$~$\lesssim$~7~days have tertiary companions \citep{Tokovinin2006}, and so most EBs also have third light contamination. 

We assess the significance of these potential biases by fitting isochrones to simulated photometry of solar-type binaries.  We download a dense grid of Dartmouth stellar evolutionary tracks \citep{Dotter2008} spanning masses $M$~=~0.15\,-\,1.7\,\Msun, ages $\tau_*$ = 1\,-\,13 Gyr, and metallicities $-$2.4~$<$~[Fe/H]~$<$~0.5.  We simulate binaries with metallicities [Fe/H] = $-$1.3, $-$0.8, $-$0.3, and +0.2 at representative ages of $\tau_*$ = 11 Gyr, 8 Gyr, 5 Gyr, and 2 Gyr, respectively.  We select G8V primaries with $T_1$~=~5,500\,K, corresponding to primary masses $M_1$ = 0.65, 0.71, 0.82, and 0.98\,\Msun\ for the four combinations of metallicities and ages.  We also consider hotter F8V primaries with $T_1$ = 6,200\,K, corresponding to slightly higher masses of $M_1$ = 0.75, 0.84, 0.99, and 1.22\,\Msun.  For different combinations of mass ratios $q$ = $M_2$/$M_1$, we add the fluxes of both binary components for all eight filters (D51grizJHK) utilized in \citet{Brown2011} and \citet{Huber2014}. We add a dust extinction of A$_{\rm r}$~=~0.2~mag and adopt a dust reddening law from \citet{Schlafly2011} such that A$_{b}$/A$_{\rm r}$ = 1.45, 1.31, 0.74, 0.55, 0.31, 0.20, 0.13 for bands $b$ = g, D51, i, z, J, H, and K, respectively.  We do not fit the distances to our simulated binaries, and so we consider only the seven unique color combinations.  \citet{Brown2011} measured the bright {\it Kepler} stars to a precision of $\approx$\,0.02~mag in the D51griz filters, and so we adopt uncertainties of 0.03~mag in all the colors.  We measure the photometric masses $M_{\rm phot}$, ages $\tau_{\rm phot}$, metallicities [Fe/H]$_{\rm phot}$, and dust extinctions $A_{\rm r, phot}$ by minimizing the $\chi^2$ statistic between the seven colors of our simulated binaries and the isochrones of {\it single stars}.  We assume uniform priors in our four photometric parameters.  In this manner, our fits are not dominated by short-lived phases of stellar evolution that provide only marginally smaller $\chi^2$ values.

\begin{figure}[t!]
\centerline{
\includegraphics[trim=3.6cm 0.3cm 4.9cm 0.4cm, clip=true, width=3.3in]{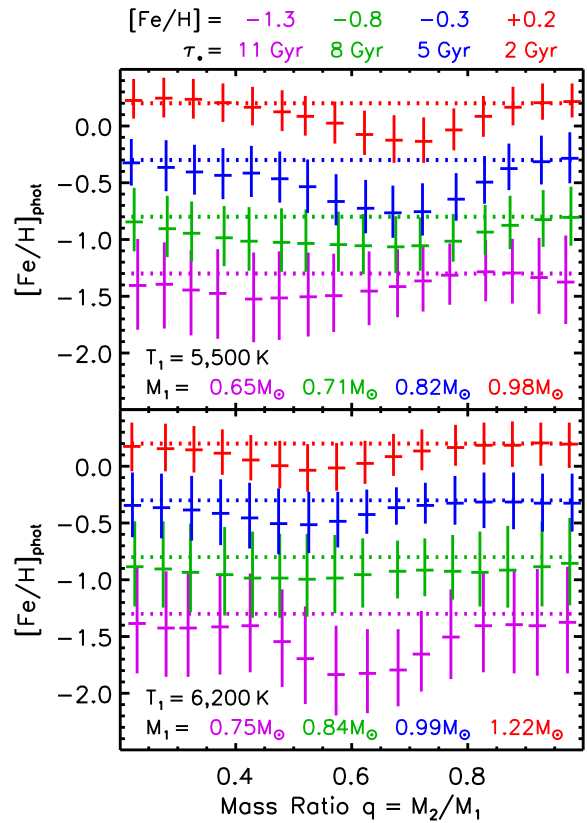}}
\caption{The photometric metallicities [Fe/H]$_{\rm phot}$ determined by fitting single-star isochrones to simulated broad-band photometry of binaries as a function of mass ratio $q$.  We consider binaries with cooler primaries ($T_1$ = 5,500\,K; top) and hotter primaries ($T_1$ = 6,500\,K; bottom) for four different metallicities [Fe/H]~=~$-$1.3 (magenta), $-$0.8 (green), $-$0.3 (blue), and +0.2 (red), where we list the corresponding ages $\tau_*$ and primary masses $M_1$.  For some combinations (e.g., $q$~$\approx$~0.6), the fitted photometric metallicities underestimate the true metallicities (dotted) by as much as $\approx$\,0.5 dex.  In general, however, the measurement uncertainties simply increase from $\approx$\,0.25~dex near [Fe/H]$_{\rm phot}$~=~0.2 to $\approx$\,0.45~dex near [Fe/H]$_{\rm phot}$~=~$-$1.3 with negligible bias between the true and photometric metallicities.}
\label{metalbias}
\end{figure}

We measure the mean and 1$\sigma$ uncertainties in the photometric metallicities [Fe/H]$_{\rm phot}$ by marginalizing across the other parameters.  We display the measured values of [Fe/H]$_{\rm phot}$ in Fig.~\ref{metalbias} for the various combinations of [Fe/H], $M_1$, and $q$.  The measurement uncertainties increase from $\delta$[Fe/H]~=~0.25~dex near [Fe/H]~=~+0.2 to $\delta$[Fe/H]~=~0.45~dex near [Fe/H]~=~$-$1.3, consistent with the average uncertainty of $\delta$[Fe/H]~=~0.3~dex reported in \citet{Huber2014}.  Compared to their primaries, low-mass companions with $q$~$<$~0.4 contribute negligible flux across the optical and near-infrared bands. For such extreme mass-ratio binaries, the photometric metallicities [Fe/H]$_{\rm phot}$ determined by fitting single-star isochrones are close to the true metallicities [Fe/H].  Similarly, companions with $q$~$>$~0.8 have SEDs similar to their primaries, and so the photometric metallicities of twin binaries are consistent with their actual values.  For $q$~$\approx$~0.4\,-\,0.8, however, there are certain combinations of [Fe/H] and $M_1$ for which the photometric metallicities underestimate the true metallicities.  In particular, Fig.~\ref{metalbias} shows that binaries with $T_1$~=~5,500\,K, [Fe/H]~$\approx$~0.0, and $q$ $\approx$ 0.6\,-\,0.8 and binaries with $T_1$~=~6,200\,K, [Fe/H]~$\approx$~$-$1.3, and $q$~$\approx$~0.5\,-\,0.7 are biased by $\Delta$[Fe/H]~$\approx$~$-$0.5~dex toward smaller metallicities.  Fortunately, only $\approx$\,20\% of close solar-type binaries have mass ratios spanning an interval of $\Delta q$~=~0.2 near $q$~$\approx$~0.6 \citep{Raghavan2010,Tokovinin2014,Moe2017}.  The photometric metallicities inferred from single-star isochrones are therefore slightly biased for only a small fraction of the close binary population.

The biases in the photometric metallicities due to eclipses and tertiary companions are even smaller.  For most {\it Kepler} stars, \citet{Brown2011} rapidly cycled through all the optical filters (D51griz) during a single pointing. The 2MASS near-infrared photometry was obtained at earlier epochs and likely coincide with different orbital phases.  Fortunately, the optical bands, especially the D51 filter, provide the most leverage in constraining the metallicities. Moreover, the majority of {\it Kepler} EBs with $P$~=~1\,-\,1,000~days have shallow eclipses, e.g., 67\% with $\Delta$m~$<$~0.1~mag and 81\% with $\Delta$m~$<$~0.2~mag \citep{Kirk2016}. The listed optical to near-infrared colors of {\it Kepler} EBs differ from their true out-of-eclipse colors by $\lesssim$~0.05~mag on average. For {\it Kepler} EBs with longer periods $P$~$\gtrsim$~20~days, the durations of the eclipses are substantially shorter than their orbital periods. The photometric colors of long-period EBs are therefore much more likely to correspond to their out-of-eclipse values.  Most importantly, the optical to near-infrared colors of EBs are randomly shifted toward either smaller or larger values relative to their out-of-eclipse colors, i.e., there is no net bias.  Regarding triple stars, the majority of tertiary companions to very close binaries are weighted toward small mass ratios $q$~=~$M_3$/$M_1$~$<$~0.5 \citep{Tokovinin2006,Moe2017}.  As demonstrated in Fig.~\ref{metalbias}, low-mass companions negligibly affect the measured photometric metallicities.  Although the majority of very close binaries have outer tertiaries, only $\approx$\,30\% of binaries with $P$~$>$~20~days are in triple systems \citep{Tokovinin2006}. We conclude the biases in the photometric metallicities of {\it Kepler} EBs, especially those with $P$~$>$~20~days, are negligible compared to the measurement uncertainties and other sources of systematic uncertainties that equally affect both EBs and single stars in the {\it Kepler} sample.  

\subsubsection{Sample with Spectroscopic Metallicities}

The metallicities measured from stellar spectra are generally more precise and less biased than photometric metallicities derived from fitting stellar isochrones. \citet{Mathur2017} compiled dozens of follow-up surveys and provided spectroscopic metallicities [Fe/H]$_{\rm spec}$ for 16,289 {\it Kepler} stars. Unfortunately, their sample of {\it Kepler} stars with spectroscopic metallicities is a heterogenous, non-random subset and therefore  cannot be utilized to investigate the EB fraction as a function of metallicity.  For example, many {\it Kepler} stars received follow-up spectroscopic observations because their light curves exhibited transiting exoplanets.  Other {\it Kepler} stars were observed spectroscopically because they displayed clean variability from pulsations that provide stringent tests for asteroseismic models. Such subsets are significantly biased against EBs. Nevertheless, the spectroscopic metallicities [Fe/H]$_{\rm spec}$ in \citet{Mathur2017} provide insight into the accuracy of the photometric metallicities.  We find 15,801 of the {\it Kepler} stars with listed spectroscopic metallicities in \citet{Mathur2017}  have photometric metallicities  $-$1.5~$<$~[Fe/H]$_{\rm phot}$~$<$~0.5 in \citet{Huber2014}. We measure a significant degree of correlation between [Fe/H]$_{\rm phot}$ and [Fe/H]$_{\rm spec}$, e.g., the Pearson correlation coefficient is $r_{\rm P}$~=~0.52.  The photometric metallicities can therefore be used to reliably measure trends between the EB fraction and metallicity.

The LAMOST spectroscopic survey ($R$~$\approx$~1,800) recently measured the metallicities of tens of thousands of {\it Kepler} stars \citep{Dong2014,DeCat2015,Ren2016,Frasca2016}.  Unlike the compilation presented in \citet{Mathur2017}, the LAMOST-{\it Kepler} project obtained spectra for a random subset of {\it Kepler} stars and is therefore not biased with respect to EBs.  The metallicities of several hundred stars in the LAMOST-{\it Kepler} field have been previously measured with high-resolution spectra and other robust techniques.  \citet{Dong2014} and \citet{Ren2016} demonstrated the metallicities derived from their low-resolution LAMOST spectra are consistent with these previous measurements.  For dwarf stars, \citet{Ren2016} reported the bias between the LAMOST and high-resolution spectroscopic metallicities is only $\delta$[Fe/H]~=~0.01~dex and that the measurement uncertainties in the LAMOST metallicities are typically $\sigma_{\rm [Fe/H]}$~$\approx$~0.1~dex.

\citet{Dong2014} and \citet{Ren2016} then compared their LAMOST spectroscopic metallicities to the photometric metallicities listed in the {\it Kepler} index catalog \citep{Brown2011}.  They both found good agreement near sub-solar metallicities [Fe/H]$_{\rm phot}$ $\approx$ [Fe/H]$_{\rm spec}$  $\approx$ $-$0.4 (see Fig.~1 in \citealt{Dong2014} and Fig.~9 in \citealt{Ren2016}).  For metal-rich dwarf stars, however, \citet{Dong2014} and \citet{Ren2016} showed the photometric metallicities systematically underestimate the true metallicities by $\delta$[Fe/H]~=~0.4~dex.  {\it Kepler} dwarfs with [Fe/H]$_{\rm phot}$~$\approx$~0.0 actually have true metallicities [Fe/H]$_{\rm spec}$~$\approx$~0.4.  The shift is likely due to the Bayesian prior metallicity distribution adopted in \citet{Brown2011} and \citet{Huber2014}, which peaks near [Fe/H]~$\approx$~$-$0.2 and is consistent with the distribution in the solar neighborhood.  Meanwhile, \citet{Dong2014} and \citet{Ren2016} found the true metallicity distribution of more distant {\it Kepler} stars peaks at [Fe/H]~$\approx$~0.0.  Nonetheless, \citet{Dong2014} and \citet{Ren2016} both confirmed [Fe/H]$_{\rm phot}$ and [Fe/H]$_{\rm spec}$ are significantly correlated. This reaffirms our conclusion that the photometric metallicities reported in \citet{Huber2014} provides leverage in measuring how the EB properties vary with metallicity.

For our {\it Kepler} sample with spectroscopic metallicities, we choose stars in the LAMOST-{\it Kepler} survey according to the same selection criteria as our photometric sample.  Specifically, we select the $N_{\rm spec}$~=~23,886 solar-type {\it Kepler} dwarfs with LAMOST spectroscopic parameters $T_{\rm eff}$~=~4,800\,-\,6,800, log\,$g$~=~4.0\,-\,5.0, and $-$1.7~$<$~[Fe/H]~$<$~+0.5 from \citet{Ren2016}.  The metallicity distribution is accurately modeled by a Gaussian with mean of $\langle$[Fe/H]$\rangle$~=~$-$0.05 and dispersion of $\sigma_{\rm [Fe/H]}$ = 0.21, which is slightly more metal-rich than our photometric sample as described above.   We find $N_{\rm EB,spec}$ = 244 of our {\it Kepler} solar-type dwarfs with spectroscopic metallicities are EBs with $P$~=~1\,-\,1,000~days \citep{Kirk2016}.  The resulting EB fraction of $F_{\rm EB,spec}$ = 244/23,866 = 1.02\%\,$\pm$\,0.07\% is consistent with the fraction $F_{\rm EB,phot}$ = 0.90\%\,$\pm$\,0.03\% measured for our {\it Kepler} sample with photometric metallicities.  This confirms the LAMOST-{\it Kepler} survey was not biased against EBs.  Although our {\it Kepler} sample of solar-type dwarfs with spectroscopic metallicities is six times smaller than our photometric sample, it is a representative subset and the stellar metallicities are measured to much higher accuracy and precision.

\subsection{Variations with Metallicity}
\label{EBmetallicities}

\begin{figure}[t!]
\centerline{
\includegraphics[trim=0.4cm 0.3cm 0.2cm 0.2cm, clip=true, width=3.4in]{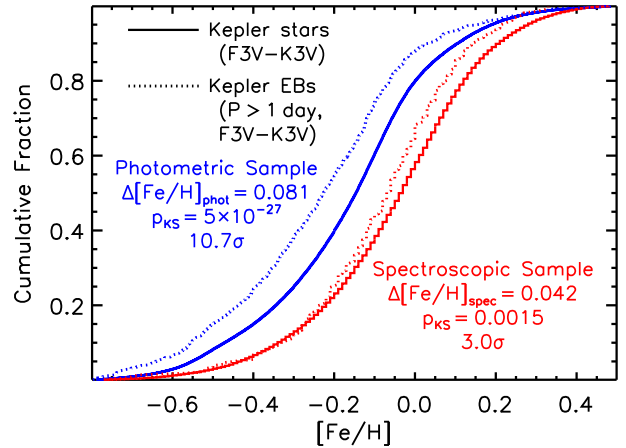}}
\caption{Cumulative metallicity distributions for our {\it Kepler} samples of solar-type dwarfs with photometric (solid blue) and spectroscopic (solid red) metallicities, and the corresponding subsets that are EBs with $P$~=~1\,-\,1,000 days (dotted). For both photometric and spectroscopic samples, the EBs are weighted toward smaller metallicities compared to their parent distributions at statistically significant levels.}
\label{cumEB}
\end{figure}

In Fig.~\ref{cumEB}, we investigate the cumulative metallicity distributions of our {\it Kepler} EBs. For visual clarity, we truncate the distributions in Fig.~\ref{cumEB} to the interval $-$0.8~$<$~[Fe/H]~$<$~0.5, but perform our statistical analysis across the full range $-$1.7~$<$~[Fe/H]~$<$~0.5. For both our photometric and spectroscopic samples, the EBs in Fig.~\ref{cumEB} are noticeably weighted toward smaller metallicities compared to their respective parent distributions.  For our {\it Kepler} sample of solar-type dwarfs with photometric metallicities, a KS test demonstrates the EBs are discrepant with the total population at the 10.7$\sigma$ significance level ($p_{\rm KS}$ = 5$\times$10$^{-27}$).  We also find the median metallicity of the EBs are shifted downward by $\Delta$[Fe/H]$_{\rm phot}$~=~0.081~dex compared to their parent distribution.  This shift is slightly larger than but consistent with the differences $\Delta$[Fe/H]~$\approx$~0.05\,-\,0.07~dex between APOGEE RV variables and their total populations as reported in \S\ref{cumAPOGEE}.  The {\it Kepler} solar-type dwarfs with measured spectroscopic metallicities are weighted toward larger metallicities compared to the photometric sample due to the biases discussed above and in \citet{Dong2014} and \citet{Ren2016}.  Nevertheless, the EBs in the more precise spectroscopic sample also have systematically lower metallicities ($\Delta$[Fe/H]$_{\rm spec}$~=~0.042~dex) than their parent distribution at the 3.0$\sigma$ confidence level ($p_{\rm KS}$ = 0.0015). Despite the smaller sample size, {\it Kepler} EBs with measured spectroscopic metallicities confirm close binaries are weighted toward lower metallicities at a statistically significant level.

\begin{figure}[t!]
\centerline{
\includegraphics[trim=0.4cm 0.3cm 0.4cm 0.2cm, clip=true, width=3.3in]{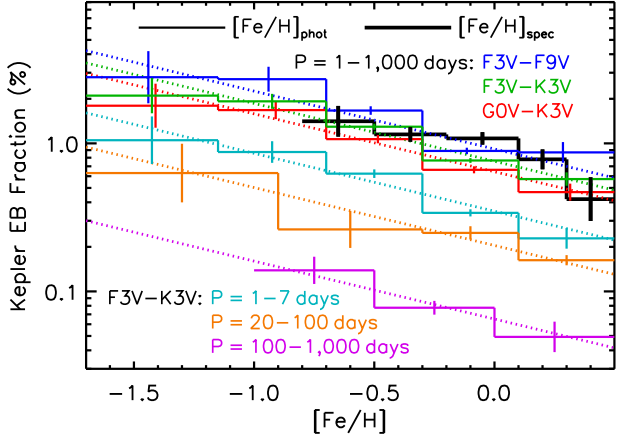}}
\caption{The fraction of {\it Kepler} solar-type dwarfs that are EBs with $P$ = 1\,-\,1,000 days within our full photometric sample (thin green) and spectroscopic sample (thick black). We divide the photometric sample according to spectral type: F3V-F9V (blue) and G0V-K3V (red).  We also compare the EB fraction within our photometric sample across different periods: $P$ = 1\,-\,7~days (cyan),  20\,-\,100~days (orange), and 100\,-\,1,000~days (magenta).  All samples show a statistically significant decrease in the EB fraction with respect to metallicity.  We show the fit log\,$F_{\rm EB}$~$\propto$~$-$0.39[Fe/H] (dotted) to the overall photometric sample scaled to the various subsamples. }
\label{EBfrac}
\end{figure}

In Fig.~\ref{EBfrac}, we next examine the {\it Kepler} EB fraction as a function of metallicity, spectral type, and orbital period.  For our full photometric sample of {\it Kepler} F3V\,-\,K3V primaries, the EB fraction across P~=~1\,-\,1,000~days decreases by a factor of 3.4\,$\pm$\,0.5 between $F_{\rm EB}$~=~1.9\%\,$\pm$\,0.2\% near [Fe/H]~=~$-$0.9 to $F_{\rm EB}$~=~0.57\%\,$\pm$\,0.06\% at [Fe/H]~=~0.3 (green histogram in Fig.~\ref{EBfrac}).  Attempting to fit a constant EB fraction to the five green metallicity bins in Fig.~\ref{EBfrac} results in a reduced $\chi^2$/$\nu$~=~25.7 with $\nu$~=~4 degrees of freedom.  A constant EB fraction with respect to metallicity can be rejected at the 9.4$\sigma$ confidence level ($p$ = 2.6$\times$10$^{-21}$), which is similar to the level of significance inferred from the cumulative metallicity distributions (see above).  We instead find the {\it Kepler} EB fraction is sufficiently modeled by a power-law such that log~$F_{\rm EB}$~$\propto$~($-$0.39\,$\pm$\,0.05)[Fe/H], which is displayed as the dotted green line in Fig.~\ref{EBfrac}.  

We then divide the photometric sample into hot ($T_{\rm eff}$~=~6,000\,-\,6,800\,K) and cool ($T_{\rm eff}$~=~4,800\,-\,6,000\,K) dwarfs, corresponding to F3V-F9V and G0V-K3V spectral types, respectively. Both the hot and cool subsamples follow the same metallicity trend (blue and red histograms in Fig.~\ref{EBfrac}, respectively).  This suggests the close binary fraction and metallicity are anti-correlated to a similar degree across the primary mass interval $M_1$~=~0.6\,-\,1.3\,\Msun.  For all metallicities, the {\it Kepler} EB fraction of F3V-F9V stars is $\approx$\,40\% larger than G0V-K3V stars for two reasons.  First, F dwarfs are larger than G/early-K dwarfs, and so their corresponding eclipse probabilities are $\approx$\,20\%\,-\,30\% larger (see \ref{EBcomp}).  Second, the intrinsic close binary fraction of F dwarfs is $\approx$\,10\%\,-\,20\% larger than that of G/early-K dwarfs \citep{Raghavan2010,Tokovinin2014,Moe2017}.  

We next compare the EB fraction as a function of metallicity for different period intervals.  Nearly half of our {\it Kepler} EBs have very short periods $P$~=~1\,-\,7~days (cyan histogram in Fig.~\ref{EBfrac}).  As discussed in \S\ref{photmetal}, such very close EBs have wide eclipses and most have tertiary companions, and so their photometric metallicities are most uncertain.  Nevertheless, EBs with $P$~=~20\,-\,100~days (orange histogram), which have narrow eclipses and are unlikely to be in triples, exhibit the same metallicity trend as the full sample.  For visual clarity, we scale the power-law fit log\,$F_{\rm EB}$~$\propto$~$-$0.39[Fe/H] to the various subsamples in Fig.~\ref{EBfrac}.  Very wide EBs with $P$~=~100\,-\,1,000~days also display the same anti-correlation between metallicity and EB fraction (magenta histogram).  The fraction of F3V-K3V {\it Kepler} stars that are EBs with $P$~=~100\,-\,1,000~days decreases from 0.14\%\,$\pm$\,0.03\% across $-$1.0~$<$~[Fe/H]~$<$~$-$0.5 to 0.05\%\,$\pm$\,0.01\% across 0.0~$<$~[Fe/H]~$<$~+0.5 at the 2.9$\sigma$ significance level.   The consistency in the metallicity trends suggests the fractions of very close binaries ($P$~$<$~7~days) and binaries with intermediate periods ($P$~=~100\,-\,1,000 days) decrease with metallicity at the same rate.  In other words, the overall close binary fraction of solar-type stars strongly decreases with metallicity, but the underlying period distribution below $P$~$\lesssim$~1,000~days is metallicity invariant.

In Fig.~\ref{EBfrac}, we also display the EB fraction for our {\it Kepler} sample of F3V-K3V stars with measured spectroscopic metallicities (thick black histogram). For this sample, the EB fraction decreases by a factor of $\approx$\,3.5 from 1.4\%\,$\pm$\,0.4\% near [Fe/H]~=~$-$0.6 to 0.4\%\,$\pm$\,0.2\% at [Fe/H]~=~+0.4. Attempting to fit a constant EB fraction to the five black metallicity bins in Fig.~\ref{EBfrac} results in a reduced $\chi^2$/$\nu$~=~4.2 with $\nu$~=~4 degrees of freedom, which can be rejected with 2.9$\sigma$ confidence ($p$~=~0.0019).  The {\it Kepler} sample of solar-type dwarfs with measured spectroscopic metallicities is fully consistent with the relation log~$F_{\rm EB}$~$\propto$~$-$0.39[Fe/H] inferred from our photometric sample.  The EB fractions based on our photometric and spectroscopic samples are nearly identical for both sub-solar metallicities [Fe/H]~=~$-$0.5 ($F_{\rm EB}$~$\approx$~1.3\%) and super-solar metallicities [Fe/H]~=~+0.3 ($F_{\rm EB}$~$\approx$~0.6\%). Our {\it Kepler} sample with spectroscopic metallicities is unfortunately too small to further divide according to spectral type or period. Nevertheless, the consistency between our overall photometric and spectroscopic EB fractions suggests the trends in period and spectral type found within our photometric sample are statistically accurate.

\begin{figure}[t!]
\centerline{
\includegraphics[trim=0.4cm 0.3cm 0.4cm 0.2cm, clip=true, width=3.3in]{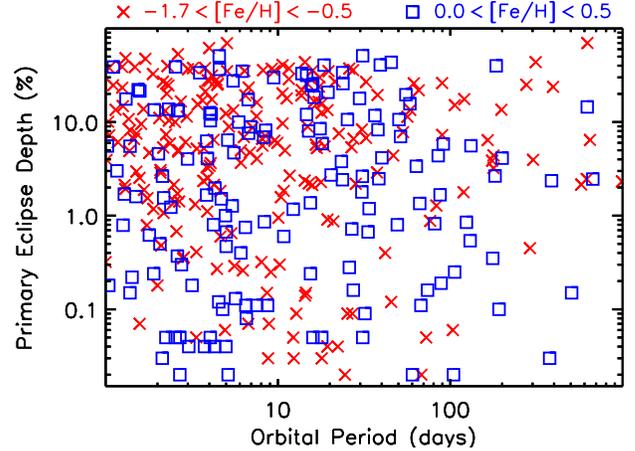}}
\caption{The measured eclipse depths versus orbital periods for the 226 metal-poor EBs ($-$1.7~$<$~[Fe/H]~$<$~$-$0.5; red crosses) and 154 metal-rich EBs (0.0~$<$~[Fe/H]~$<$~0.5; blue squares) within our photometric sample of solar-type dwarfs. Our younger, metal-rich sample exhibits a statistically significant excess of eccentric heartbeat binaries and contamination by transiting planets toward short periods $P$~$<$~10~days and small amplitudes $d_{\rm p}$~$<$~0.2\%. Outside this parameter space, the two samples have consistent period and eclipse depth distributions.  Although the close binary fraction is anti-correlated with metallicity, the period and mass-ratio distributions of close solar-type binaries are metallicity invariant. }
\label{PvsF}
\end{figure}

We perform additional KS tests to determine if the period and mass-ratio distributions of EBs within our photometric sample vary with metallicity. We compare the 226 solar-type EBs with photometric metallicities $-$1.7~$<$~[Fe/H]~$<$~$-$0.5 to the 154 EBs with 0.0~$<$~[Fe/H]~$<$~0.5.  The EB fraction of our metal-poor sample ($F_{\rm EB}$~=~1.60\%\,$\pm$\,0.11\%) is $\approx$\,3.0 times the EB fraction of the metal-rich sample ($F_{\rm EB}$~=~0.54\%\,$\pm$\,0.04\%) at the 9.2$\sigma$ significance level, consistent with the green histogram in Fig.~\ref{EBfrac}.  In Fig.~\ref{PvsF}, we plot the measured primary eclipse depths $d_{\rm p}$ as a function of orbital period $P$ for both our metal-poor and metal-rich photometric samples.  

Across the full period interval $P$~=~1\,-\,1,000~days, the metal-poor and metal-rich EBs have marginally consistent period distributions at the 1.8$\sigma$ level ($p_{\rm KS}$~=~0.065).  Metal-poor systems with [Fe/H]~$<$~$-$1.0, which are likely to be old halo or thick disk stars, may exhibit a slight deficit of long-period EBs with $P$~=~100\,-\,1,000 days. In \S\ref{EBcomp}, we attribute this to tidal evolution toward smaller eccentricities and hence a smaller probability of producing eclipses, rather than a shift in the period distribution.  In any case, it is only a 1.8$\sigma$ effect. The 188 metal-poor EBs and 112 metal-rich EBs with $P$~=~1\,-\,30~days, which all have small enough eccentricities to negligibly affect the eclipse probabilities, exhibit nearly the identical period distribution ($p_{\rm KS}$~=~0.72).

The primary eclipse depth distribution maps to the mass-ratio distribution \citep{Moe2013}.  For MS components, EBs with deep eclipses $d_{\rm p}$~$>$~25\% must have large mass ratios $q$~$\gtrsim$~0.7. EBs with shallower eclipses $d_{\rm p}$~=~1\%\,-\,25\% may have large companions in grazing, inclined orbits, but more likely contain small, low-mass companions. In general, systems with $d_{\rm p}$~$<$~1\% not only include true EBs, but also ellipsoidal binaries, transiting planets, and heartbeat stars, which are eccentric binaries that induce tidal distortions and dynamical oscillations near periastron \citep{Thompson2012}. 

A KS test demonstrates the eclipse depth distribution of our metal-poor and metal-rich subsamples are inconsistent with each other at the 3.7$\sigma$ significance level ($p_{\rm KS}$~=~1.3$\times$10$^{-4}$).  As shown in Fig.~\ref{PvsF}, our metal-rich subsample exhibits an excess of EBs with short periods $P$~$<$~10~days and shallow eclipses $d_{\rm p}$~$<$~0.2\%.  We inspected the individual light curves of these 22 systems, and found most were not true EBs. Three were ellipsoidal binaries showing sinusoidal light curves. Eight exhibited peculiar non-sinusoidal variability, six of which were flagged as heartbeat stars by \citet{Kirk2016}. An additional six did not have definitive secondary eclipses, indicative of a transiting planet, four of which were flagged by \citet{Kirk2016} as also having flat-bottomed primary eclipses.  Flat-bottomed eclipses further suggests they are transiting planets as opposed to grazing EBs.  Only five of the metal-rich systems with short periods and small amplitudes appear to be genuine EBs. Heartbeat binaries with $P$~$<$~10~days are likely to be relatively young, and therefore metal rich, to still be eccentric enough to induce strong tidal distortions at periastron \citep{Shporer2016}.  Hot Jupiters, Neptunes, and super-Earths with $P$~$<$~10~days are all significantly weighted toward metal-rich hosts with [Fe/H]~$>$~0.0 \citep{Fischer2005,Mulders2016,Owen2018}.  It is therefore not surprising that our metal-rich EB sample is contaminated more by both heartbeat stars and transiting planets. This provides further confirmation that the photometric metallicities from \citet{Huber2014} can reliably distinguish metal-poor from metal-rich systems. 

We therefore restrict our eclipse depth analysis to the 171 metal-poor and 91 metal-rich systems with $d_{\rm p}$~$>$~1.0\% that are most likely genuine EBs. For $d_{\rm p}$~$>$~1.0\%, the EB fraction of our metal-poor sample ($F_{\rm EB}$~=~1.20\%\,$\pm$\,0.09\%) is $\approx$\,3.6 times the EB fraction of the metal-rich sample ($F_{\rm EB}$~=~0.33\%\,$\pm$\,0.03\%) at the 8.8$\sigma$ level.  Focusing on genuine EBs with deeper eclipses accentuates the anti-correlation between the EB fraction and metallicity. The metal-poor and metal-rich EBs have eclipse depth distributions above $d_{\rm p}$~$>$~1.0\% that are fully consistent with each other ($p_{\rm KS}$ = 0.52).  Although the close binary fraction decreases significantly with metallicity, both the period and mass-ratio distributions of close solar-type binaries are metallicity invariant.

\subsection{Corrections for Selection Effects}
\label{EBcomp}

We calculate the eclipse probabilities $p_{\rm EB}$ to recover the intrinsic close binary fraction from the observed EB fraction.  For the full {\it Kepler} EB sample, \citet{Kirk2016} utilized the stellar radii reported in the {\it Kepler} input catalog \citep{Brown2011} to calculate $p_{\rm EB}$ as a function of period (see their Fig.~11).  Across $P$~$\approx$~3\,-\,20~days, \citet{Kirk2016} found the eclipse probabilities decrease from $p_{\rm EB}$~$\approx$~0.17 to 0.05 as expected from the geometry of circular orbits, i.e., $p_{\rm EB}$ = ($R_1$+$R_2$)/$a$.  Toward very short periods $P$~$<$~3~days,  non-eclipsing ellipsoidal binaries are detected across a wider range of inclinations compared to true EBs. 
 
Toward longer periods $P$~$>$~20~days, three additional effects modify the eclipse probabilities.  First, the majority of solar-type binaries with $P$~$>$~20~days are in eccentric orbits with $e$~$>$~0.3 \citep{Meibom2005,Raghavan2010,Tokovinin2014,Moe2017}.  For an eccentric binary, there are certain combinations of inclination and argument of periastron such that there is only one eclipse per orbit \citep{Moe2015}.  In these cases, the projected separation at the conjunction closest to periastron is small enough to produce an eclipse  while the projected separation at conjunction nearest apastron is too wide. \citet{Kirk2016} includes EBs with only one eclipse per orbit in their catalog, and so the probability of detecting eccentric EBs is larger than that of their circular counterparts.  Second, the main {\it Kepler} mission observed continuously for 17 $\approx$\,90-day quarters with small gaps between the quarters to roll the spacecraft.  A non-negligible fraction of {\it Kepler} stars fell in the chip gaps or on bad pixels during one or multiple quarters. Some EBs with long periods were therefore missed due to the duty cycle of the {\it Kepler} observations.  Finally, EBs with especially long periods $P$~$\gtrsim$~500~days were difficult to detect given the four-year timespan of the main {\it Kepler} mission.  \citet{Kirk2016} estimated only $\approx$20\% of {\it Kepler} EBs with $P$~$\approx$~1,000 days were actually identified.

\citet{Kirk2016} measured $p_{\rm EB}$($P$) for the full {\it Kepler} sample by averaging across various stellar and orbital properties.  Our culled {\it Kepler} sample contains exclusively solar-type dwarfs, which are on average smaller than the mean radii of {\it Kepler} stars as a whole. Most importantly, stellar radii depend on metallicity, and so we must account for the eclipse probabilities as a continuous function of metallicity. We therefore utilize a Monte Carlo technique to calculate $p_{\rm EB}$($P$,\,$T_{\rm eff}$,\,[Fe/H]) for our {\it Kepler} sample of solar-type dwarfs.  For a given combination of $T_{\rm eff}$ and [Fe/H], we estimate the primary mass $M_1$ and radius $R_1$ from the Dartmouth stellar evolutionary tracks \citep{Dotter2008}.  We adopt an age-metallicity relation as done in \S\ref{photmetal} and Fig.~\ref{metalbias}.  Specifically, stars with [Fe/H]~$>$~0.2 have ages $\tau_*$~=~2~Gyr, stars with [Fe/H]~$<$~$-$1.3 are $\tau_*$~=~11~Gyr old, and we linearly interpolate between these two regimes.  

In the previous sections, we adopted a uniform mass-ratio distribution, which adequately describes the overall population of close solar-type binaries with $a$~$\lesssim$~10~AU.  However, the majority of {\it Kepler} EBs have very short periods $P$~$<$~10~days ($a$~$\lesssim$~0.1~AU).  Very close solar-type binaries exhibit an excess fraction of twins with $q$ = 0.95\,-\,1.00 \citep{Tokovinin2000,Moe2017}.  We therefore adopt a twin fraction that decreases linearly with respect to log\,$P$ from $F_{\rm twin}$~=~0.30 at log\,$P$\,(days)~=~0 to $F_{\rm twin}$~=~0.15 at log\,$P$~=~3.  We generate a fraction $F_{\rm twin}$ of binaries to be uniformly distributed across $q$~=~0.95\,-\,1.00 while the remaining fraction 1$-F_{\rm twin}$ of binaries are uniformly distributed across $q$~=~0.10\,-\,0.95.  We then select $M_2$ and $R_2$ from the Dartmouth tracks accordingly.  

We adopt circular orbits below $P$~$<$~$P_{\rm circ}$~=~10~days and a uniform eccentricity distribution across 0~$<$~$e$~$<$~$e_{\rm max}$($P$) toward longer periods (see Eqn.~\ref{tide}).  We assume random orientations so that the arguments of periastron $\omega$ follow a uniform distribution.  The eclipse probability at superior and inferior conjunction is $p_{\rm sup,inf}$ = ($R_1$+$R_2$)(1\,$\pm$\,$e$\,sin\,$\omega$)/[$a$(1$-e^2$)] \citep{Kirk2016}.  By requiring only one eclipse per orbit, we adopt the larger of the two eclipse probabilities.  According to our Monte Carlo model, a population of wide binaries with $P$~=~1,000~days that are evenly distributed across 0~$<$~$e$~$<$~$e_{\rm max}$~=~0.98 are $\approx$\,3.3 times more likely to produce eclipses than binaries in circular orbits. 

Due to the (1$-e^2$) term in the denominator of the eclipse probability, the frequency of highly eccentric, long-period binaries with $e$~$>$~0.9 and $P$~$>$~100~days strongly affects the inferred close binary fraction.  In~\ref{EBmetallicities}, we noticed a small 1.8$\sigma$ discrepancy whereby our metal-poor sample exhibited a slight deficit of long-period EBs, possibly due to tidal evolution. The population of solar-type binaries in the old, metal-poor halo indeed has a slightly longer circularization period of $P_{\rm circ}$~$\approx$~15~days \citep{Meibom2005}.  Adopting a longer circularization period for our metal-poor simulations would reduce the eclipse probabilities and increase the inferred close binary fraction, thereby strengthening our main conclusion.  However, tidal evolution of binaries with long periods and large eccentricities is highly uncertain \citep{Moe2018}.  We therefore adopt $P_{\rm circ}$~=~10~days for all metallicities, and compare the corrected close binary fractions inferred from the population of EBs with $P$~$<$~1,000~days and $P$~$<$~100~days (see below).  

For $P$~=~3\,-\,20~days, the eclipse probabilities $p_{\rm EB}$ are completely described by the geometry of the orbits.  Toward shorter periods, we account for the enhanced probability of detecting ellipsoidal binaries, whereby $p_{\rm EB}$ reaches 1.2 times the pure eclipse probability at $P$~=~1~day. Toward longer periods, we assume the probabilities are suppressed by a reduction factor of 80\% at $P$~=~300~days and 20\% at $P$~=~1,000~days to correct for the duty cycle and four-year timespan of the {\it Kepler} observations (see Fig.~11 in \citealt{Kirk2016}). We linearly interpolate these correction factors with respect to log\,$P$.

Because a significant fraction of very close EBs are twins, we must also account for Malmquist bias.  Given the same magnitude limit, twin binaries are observed up to $\sqrt{2}$~$\approx$~1.4 times the distance and are therefore overrepresented by a factor of 2$^{\nicefrac{3}{2}}$~$\approx$~2.8 compared to a volume-limited sample.  We weight $p_{\rm EB}$ according to the combined luminosities $L_1$+$L_2$ so that twin binaries have 2.8 times the probability than single stars and binaries with faint companions.  

\begin{figure}[t!]
\centerline{
\includegraphics[trim=0.4cm 0.3cm 0.4cm 0.2cm, clip=true, width=3.3in]{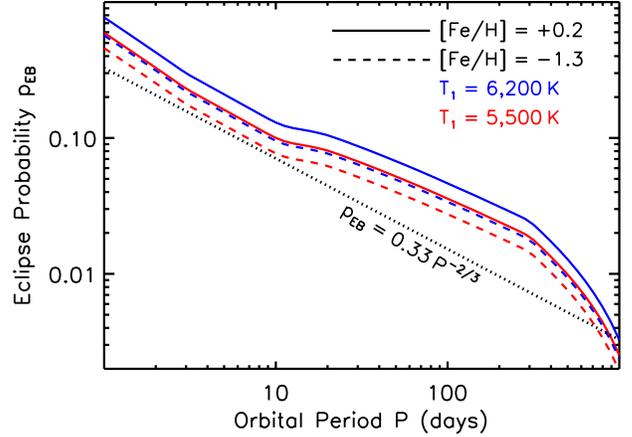}}
\caption{The eclipse probabilities $p_{\rm EB}$ of solar-type binaries as a function of orbital period for metallicities [Fe/H] = $-$1.3 (dashed) and +0.2 (solid) and for primary temperatures $T_{\rm eff}$ = 5,500\,K (red) and 6,200\,K (blue). We also show the eclipse probability $p_{\rm EB}$~=~0.33\,($P$/day)$^{\nicefrac{-2}{3}}$ based solely on geometrical selection effects for a solar-metallicity binary with $M_1$~=~1\,\Msun, $M_2$~=~0.5\,\Msun, and $e$~=~0.0 (dotted).  Compared to this simple power-law approximation, ellipsoidal variability and Malmquist bias increase $p_{\rm EB}$ at short periods, eccentric orbits further increase $p_{\rm EB}$ across intermediate periods, and the duty cycle and four-year timespan of the {\it Kepler} observations reduce $p_{\rm EB}$ toward long periods.}
\label{EBprob}
\end{figure}

We present our results for $p_{\rm EB}$($P$,\,$T_{\rm eff}$,\,[Fe/H]) in Fig.~\ref{EBprob} for the same combinations of primary temperatures $T_{\rm eff}$~=~5,500\,K and 6,200\,K and metallicities [Fe/H] = $-$1.3 and +0.2 investigated in \S\ref{photmetal} and Fig.~\ref{metalbias}.  The eclipse probabilities dramatically decrease with orbital period as expected, but there are also noticeable variations with respect to metallicity and primary temperature.  Given the same metallicity, F~dwarfs are larger than G~dwarfs, and so the eclipse probabilities of binaries containing $T_{\rm eff}$ = 6,200\,K primaries are $\approx$\,20\%\,-\,30\% larger than those with $T_{\rm eff}$~=~5,500\,K. Similarly, metal-rich dwarfs are larger given the same effective temperatures, and so the eclipse probabilities of metal-rich binaries with [Fe/H]~=~0.2 are $\approx$\,25\%\,-\,30\% larger than those of metal-poor binaries with [Fe/H]~=~$-$1.3.

For comparison, we also display in Fig.~\ref{EBprob} the eclipse probabilities $p_{\rm EB}$($P$) for a solar-metallicity binary with $M_1$~=~1.0\,\Msun, $M_2$~=~0.5\,\Msun, and $e$~=~0.0.  In this case, we do not account for ellipsoidal variability, Malmquist bias, or the duty cycle of the {\it Kepler} observations, and therefore the eclipse probabilities follow $p_{\rm EB}$~=~0.33\,($P$\,[day])$^{\nicefrac{-2}{3}}$. Toward very short periods $P$~$<$~10~days, the Malmquist bias associated with the excess twin population substantially elevates the true eclipse probabilities above the simple model.  Across intermediate periods $P$~$\approx$~10\,-\,300~days, eccentric EBs further increase $p_{\rm EB}$.  Only toward the longest periods do the duty cycle and timespan of the {\it Kepler} observations reduce $p_{\rm EB}$ below the simple power-law approximation.  

For each EB, we compute the eclipse probability $p_{\rm EB}$($P$,\,$T_{\rm eff}$,\,[Fe/H]) based on its measured period, primary temperature, and metallicity.  We calculate the corrected binary fraction below $P$~$<$~1,000~days by summing the inverse of the eclipse probabilities $p_{\rm EB}$ for both our photometric and spectroscopic samples according to the metallicity intervals investigated in Fig.~\ref{EBfrac}.  Specifically, we measure:

\begin{align}
F_{\rm P<1000d}&({\rm [Fe/H]}) = \frac{1}{N({\rm [Fe/H]})} \times \nonumber \\
 & \sum_i^{N_{\rm EB}({\rm [Fe/H]})} \frac{1}{p_{{\rm EB},i}(P_i,\,T_{{\rm eff},i},\,{\rm [Fe/H}]_i)}
\end{align}

\noindent where $N$([Fe/H]) is the total number of solar-type dwarfs in a specific metallicity interval and $N_{\rm EB}$([Fe/H]) is the number of those stars that have eclipsing companions across $P$~=~1\,-\,1,000~days. We perform jackknife resamplings of our systems to measure the uncertainties in $F_{\rm P<1000d}$([Fe/H]).  

\begin{figure}[t!]
\centerline{
\includegraphics[trim=0.4cm 0.3cm 0.4cm 0.2cm, clip=true, width=3.5in]{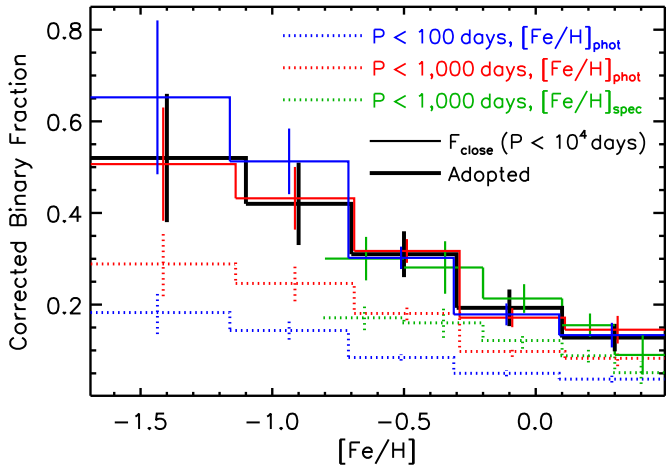}}
\caption{As a function of metallicity, the corrected binary fraction $F_{\rm P<1000d}$ below $P$~$<$~1,000~days for our photometric (dotted red) and spectroscopic (dotted green) samples of {\it Kepler} solar-type dwarfs, and the corrected binary fraction $F_{\rm P<100d}$ below $P$~$<$~100~days for the photometric sample (dotted blue).  We divide $F_{\rm P<1000d}$ and $F_{\rm P<100d}$ by 0.57 and 0.27, respectively, to recover the intrinsic close binary fraction $F_{\rm close}$ below $P$~$<$~10$^4$~days (solid colored histograms). All three histograms for $F_{\rm close}$ are consistent with each other, and so we adopt a moving weighted average (thick black) that decreases from $F_{\rm close}$~=~52\%\,$\pm$\,14\% across $-$1.7~$<$~[Fe/H]~$<$~$-$1.1 to $F_{\rm close}$~=~13\%\,$\pm$3\% across 0.1~$<$~[Fe/H]~$<$~0.5.}
\label{binfrac_EB}
\end{figure}

We present $F_{\rm P<1000d}$([Fe/H]) for both our photometric and spectroscopic samples of {\it Kepler} solar-type dwarfs in Fig.~\ref{binfrac_EB} (dotted red and green histograms, respectively).  According to our sample with photometric metallicities, the corrected binary fraction below $P$~$<$~1,000~days decreases from $F_{\rm P<1000d}$~=~0.29\,$\pm$\,0.07 near [Fe/H]~=~$-$1.4 to $F_{\rm P<1000d}$~=~0.08\,$\pm$\,0.02 at [Fe/H]~=~0.3.  The {\it Kepler} sample with spectroscopic metallicities exhibits a consistent trend, whereby the corrected binary fraction decreases from $F_{\rm P<1000d}$~=~0.17\,$\pm$\,0.03 near [Fe/H]~=~$-$0.6 to $F_{\rm P<1000d}$~=~0.05\,$\pm$\,0.02 at [Fe/H]~=~0.4.  The {\it Kepler} sample of EBs with $P$~=~100\,-\,1,000 days is relatively small, and the uncertainties in their eclipse probabilities may be relatively large (see above).  For our {\it Kepler} sample with photometric metallicities, we therefore also compute $F_{\rm P<100d}$([Fe/H]) by summing $p_{EB}^{-1}$ for only those EBs with $P$~=~1\,-\,100~days. The resulting corrected binary fraction below $P$~$<$~100~days decreases from $F_{\rm P<100d}$ = 0.18\,$\pm$\,0.05 near [Fe/H]~=~$-$1.4 to $F_{\rm P<100d}$ = 0.04\,$\pm$\,0.01 at [Fe/H]~=~0.3 (dotted blue histogram in Fig.~\ref{binfrac_EB}).  

According to our adopted short-end tail of a log-normal period distribution, 57\% of close solar-type binaries with $P$~$<$~10$^4$ days have $P$~$<$~1,000~days. We therefore divide $F_{\rm P<1000d}$ by 0.57 to recover the intrinsic close binary fraction $F_{\rm close}$.  Similarly, 27\% of close solar-type binaries have short periods $P$~$<$~100~days, so we divide $F_{\rm P<100d}$ by 0.27 to measure $F_{\rm close}$.  The three methods for measuring $F_{\rm close}$ from the {\it Kepler} sample of solar-type EBs are all consistent with each other (see thin colored histograms in Fig.~\ref{binfrac_EB}).  The consistency between our photometric and spectroscopic samples further demonstrates the metallicities of our {\it Kepler} solar-type dwarfs are sufficiently calibrated to reliably measure $F_{\rm close}$([Fe/H]).  In addition, the similarity in $F_{\rm close}$ inferred from $F_{\rm P<1000d}$ and $F_{\rm P<100d}$ confirms both metal-poor and metal-rich solar-type binaries follow the same short-end tail of a log-normal period distribution. 

We calculate a moving weighted average utilizing the three histograms for $F_{\rm close}$([Fe/H]) in Fig.~\ref{binfrac_EB}. We adopt the measurement uncertainties according to the photometric sample of EBs with $P$~=~1\,-\,1,000~days.  Given the model uncertainties in the eclipse probabilities $p_{\rm EB}$ and the extension of the period distribution beyond $P$~$>$~1,000~days, we also add a systematic uncertainty of $\delta F_{\rm close}$/$F_{\rm close}$ = 15\% in quadrature with the measurement uncertainties.  We show our final $F_{\rm close}$([Fe/H]) based on {\it Kepler} EBs as the thick black histogram in Fig.~\ref{binfrac_EB}.  The corrected close binary fraction decreases from $F_{\rm close}$~=~0.52\,$\pm$\,0.14 for [Fe/H]~=~$-$1.4\,$\pm$\,0.3 to $F_{\rm close}$~=~0.13\,$\pm$\,0.03 for [Fe/H]~=~0.3\,$\pm$\,0.2.  The relative decrease in the corrected close binary fraction (0.52/0.13 = 4.0) is slightly larger than the decrease in the observed EB fraction (factor of 3.4 across the same metallicity interval; see \S\ref{EBmetallicities}).  This is because the eclipse probabilities of metal-poor binaries are smaller  (see above and Fig.~\ref{EBprob}), and so their intrinsic close binary fractions are even larger. Correcting for incompleteness further strengthens our conclusion that the close binary fraction of solar-type stars decreases with metallicity.

\section{Summary of Observational Constraints}
\label{Summary}

\subsection{Close Binary Fraction of Solar-type Stars}

\begin{figure*}[t!]
\centerline{
\includegraphics[trim=0.3cm 0.1cm 0.3cm 0.2cm, clip=true, width=5.6in]{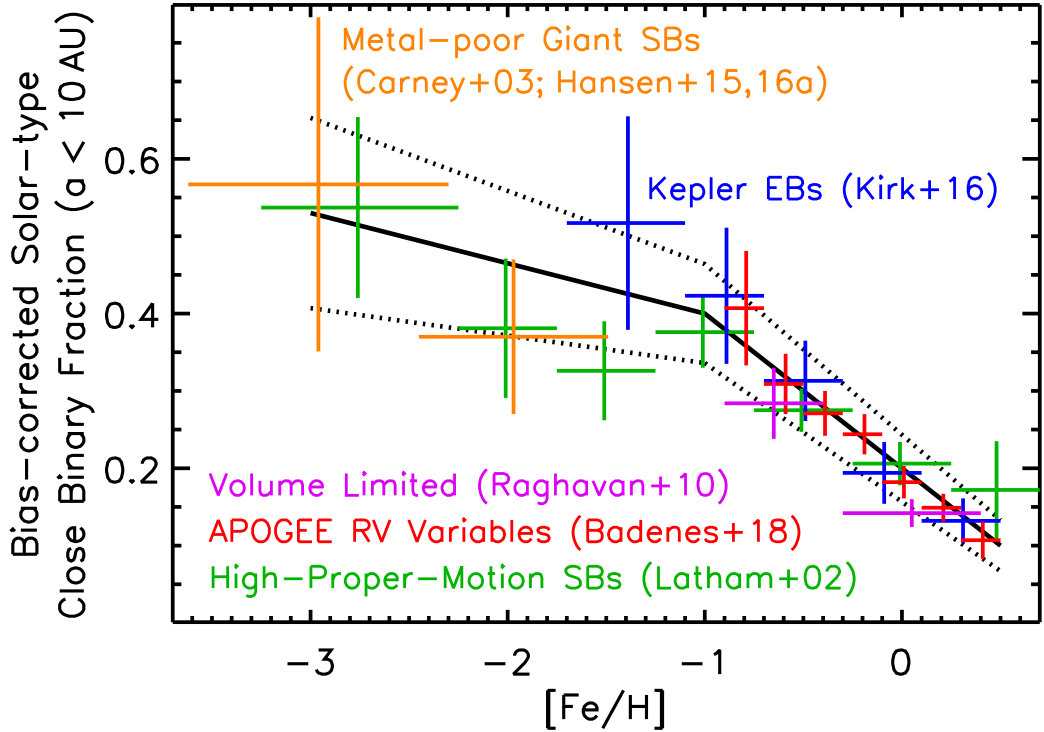}}
\caption{The intrinsic close binary fraction ($P$~$<$~10$^4$~days; $a$~$<$~10~AU) of $M_1$~$\approx$~1\,\Msun\ primaries as a function of metallicity {\it after} correcting for incompleteness and other selection biases.  We compare the measurements from: (1) SBs in samples of metal-poor giants (orange), (2) {\it Kepler} EBs with solar-type dwarf primaries (blue), (3) a volume-limited sample of solar-type primaries (magenta), (4) RV variables in the APOGEE survey of GK\,IV/V stars (red), and (5) SBs in the Carney-Latham survey of high-proper-motion stars (green).  All five samples / methods show a consistent metallicity trend that can be fitted by two line segments (black) in which the close binary fraction decreases from $F_{\rm close}$~=~53\%\,$\pm$\,12\% at [Fe/H]~=~$-$3.0 to $F_{\rm close}$~=~40\%\,$\pm$\,6\% at [Fe/H]~=~$-$1.0 and then to $F_{\rm close}$~=~10\%\,$\pm$\,3\% at [Fe/H]~=~+0.5. Even after accounting for systematic uncertainties, the close binary fraction of solar-type stars is anti-correlated with metallicity at the $\approx$\,9$\sigma$ significance level.}
\label{allbin}
\end{figure*}

A variety of observational techniques all confirm the close binary fraction of solar-type stars dramatically decreases with metallicity.  In Fig.~\ref{allbin}, we display the bias-corrected close binary fraction $F_{\rm close}$ across log\,$P$\,(days) = 0\,-\,4 ($a$ $\lesssim$ 10 AU) as a function of metallicity determined from SBs in the Carney-Latham survey of high-proper-motion stars (\S\ref{Latham}), SBs in samples of metal-poor giants (\S\ref{Giants}), RV variables in the APOGEE survey of GK\,IV/V stars (\S\ref{APOGEE}), and {\it Kepler} EBs with F3V-K3V primaries (\S\ref{Kepler}).  Based on the \citet{Raghavan2010} volume-limited sample of solar-type stars, we also showed in \S\ref{Overview} that the binary fraction below log\,$P$\,(days)~$<$~6 ($a$ $\lesssim$ 200 AU) is 50\%\,$\pm$\,8\% across $-$0.9~$<$~[Fe/H]~$<$~$-$0.4 and 25\%\,$\pm$\,2\% across $-$0.3~$<$~[Fe/H]~$<$~0.4. According to our adopted log-normal period distribution, 55\% of binaries below log\,$P$\,(days)~$<$~6 are close binaries with log\,$P$\,(days)~$<$~4.  This provides close binary fractions of $F_{\rm close}$ = 28\%\,$\pm$\,5\% and 14\%\,$\pm$\,2\% across $-$0.9~$<$~[Fe/H]~$<$~$-$0.4 and $-$0.3~$<$~[Fe/H]~$<$~0.4, respectively, which we also show in Fig.~\ref{allbin}.

All five samples / methods presented in Fig.~\ref{allbin} exhibit a quantitatively consistent anti-correlation between $F_{\rm close}$ and [Fe/H]. Because of the different methods used to identify binaries in the various samples, it is difficult for them to conspire to produce consistent results erroneously.  The error bars for each of the data points in Fig.~\ref{allbin} not only incorporate the measurement uncertainties according to their respective sample sizes, but also the systematic uncertainties in transforming the observed (incomplete) close binary fractions into intrinsic bias-corrected close binary fractions.  Attempting to fit a constant $F_{\rm close}$ to the 23 independent measurements in Fig.~\ref{allbin} results in a reduced $\chi^2$/$\nu$~=~6.2 with $\nu$~=~22  degrees of freedom.  Even after considering systematic uncertainties, we can reject the null hypothesis that the close binary fraction of solar-type stars is invariant with respect to metallicity at the 8.7$\sigma$ significance level ($p$~=~2.2$\times$10$^{-18}$).  

We instead adopt a weighted moving average for $F_{\rm close}$([Fe/H]) that can be accurately fitted by two line segments. The corrected close binary fraction of solar-type stars decreases from $F_{\rm close}$~=~53\%\,$\pm$\,12\% at [Fe/H]~=~$-$3.0 to $F_{\rm close}$~=~40\%\,$\pm$\,6\% at [Fe/H]~=~$-$1.0, and then to $F_{\rm close}$~=~10\%\,$\pm$\,3\% at [Fe/H]~=~+0.5.  We display our two-segment fit to the various observations in Fig.~\ref{allbin}.  Across the full metallicity interval $-$3.0~$<$~[Fe/H]~$<$~0.5, the close binary fraction of solar-type stars decreases by a factor of $\approx$\,5. Metal-poor halo stars clearly have a higher close binary fraction than metal-rich disk stars. Most of the variation in $F_{\rm close}$ occurs across the narrower interval $-$1.0~$<$~[Fe/H]~$<$~0.5, whereby the close binary fraction decreases by a factor of $\approx$\,4.  Even within the galactic disk, the close binary fraction of solar-type stars decreases dramatically with metallicity. By interpolating our fit at the mean metallicity of the field, i.e., [Fe/H]~$\approx$~$-$0.2, we measure a close binary fraction of $F_{\rm close}$ = 24\%\,$\pm$\,4\%.  This matches the close binary fraction inferred from volume-limited samples of solar-type stars in the solar neighborhood \citep{Duquennoy1991,Raghavan2010,Tokovinin2014,Moe2017}.

\subsection{Binary Period Distributions}

Solar-type binaries in the field follow a log-normal companion period distribution that peaks at log\,$P$\,(days) = 4.9 ($a_{\rm peak}$~$\approx$~40~AU) with a dispersion of $\sigma_{\rm logP}$~=~2.3 \citep{Duquennoy1991,Raghavan2010,Tokovinin2014}. After making small corrections for incompleteness \citep{Chini2014,Moe2017}, the single, binary, triple, and quadruple star fractions are $F_{\rm single}$~$\approx$~51\%, $F_{\rm binary}$~$\approx$~34\%, $F_{\rm triple}$~$\approx$~12\%, and $F_{\rm quadruple}$~$\approx$~3\%, respectively.  These fractions provide the average multiplicity frequency of companions per primary of $f_{\rm mult}$ = $F_{\rm binary}$ + 2$F_{\rm triple}$ + 3$F_{\rm quadruple}$ = 0.67\,$\pm$\,0.05. We define the frequency $f_{\rm logP}$ of stellar companions per decade of orbital period such that:

\begin{equation}
f_{\rm mult} = \int_0^9 f_{\rm logP}\,d{\rm log}P.
\end{equation}

\begin{figure}[t!]
\centerline{
\includegraphics[trim=0.3cm 0.1cm 0.3cm 0.2cm, clip=true, width=3.5in]{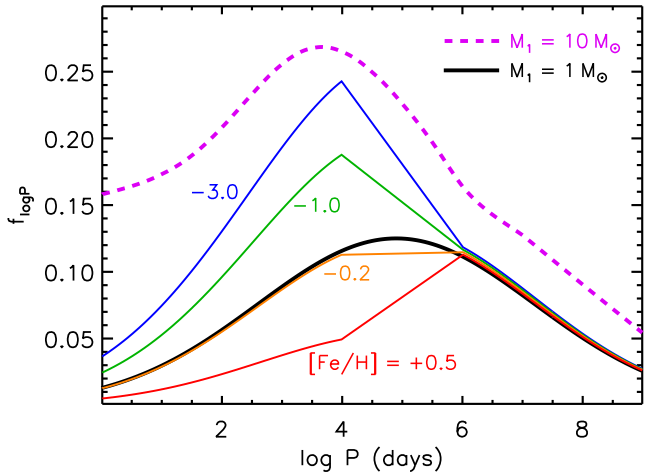}}
\caption{The frequency $f_{\rm logP}$ of stellar companions per decade of orbital period.  We compare the canonical log-normal period distribution of solar-type multiples in the solar neighborhood (thick black) to the companion distribution of early-B stars (thick dashed magenta).  We also show the metallicity-dependent period distributions for solar-type primaries with [Fe/H] = $-$3.0 (blue), $-$1.0 (green), $-$0.2 (orange), and +0.5 (red).  The close binary fraction (log\,$P$~$<$~4; $a$~$<$~10~AU) of solar-type stars is significantly anti-correlated with metallicity while the frequency of wide companions (log\,$P$~$>$~6; $a$~$>$~200~AU) is metallicity invariant.  As solar-type stars decrease in metallicity, both their binary fraction and binary period distribution approaches that of early-B stars.}
\label{Pdist}
\end{figure}

\noindent In Fig.~\ref{Pdist}, we plot the log-normal period distribution $f_{\rm logP}$ of solar-type multiples in the solar neighborhood scaled to $f_{\rm mult}$~=~0.67 across log\,$P$\,(days)~=~0\,-\,9 (thick black curve).

We found five lines of evidence that the period distribution of solar-type binaries across log\,$P$\,(days)~=~0\,-\,4 ($a$~$<$~10~AU) is relatively independent of metallicity  but simply scales according to $F_{\rm close}$.  First, the anti-correlation between the SB fraction and metallicity occurs across a broad range of periods $P$~=~20\,-\,2,000~days (Fig.~\ref{Latham_fM}). Second, the RV variability fraction decreases with metallicity at the same rate for both close companions to GK dwarfs and wide companions orbiting giants (see Fig.~\ref{DeltaRV} and \citealt{Badenes2018}).  Third, the observed distribution of RV amplitudes across $\Delta$RV$_{\rm max}$~=~1\,-\,10~km~s$^{-1}$ is independent of metallicity and consistent with the short-period tail of our adopted log-normal period distribution (\S\ref{APOGEE}).  Fourth, the same anti-correlation between the {\it Kepler} EB fraction and metallicity is observed across a wide range of periods $P$~$\approx$~1\,-\,1,000~days (Fig.~\ref{EBfrac}).  Finally, both metal-poor and metal-rich {\it Kepler EBs} have the same period and eclipse-depth distributions, suggesting the period and mass-ratio distributions of close solar-type binaries are metallicity invariant (Fig.~\ref{PvsF}).  In Fig.~\ref{Pdist}, we display the short-period tail (log $P$~=~0\,-\,4) of our adopted log-normal period distribution scaled to $F_{\rm close}$ for the four metallicities [Fe/H] = $-$3.0, $-$1.0, $-$0.2, and +0.5 evaluated above (solid colored curves).

Meanwhile, as discussed in \S\ref{Overview}, observations of visual and common-proper-motion binaries demonstrate that the wide binary fraction of solar-type stars is relatively independent of metallicity \citep{Chaname2004,Zapatero2004}.  We also showed in \S\ref{Overview} that the frequency of wide companions with log\,$P$\,(days)~$>$~6 ($a$ $\gtrsim$ 200 AU) in the \citet{Raghavan2010} sample is independent of metallicity.  Based on volume-limited samples of solar-type stars \citep{Duquennoy1991,Raghavan2010,Tokovinin2014}, and after making small corrections for incompleteness \citep{Chini2014,Moe2017}, we estimate the frequency of companions across log\,$P$\,(days) = 6\,-\,9 ($a$~=~200\,-\,20,000~AU) is $f_{\rm wide}$ = 0.21\,$\pm$\,0.03. As shown in Fig.~\ref{Pdist}, the long-period tail of companions to solar-type stars follows our adopted log-normal period distribution scaled to $f_{\rm wide}$~=~0.21 across log\,$P$\,(days) = 6\,-\,9, independent of metallicity.

There is a transition region across intermediate periods log\,$P$\,(days)~=~4\,-\,6 ($a$~$\approx$~10\,-\,200~AU).  For simplicity, we linearly interpolate the period distribution with respect to log\,$P$ between close binaries (log\,$P$~$<$~4) that exhibit a strong metallicity dependence and very wide binaries (log\,$P$~$>$~6) that are metallicity invariant.   Our distribution for [Fe/H]~=~$-$0.2 in Fig.~\ref{Pdist} nearly coincides with the log-normal distribution of solar-type binaries in the solar neighborhood, which also have $\langle$[Fe/H]$\rangle$~$\approx$~$-$0.2.  Metal-poor solar-type binaries peak at log\,$P$\,(day)~$\approx$~4 ($a_{\rm peak}$ $\approx$ 10 AU) while solar-type binaries with super-solar metallicity peak at log\,$P$\,(day) $\approx$ 6 ($a_{\rm peak}$~$\approx$~200 AU).  This is consistent with the results in \citet{Rastegaev2010}, who also found metal-poor solar-type binaries peak at shorter separations compared to solar-type binaries in the solar neighborhood.  

By integrating $f_{\rm logP}$, we measure multiplicity frequencies of $f_{\rm mult}$ = 1.11, 0.92, 0.66, and 0.47 for solar-type primaries with [Fe/H] = $-$3.0, $-$1.0, $-$0.2, and +0.5, respectively. Our [Fe/H]~=~$-$0.2 multiplicity frequency of $f_{\rm mult}$~=~0.66 nearly matches the measured value $f_{\rm mult}$~=~0.67\,$\pm$\,0.05 for solar-type systems in the field.   As the close binary fraction of solar-type stars increases toward smaller metallicities, the triple star fraction also increases. For solar-type stars in the field, about half of wide companions are outer tertiaries in hierarchical triples, and the overall triple/quadruple star fraction is $F_{\rm triple}$\,+\,$F_{\rm quadruple}$~$\approx$~15\% \citep{Raghavan2010,Tokovinin2014,Chini2014,Moe2017}.  If the close binary fraction doubles toward  decreasing metallicity compared to the field population, then nearly all wide companions to metal-poor stars are outer tertiaries.  A similar effect is observed for massive OB stars, which also have a large close binary fraction (see below), whereby nearly all wide companions ($a$~$\gtrsim$~100~AU) are outer tertiaries in triples \citep{Sana2014,Moe2017}.  Not only are half of extremely metal-poor solar-type stars in close binaries ($F_{\rm close}$ $\approx$ 50\%), but a substantial fraction are also in triples and quadruples, i.e., $F_{\rm triple}$\,+\,$F_{\rm quadruple}$~$\approx$~35\%. 

\subsection{Comparison to Massive Binaries}

We next investigate the multiplicity properties of early-B stars with $M_1$~$\approx$~6\,-\,17\,\Msun\ ($\langle M_1 \rangle$ $\approx$ 10\,\Msun).  \citet{Moe2017} compiled several surveys of early-B MS stars in the Milky Way and Magellanic Clouds ($-$0.7~$\lesssim$ [Fe/H]~$\lesssim$~0.1) to fit $f_{\rm logP}$ across all periods (see green and blue data points in their Fig.~37).  The measured companion frequency is $f_{\rm logP}$~$\approx$~0.15\,-\,0.20 across log\,$P$\,(days)~=~0\,-\,2 according to observations of spectroscopic \citep{Levato1987,Abt1990,Kobulnicky2014} and eclipsing \citep{Moe2013,Moe2015} early-B binaries.  The period distribution then peaks across log\,$P$\,(days)~=~3\,-\,4 ($a$~$\approx$~10~AU) at $f_{\rm logP}$~$\approx$~0.25\,-\,0.30 based on long-baseline interferometry of early-B primaries \citep{Rizzuto2013} and spectroscopic RV observations of Cepheids, which evolved from early-B primaries \citep{Evans2015}. The frequency then declines to  $f_{\rm logP}$~$\approx$~0.10\,-\,0.20 across log\,$P$\,(days)~=~5\,-\,7 according to adaptive optics, speckle imaging, visual observations, and common-proper-motion astrometry of wide companions to early-B stars \citep{Abt1990,Duchene2001,Shatsky2002,Peter2012}. The dashed magenta curve in Fig.~\ref{Pdist} is consistent with all of these observational constraints.  

Integrating the dashed magenta curve in Fig.~\ref{Pdist} yields a multiplicity frequency of $f_{\rm mult}$ = 1.62 for $M_1$~=~10\,\Msun.  This is consistent with the value of $f_{\rm mult}$~=~1.6\,$\pm$\,0.2 reported in \citet{Moe2017} for early-B primaries (see their Table 13).  Integrating $f_{\rm logP}$ across 0~$<$~log\,$P$\,(days)~$<$~4 results in a close companion {\it frequency} of $f_{\rm close}$ = 0.85. The majority of these companions are in close binaries, i.e., $F_{\rm close}$~=~70\%\,$\pm$\,11\% of $M_1$~=~10\,\Msun\ primaries have stellar companions below log\,$P$\,(days)~$<$~4.  The remaining companions are outer tertiaries in compact triples, i.e., $\approx$\,15\% of $M_1$~=~10\,\Msun\ primaries are in compact triples in which the outer tertiary is below log\,$P_{\rm outer}$\,(days)~$<$~4 \citep[see][]{Moe2017}. 

The close binary fraction of early-B primaries ($F_{\rm close}$~=~70\%\,$\pm$\,11\%) is considerably larger than that of solar-type stars in the field with $\langle$[Fe/H]$\rangle$ $\approx$ $-$0.2 (24\%\,$\pm$\,4\%), but is only slightly larger than that of extremely metal-poor FGK stars with [Fe/H]~$\approx$~$-$3.0 (53\%\,$\pm$\,12\%).  The separation distribution of companions to early-B primaries peaks at $a_{\rm peak}$~$\approx$~10~AU \citep{Rizzuto2013,Evans2015,Moe2017}. This is shorter than the peak in the field solar-type binary period distribution ($a_{\rm peak}$~$\approx$~40~AU), but is consistent with the peak for metal-poor solar-type binaries ($a_{\rm peak}$~$\approx$~10~AU).  As solar-type stars decrease in metallicity, both their binary fraction and binary period distribution approaches that of early-B stars (see Fig.~\ref{Pdist}). 

We divided our APOGEE RV and {\it Kepler} EB samples according to spectral type, and we found the same degree of anti-correlation between the close binary fraction and metallicity across a broad range of primary masses $M_1$~$\approx$~0.6\,-\,1.5\,\Msun.  Meanwhile, as discussed in \S\ref{Overview}, the multiplicity properties of massive stars are relatively independent of metallicity \citep{Moe2013,Dunstall2015,Almeida2017}. In particular, \citet{Moe2013} found the close binary fraction of early-B primaries with $M_1$~$\approx$~6\,-\,16\,\Msun\ decreases by less than $\Delta F_{\rm close}$/$F_{\rm close}$~$<$~20\% across $-$0.7~$<$~[Fe/H]~$<$~0.1.  Across this same metallicity interval, the close binary fraction of solar-type stars decreases by a factor of $\approx$\,1.9 from $F_{\rm close}$~=~34\%\,$\pm$\,5\% to 18\%\,$\pm$\,4\% (see \ref{allbin}).  In \S\ref{Models}, we discuss disk fragmentation models that explain why the close binary fraction of solar-type stars is strongly anti-correlated with metallicity while the close binary fraction of massive stars is higher but relatively insensitive to metallicity.

\subsection{Implications for Binary Evolution}

The anti-correlation between metallicity and the close binary fraction of solar-type stars has profound implications for binary evolution.  All close solar-type binaries with $P$~$<$~10$^4$~days ($a$~$\lesssim$~10~AU) will interact in some manner, either through Roche-lobe overflow or wind accretion.  Companions to blue stragglers have been observed up to $P$~$\approx$~3,000 days ($a$~$\approx$~5~AU; \citealt{Mathieu2009}), companions to barium stars extend to $P$~$\approx$~20,000 days ($a$~$\approx$~20~AU; \citealt{Jorissen1998,VanderSwaelmen2017}), and the widest known symbiotic, Mira, has an orbital period of $P$~$\approx$~500~years ($a$~$\approx$~80\,AU; \citealt{Prieur2002,Sokoloski2010}). Future studies of blue stragglers, barium stars, cataclysmic variables, novae, and symbiotics must consider the effects of a metallicity-dependent close binary fraction.  The metallicity trend likely extends to intermediate masses $M_1$~$\approx$~2\,-\,5\,\Msun\ (at least to some extent), and therefore is also important for Type Ia supernovae.  

More than half of solar-type stars with [Fe/H]~$\lesssim$~$-$1.0 will interact with a stellar companion.  The fraction of solar-type stars that experience significant binary evolution in metal-poor environments, e.g., the galactic halo, dwarf galaxies, and high-redshift universe, is more than double the fraction in the field.  About 20\% of stars in the galactic bulge \citep{Ness2016,GarciaPerez2018} and most of the stars in the thick disk \citep{Ruchti2011,Beers2014} also have [Fe/H]~$\lesssim$~$-$1.0, and therefore have higher rates of binary interactions.  Although the binary fraction in dense globular clusters has significantly evolved due to dynamical interactions, the initial close binary fraction of metal-poor solar-type stars in globular clusters must have been large, consistent with the results of N-body simulations \citep{Ivanova2005}.  The metallicity distribution of all stars that have ever formed, including the progenitors of compact remnants, are weighted toward lower metallicities than systematically younger stars still on the MS.  The number of compact remnants in binaries is therefore larger than previously anticipated due to the larger binary fraction at lower metallicities. For example, $\approx$\,20\% of close solar-type binaries contain WD secondaries \citep{Moe2017,Murphy2018}, which is slightly larger than that predicted by population synthesis studies. 

A substantial fraction of metal-poor stars that have recently evolved off the MS, e.g., giants and planetary nebulae (PN), have been influenced by binary interactions. The IMF is significantly weighted toward low-mass stars \citep{Bastian2010,Kroupa2013} and the Milky Way star formation rate was $\approx$\,3 times larger $\approx$\,10~Gyr ago than it is now \citep{Governato2007,DeLucia2014}.  Based on the measured IMF and modeled galactic star formation history, we estimate $\approx$\,55\% of Milky Way giants and PN have old, solar-type progenitors ($\tau_*$~$>$~7~Gyr, $M$~$\approx$~0.8\,-\,1.2\,\Msun).  Such old, low-mass giants tend to be metal poor \citep{Ratnatunga1991,Carollo2010,Mackereth2017}. The metallicity trend therefore dramatically affects the properties of low-mass evolved stars.  For example, the enhanced close binary fraction of metal-poor solar-type stars substantially strengthens the conclusion that the shaping of PN morphologies is the result of binary interactions \citep{Moe2006,DeMarco2009,Jones2017}. Providing further corroboration, \citet{Badenes2015} measured the delay-time distribution of bright PN in the LMC and discovered two distinct populations of PN progenitors: an old channel ($\tau_*$~=~5\,-\,8~Gyr) deriving from solar-type stars ($M$~$\approx$~1.0\,-\,1.2\,\Msun) and a young channel (35\,-\,800~Myr) evolving from late-B/early-A stars ($\approx$\,2\,-\,8\,\Msun).  According to the measured age-metallicity relation of the LMC \citep{Olszewski1991,Pagel1998,Cole2005,Carrera2011,Piatti2013}, the old, solar-type progenitors are metal-poor ([Fe/H]~$\lesssim$~$-$1.0) and hence have a large close binary fraction of $F_{\rm close}$~=~40\%\,-\,50\%.  The young progenitors have a higher metallicity of [Fe/H]~$\approx$~$-$0.4, but are sufficiently massive so that they also have a large close binary fraction of $F_{\rm close}$~=~40\%\,-\,60\%.  Meanwhile, evolved stars with intermediate masses ($M$~$\approx$~1.2\,-\,2.0\,\Msun) in the LMC have intermediate metallicities, and therefore have a smaller close binary fraction of $F_{\rm close}$~$\approx$~30\%.  If PN derive from interactions in close binaries, then the variations in $F_{\rm close}$ with respect to mass and metallicity can explain the observed bimodal mass/age distribution of PN progenitors in the LMC.

\section{Fragmentation Models}
\label{Models}

Binary star formation is thought to occur through two primary channels. On large scales, turbulent core fragmentation creates binaries originally separated by 1000s of AU \citep{Fisher2004,Bate2008,Offner2010}. On smaller scales, individual disks around young stars can become unstable due to strong self-gravity and fragment into multiple stellar or sub-stellar mass objects on scales of 10s\,-\,100s of AU \citep{ARS89,Bonnell1994}. Previous work has shown that the enhanced multiplicity of higher mass stars, particularly at close separations, likely derives from the increased likelihood of disk fragmentation \citep{Kratter2006, Kratter2008, Krumholz2007a, Moe2017, Moe2018}. The observed close binary fraction versus metallicity anti-correlation (Fig.~\ref{allbin}) suggests that disk fragmentation should occur more frequently for solar-type protostars as the metallicity decreases. Since the IMF and wide binary fraction do not change within the measurement uncertainties, we expect core fragmentation to be relatively independent of metallicity.   We review previous models of the metallicity dependence below, and subsequently present a simple argument as to why enhanced disk fragmentation in low-mass protostars should be a consequence of low metallicity. 

\subsection{Previous Models of Fragmentation \\ at Low Metallicity}

Previous models are in tension regarding the effect of metallicity on stellar populations.  Given the same initial conditions but varying the metallicity across $-$2.0~$<$~log(Z/\Zsun)~$<$~0.5,  \citet{Bate2005} and \citet{Bate2014} simulated the same IMF, binary fraction, period distribution, and mass-ratio distribution.  They concluded the differences in opacity arising from differences in metallicity have a negligible effect on the processes of protobinary fragmentation and accretion.   However, the hydrodynamic simulations conducted by \citet{Bate2005} and \citet{Bate2014} had a resolution limit of $\approx$\,1\,AU, and so they could not directly probe trends with metallicity at very short separations.  Moreover, their low-metallicity simulations produced significantly more binary mergers, which might be unresolved close binaries. Most important (see below), these papers only changed the opacity from one calculation to the next, not the initial conditions. These simulations also neglected the intrinsic stellar and accretion luminosity of stars, which affects the temperatures, disk masses, and radii at which disk fragmentation occurs \citep{KMC11}. 

\citet{Glover2012} explored the onset of star formation in molecular clouds across $-$2~$<$~log(Z/\Zsun)~$<$~0. As expected, they found that gas temperatures in optically-thin cores rise as metallicity declines, thereby increasing the typical Jeans mass. However, they did not report substantial changes in the star formation outcome on large scales. \citet{Dopcke2011} and \citet{Dopcke2013} followed the thermal evolution and fragmentation of collapsing cores as a function of metallicity, and concluded differences only became pronounced at Z~$<$~10$^{-5}$\,\Zsun.  \citet{Myers2011} included the effects of radiative feedback, and still found that dust opacity negligibly affects the temperatures and fragmentations of cores as they collapse. \citet{Myers2011} also presented simple analytic models illustrating why the IMF is insensitive to metallicity.  Like the \cite{Bate2014} models, the simulations by \citet{Myers2011} and \citet{Dopcke2013} are limited by resolution, and therefore cannot reliably characterize disk properties on small scales.  Nevertheless, we conclude their results are robust on large scales.  Core fragmentation is relatively independent of metallicity, which is why the observed IMF and wide binary fraction are invariant across $-$1.5~$\lesssim$~log(Z/\Zsun)~$<$~0.5.

\citet{Machida2008} and \citet{Machida2009} argued that the alteration of the cloud initial conditions do affect fragmentation on smaller scales. In their low-metallicity models,  hotter cloud temperatures translate to larger mass accretion rates, making the disks  more susceptible to fragmentation.  In their simulations, which cover a broad range of metallicities $-$6~$<$~log(Z/\Zsun)~$<$~0, \citet{Machida2009} found the binary fraction measurably decreases with metallicity.  They also found the peak in the fragmentation separation transitions from $a_{\rm peak}$~$\approx$~1~AU for Z~=~10$^{-6}$\,\Zsun\ to $a_{\rm peak}$~$\approx$~100~AU for Z~=~\Zsun.  

More recently, \citet{Tanaka2014} expanded on these models by studying the changes in protostellar disk properties as a function of metallicity and primary mass.  They found disks of massive protostars ($M_1$~$\approx$~\,10\,\Msun) are gravitationally unstable and susceptible to fragmentation, even at solar-metallicity (see their Fig.~7).  This is consistent with previous models that showed the likelihood of disk fragmentation increases with final stellar mass as a result of the higher mass accretion rates \citep{Kratter2006, Kratter2008, Krumholz2007a}.  At solar-metallicity, the observed binary fraction of massive stars is already large, i.e., $\approx$\,70\% below $a$~$<$~10\,AU and nearly 100\% within $a$~$<$~100\,AU \citep[][\S\ref{Summary}]{Sana2012,Sana2014,Moe2017}.  Decreasing the metallicity can only marginally increase the close binary fraction of massive stars.

For low-mass stars, \citet{Tanaka2014} showed solar-metallicity disks are unlikely to fragment, consistent with previous results \citep{Kratter2008}.  Below Z~$<$~$10^{-3}$\Zsun, \citet{Tanaka2014} also found disk fragmentation is  more probable due to both increasing infall rates and more efficient disk cooling (see their Fig.~7).  Similarly, \citet{Clark2011a,Clark2011b} demonstrated the disks of primordial Population III stars are highly susceptible to fragmentation. The \citet{Machida2009} and \citet{Tanaka2014} models of disk fragmentation are qualitatively consistent with two observed trends:  (1) the anti-correlation between the close binary fraction and metallicity of solar-type stars (Fig.~\ref{allbin}), and (2) the shift in the binary period distribution toward smaller separations as the metallicity decreases (Fig.~\ref{Pdist}).  

Quantitatively, however, there is a large disagreement between the observations and previous simulations.  \citet{Tanaka2014} found only extremely metal-poor solar-type stars with log(Z/\Zsun)~$<$~$-$3 are more likely to have experienced disk fragmentation.  Meanwhile, we found the close binary fraction increases by a factor of $\approx$\,4 from [Fe/H]~=~+0.5 to $-$1.0 and then only slightly increases below [Fe/H]~$<$~$-$1.0 (see Fig.~\ref{allbin}).  We note \citet{Tanaka2014} neglected the impact of protostellar luminosity on disk temperatures, and also assumed that core radii, and thus disk radii, decrease with decreasing metallicity. For the parameters chosen in their models, low-mass solar-metallicity stars have disk radii of order $\approx$\,1,000\,AU, which are large compared to our best observational constraints of $\approx$\,100\,-\,300 AU \citep{Ansdell2018}. In the following, we address these concerns and present our own toy model of disk fragmentation for solar-type stars as a function of metallicity.

\subsection{A Simple Model for Disk Fragmentation}

Stellar binary formation via disk fragmentation requires the attainment of two conditions. First, the disk must be driven to be strongly self-gravitating, with Toomre parameter $Q$~=~$c_s \Omega / \pi G \Sigma$~$\approx$~1. Second, for gravitational instability to lead to the formation of bound clumps, gas must cool quickly so that the instability does not saturate in a gravito-turbulent state \citep{Kratter2016}. We can understand how decreased metallicity leads to enhanced disk fragmentation through the examination of a single dimensionless number:

\begin{equation}
\label{eq:xi}
\xi = \frac{G\dot{M}_{\rm in}}{{c_{\rm s,d}^3}},
\end{equation}

\noindent where $\dot{M}_{\rm in}$ is the infall rate onto the disk and $c_{\rm s,d}$ is the sound speed in the disk. \cite{Kratter2010} showed that disk fragmentation becomes prevalent when $\xi$~$\gtrsim$~1, with a weak dependence on cloud angular momentum. In the following, we show that as the metallicity decreases, $\xi$ increases due to the differential influence of metallicity on gas cooling in the optically thin cores versus optically thick disks.

First consider the scaling of the numerator, $\dot{M}_{\rm in}$.  It should scale with the core temperature, roughly as $c_{\rm s,c}^3/G$ or core sound speed cubed, which is the characteristic infall rate of an isothermal sphere \citep{Larson1969,Shu77}. While real infall rates are not constant in time, the sound speed sets the scale parameter around which excursions of order a few are expected. The ratio ${c_{\rm s,d}^3}/G$ in Eq.~\ref{eq:xi} parameterizes accretion through a self-gravitating disk. For a steady-state,  $\alpha$-disk model:

\begin{equation}
\label{eq:mdotd}
\dot{M} = 3 \pi \nu \Sigma = \frac{3 \alpha c_{s,d}^3}{GQ},
\end{equation}

\noindent where $\nu = \alpha c_s H$.  Even when global transport through spiral arm torques is poorly described by simple viscous $\alpha$ models, one still expects that the above equation, evaluated as $\alpha$\,$\rightarrow$\,1, represents an upper limit to the rate at which material can be processed through the accretion disk. With all other parameters held fixed, we see that $\xi$~$\propto$~$c_{\rm s, c}^3/c_{\rm s,d}^3$. Thus $\xi$ will increase if core temperatures rise or disk temperatures fall. Lowering the metallicity induces both affects simultaneously. 

Metallicity affects star formation by altering the cooling rates of gas. In low-density, optically thin gas, e.g. cores, the removal of metals decreases cooling rates, leading to systematically higher cloud temperatures, and thus infall rates. In contrast, protostellar disks are often optically thick to their own cooling radiation when $Q$~$\sim$~1, at least at metallicities near \Zsun. In this limit, gas cools predominantly through coupling with the dust, which radiates efficiently. Reducing the metallicity reduces the dust opacity by changing the gas-to-dust ratio. Thus when $\tau$~$>$~1, lowering the metallicity reduces the optical depth and thus enhances disk cooling rates at fixed temperatures and surface densities. In this regime, $c_{\rm s,c}$ rises while $c_{\rm s,d}$ falls, driving $\xi$ to higher values, and increasing the propensity of disks to fragmentation. 

There is a complication, however, which is that for sufficiently low metallicities, the disk becomes optically thin, and therefore further decreasing the metallicity would have the opposite effect. Even though core temperatures, and thus infall rates, continue to rise, disks temperatures should also rise. Thus at some metallicity, disk fragmentation should level off. In fact, the observed solar-type close binary fraction in Fig.~\ref{allbin} dramatically increases by a factor of $\approx$4 from [Fe/H]~=~+0.5 to $-$1.0, and then increases only by an additional $\approx$\,20\% toward smaller metallicities [Fe/H]~$<$~$-$1.0.   We partially attribute this break to the metallicity at which disk fragmentation transitions from the optically thick ([Fe/H]~$\gtrsim$~$-$1.0) to optically thin ([Fe/H]~$\lesssim$~$-$1.0) regimes.  We now present a simple model in which the combination of these affects can explain the rapid increase in the close binary fraction via disk fragmentation down to metallicities of Z~$\sim$~0.1\Zsun.

\subsection{Limitations on Fragmentation \\ as a Function of Metallicity}

We construct a quantitative model for when disk fragmentation should occur at a range of metallicities for forming solar-mass stars. We can place limits on disk fragmentation by constructing self-consistent models for self-gravitating disks undergoing rapid infall. We begin with an expression for the disk midplane equilibrium temperature (see \citealt{Kratter2008,KMCY10}):

\begin{equation}
\label{eqT}
\sigma T^4 = F_{\rm visc}\left(\frac{3\tau}{8}+\frac{1}{2\tau}\right) + F_{\rm irrad},
\end{equation}

\noindent where:

\begin{eqnarray}
F_{\rm visc} &=& \frac{3 \dot{M} \Omega^2}{4 \pi}, \\
F_{\rm irrad} &=& \sigma \left[\left(\frac{k_b}{G^3 M_* \mu}\right)^{1/7} \left(\frac{L_*}{4\pi}\right)^{2/7} \frac{1}{r^{3/7}}\right]^4, \\
L_* &=& \frac{1}{2}\frac{G M \dot{M}}{R_*}, {\rm and} \\ 
\tau &=& \frac{\kappa\Sigma}{2}.
\end{eqnarray}

\noindent We set $L_*$ to be the accretion luminosity, which dominates over gravitational contraction during the earliest phases of star formation. In order to determine the opacities as a function of temperature, we fit a polynomial to the \citet{Semenov2003} opacities in the range of 10\,-\,400\,K and adopt a constant value of $\kappa$~=~9.5~cm$^2$/g above $>$\,400\,K for solar metallicity. We decrease the opacity $\kappa$~$\propto$~Z in direct proportion to the metallicity as done in \citet{Bate2014}. Our results are only weakly dependent on the exact fit used for the opacities.

We now proceed to solve Eqn.~\ref{eqT} under a series of constraints: 
\begin{enumerate}
\item $Q$~=~1. This ensures that the disk is susceptible to fragmentation.
\item $\dot{M}$~=~3$\alpha c_{c,s}^3/(GQ)$, where $\alpha$~=~0.2. We set the accretion rate through the disk to be consistent with values expected for a strongly self-gravitating disk \citep{Kratter2010}. Because disks are driven unstable by rapid infall with $\xi$~$\ge$~1, we expect an unstable disk to process material at roughly this rate. This relationship is the standard viscous accretion rate expressed as a function of sound speed and Q.
\item $t_{\rm cool} \Omega$~$\le$~7. We require that the disk be able to radiate efficiently so that gravitational instability can lead to fragmentation, rather than gravitoturbulence or spiral mode saturation \citep{Gam2001,Kratter2016}. The cooling time indicates how long it takes a perturbation in temperature to radiatively cool from the midplane \citep{KMCY10}:
\begin{equation}
t_{\rm cool} = \frac{3\gamma\Sigma c_s^2}{32(\gamma-1)}\left(\tau+\frac{1}{\tau}\right){\sigma T^4}.
\label{eq_cool}
\end{equation}
\end{enumerate}

We consider a solar-type protostar with mass $M_*$~=~0.75\,\Msun\ and radius $R_*$~=~4\,\Rsun. Eqn.~\ref{eqT} can therefore be written as a function of accretion rate, disk radius, and metallicity. We solve for the critical accretion rate $\dot{M}_{\rm crit}$ at which all of the above constraints are satisfied simultaneously for a wide range of disk radii between $r_{\rm d}$~=~10\,-\,300\,AU and metallicities $-$3.0~$<$~log(Z/\Zsun)~$<$~0.5. We do note assume a scaling of the size of disks with metallicity, and therefore leave it as a free parameter in our model. Because disks are most unstable at their outer edge, our models are described by a single number rather than a disk profile. This solution provides viable combinations of $T$, $\Sigma$, $\dot{M}$, $Z$, and $r_d$ that could describe fragmenting disks. There is no guarantee of solutions for arbitrary combinations of temperature and metallicity. Moreover, the existence of a solution does not guarantee that real, astrophysical disks will achieve such disk properties in a given environment.

\begin{figure}[t!]
\centerline{
\includegraphics[trim=-0.2cm 0.0cm 0.1cm 0.1cm, clip=true, width=3.7in]{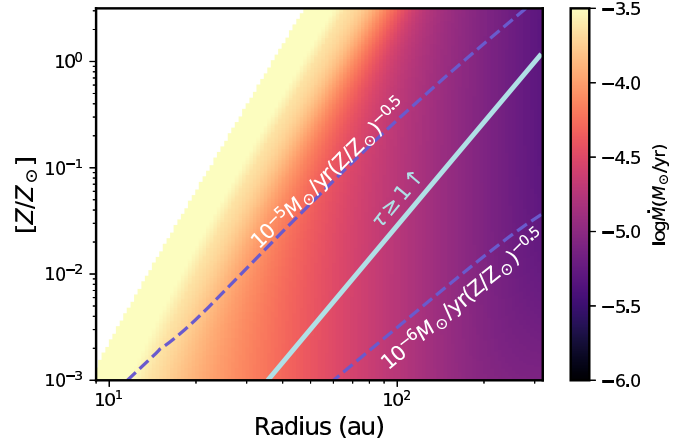}}
\caption{The color scale indicates the critical accretion rate, $\dot{M}_{\rm crit}$, required to drive a solar-type disk of a given radius and metallicity to fragment. In our model, fragmentation requires that the disk reach $Q$~=~1 and $t_{\rm cool}\Omega$~$<$~7, assuming that gravitational instability processes material at $\alpha$~$\approx$~0.2. The white line indicates the point at which disks transition from optically thick to thin. The bottom dashed line indicates the expected mass-weighted average infall rate $\langle \dot{M}_{\rm in} \rangle$ as a function of metallicity from \citet{Tanaka2014}, and the top dashed line represents a factor of ten excursion higher due to stochastic variations. All disks achieve accretion rates of $\dot{M}$~=~$\langle \dot{M}_{\rm in} \rangle$ while only a small fraction reach 10$\langle \dot{M}_{\rm in} \rangle$.  Given a maximum disk size of $r_d$~$\lesssim$~300~AU, the propensity for disk fragmentation increases, especially at smaller separations, as the metallicity decreases.}
\label{diskMdot}
\end{figure}

In Fig.~\ref{diskMdot}, we show the critical mass accretion rates $\dot{M}_{\rm crit}$ that satisfy $Q$~=~1 and $t_{\rm cool} \Omega$~$\le$~7 as a function of $r_d$ and Z for our self-consistent models. We also demarcate the radius at which $Q$~=~1 coincides with an optical depth of $\tau$~=~1, which decreases from $r_d$~=~300~AU near Z~=~\Zsun\ to $r_d$~=~40~AU near Z~=~10$^{-3}$\Zsun.  For solar metallicity, no solution exists below $r_d$~$<$~40~AU because the disks are too optically thick and therefore the disk cooling timescale according to Eqn.~\ref{eq_cool} is longer than $t_{\rm cool}$~$>$~7/$\Omega$.  Meanwhile, metal-poor disks, in principle, can fragment at slightly smaller separations, but only down to $r_d$~$\approx$~10\,AU at Z~=~10$^{-3}$\Zsun. The inability to directly fragment at small separations is consistent with previous studies that demonstrated close binaries ($a$~$<$~10~AU) could not have formed {\it in situ} \citep{Boss1986,Bate1998,Bate2009}.  Instead, close binaries initially fragmented on larger scales and then migrated inward, probably via interactions with the disk and/or external companions \citep{Artymowicz1983,Artymowicz1991,Bate1995,Bate1997,Bate2002,Moe2018}.

To estimate the parameter space that disks might inhabit, we consider the expected infall rates from cores of different metallicities. Following \citet{Tanaka2014}, we consider:

\begin{equation}
\label{mdotz}
\langle \dot{M}_{\rm in} \rangle = 10^{-6} {\rm M}_{\odot}\,{\rm yr}^{-1} \left(\frac{\rm Z}{{\rm Z}_{\odot}}\right)^{-1/2}.
\end{equation}

\noindent We display the combination of metallicities and disk radii that satisfy this mass accretion rate as the bottom dashed line in Fig.~\ref{diskMdot}.  For solar-type stars with solar-metallicity, an accretion rate of 10$^{-6}$\,\Msun\,yr$^{-1}$ is consistent with the mass-weighted average accretion rate during the earliest phases of growth. However, the typical accretion rates are likely variable during the first $\approx$\,0.5~Myr, and thus most objects experience excursions above (or well above) $\langle \dot{M}_{\rm in} \rangle$ \citep{Hartmann2001a,Evans2009,Offner2011,Hartmann2016}.  Moreover, because accretion is likely stochastic, driven by non-uniform turbulent molecular clouds, not all solar-type stars of the same metallicity experienced the same accretion history \citep{Nguyen2009,Cody2014,Bate2018}.  An increase in infall to even a few times the average accretion rate can greatly increase the propensity for disk fragmentation. We therefore display the solution for $\dot{M}_{\rm crit}$~=~10$\langle \dot{M}_{\rm in} \rangle$ as the top dashed line in Fig.~\ref{diskMdot}. Note that a very brief increase in the accretion rate above some threshold may not always trigger fragmentation, as the disk in some cases can quickly redistribute mass to remain stable.

According to Fig.~\ref{diskMdot}, it is quite difficult for metal-rich solar-type stars with Z~=~3\Zsun\ to have formed close binaries via disk fragmentation. If such stars accrete constantly at their mass-weighted average rates and the sizes of their gaseous disks extend only to $r_{\rm d}$~$\approx$~100\,-\,300~AU \citep{Ansdell2018}, then disk fragmentation would be impossible.  Instead, the small fraction of metal-rich solar-type protostars that undergo significant stochastic excursions up to $\dot{M}$~$\approx$~20$\langle \dot{M}_{\rm in} \rangle$~$\approx$~10$^{-5}$\,\Msun~yr$^{-1}$ are capable of disk fragmentation.  Even then, their disks are likely to fragment at large separations $r_d$~$\approx$~200~AU.  Meanwhile, disk fragmentation is highly more probable with decreasing metallicity, especially at smaller separations.  For Z~=~10$^{-3}$\Zsun, disks can fragment at $r_d$~$\approx$~60~AU given their nominal infall rate of $\langle \dot{M}_{\rm in} \rangle$.  If the disks accrete at 10$\langle \dot{M}_{\rm in} \rangle$, then fragmentation can occur at separations as small as $r_d$~$\approx$~10~AU. The shift in the minimum allowed fragmentation radius with decreasing metallicity is consistent with the inward shift in the peak of the binary distribution; metal-poor binaries peak at separations of only $a_{\rm peak}$~$\approx$~10~AU, while metal-rich binaries peak at wide separations $a_{\rm peak}$~$\approx$~200~AU  (Fig.~\ref{Pdist}). Although the location at which fragmentation occurs does not dictate the final binary period, correlations are to be expected \citep{Moe2018}.

Considering most disks will achieve at least a few times $\langle \dot{M}_{\rm in} \rangle$ at some time in their accretion history,  the majority of solar-type stars with intermediate metallicity Z~=~10$^{-1}$\Zsun\ should experience disk fragmentation. We therefore expect a rapid change in the probability of disk fragmentation across $-$1.0~$<$~[Fe/H]~$<$~0.5, consistent with the observed factor of $\approx$\,4 change in the close binary fraction across this same metallicity interval.  Below [Fe/H]~$<$~$-$1.0, the observed flattening in the slope of the close binary fraction versus metallicity anti-correlation (Fig.~\ref{allbin}) is due to two effects. First, as the fraction of disks undergoing fragmentation increases, at some point nearly all disks experience fragmentation.  According to Fig.~\ref{diskMdot},  essentially all disks with log(Z/\Zsun)~$\approx$~$-$2.0 will experience disk fragmentation.  Further decreasing the metallicity can only slightly increase the close binary fraction. The universality of disk fragmentation at higher stellar masses, even at \Zsun, may similarly explain the insensitivity of the close massive binary fraction to metallicity (see above). 

Second, depending on the variation in disk size with metallicity, disk optical depth may also contribute to the slope change. Across the interval $-$0.5~$\lesssim$~log(Z/\Zsun)~$<$~0.5, fragmentation likely occurs in the optically thick regime (see Fig.~\ref{diskMdot}). Thus decreasing the metallicity decreases the disk temperatures and cooling rates, which increases the probability of disk fragmentation.  If disk sizes remain large across $-$3~$<$~log(Z/\Zsun)~$\lesssim$~$-$1.5, (contrary to the models of \citealt{Tanaka2014}),  fragmentation instead occurs in the optically thin regime, wherein the decline in disk metallicity tends to stabilize disks. Thus one might expect this shift from optically thick to thin disks to temper the increase in binary formation. 
 
The consistency between the period distribution of early-B stars and low metallicity solar-type stars also supports a model in which enhanced disk fragmentation is responsible for the increase in close binaries (see Fig.~\ref{Pdist}). Disk fragmentation is thought to become more prominent for higher masses due to the increased infall rates and correspondingly higher $\xi$ associated with high mass star formation \citep{Kratter2006,Kratter2010}. We note that our models are substantially in agreement with the those of \citet{Tanaka2014} in terms of the critical accretion rates $\dot{M}_{\rm crit}$ required to drive disks unstable. Our conclusions regarding the metallicity at which disk fragmentation occurs for solar-type binaries differ because we account for the expected stochastic excursions in infall rate above $\langle \dot{M}_{\rm in} \rangle$. These fluctuations are responsible for the instability across a wide range of metallicites in our model. There is now compelling observational evidence that disk fragmentation may occur for low-mass stars near solar metallicity \citep{Tobin2016}, which boosts our confidence in this interpretation.

The increased probability for metal-poor disks to fragment must alter the IMF, at least to some extent.  The IMF can actually describe three different parameters: (1) the total IMF, $f$($M$), of all stars, including all companions in multiple systems, (2) the primary star IMF, $f$($M_1$), and (3) the system IMF, $f$($M_1$+$M_2$+...+$M_n$).  \citet{Chabrier2003} and \citet{Kroupa2013} discussed the differences in these distributions, noting that the primary star IMF derives most directly from the observations. At least one, possibly all three, of these distributions change with metallicity. Nevertheless, the effect is relatively small.  According to Fig.~\ref{Pdist}, the solar-type binary fraction below $a$~$<$~100~AU is $\approx$90\% for [Fe/H]~=~$-$3.0 and  $\approx$\,30\% for [Fe/H]~=~+0.5, a net change of $\approx$\,60\%.  The average mass ratio of solar-type binaries is $q$~$\approx$~0.5, relatively independent of metallicity (\S\ref{Summary}).  Hence, extremely metal-poor systems are on average $\approx$\,30\% more massive than their metal-rich counterparts.  Such a small change in the characteristic system mass is well within the observational measurement uncertainties and the resolution limit of simulations.  We therefore do not expect the system IMF to vary significantly across $-$1.5~$\lesssim$~[Fe/H]~$<$~0.5.  The effect of a metallicity-dependent close binary fraction on the three different IMFs needs to be studied in more detail.  

\section{Conclusions}
\label{Conclusions}

We have thoroughly examined the selection biases in various samples of solar-type stars and measured the intrinsic close binary fraction ($a$~$<$~10~AU) as a continuous function of metallicity.  We investigated multiple samples of SBs~(\S\ref{Spectroscopic}), APOGEE RV variables~(\S\ref{APOGEE}), and {\it Kepler} EBs~(\S\ref{Kepler}), all of which exhibit the same anti-correlation between $F_{\rm close}$ and [Fe/H]~(\S\ref{Summary}). We discussed and presented our own analytic models of fragmentation that reconcile the observed trends in binary properties as a function of mass, period, and metallicity~(\S\ref{Models}). We summarize the main results in the following. 

\vspace*{0.2cm}

\noindent {\it Spectroscopic Binaries}. Although the observed SB fraction appears to be constant with metallicity, metal-poor stars have weaker absorption lines, making it more difficult to identify SBs (Fig.~\ref{Latham_RV}).  After correcting the \citet{Latham2002} sample of high-proper-motion FGK stars for incompleteness, the intrinsic close binary fraction decreases from $F_{\rm close}$~=~54\%\,$\pm$\,12\% near [m/H]~=~$-$2.7 to $F_{\rm close}$~=~17\%\,$\pm$\,6\% at [m/H]~=~+0.5 (Fig.~\ref{Latham_binfrac}).  Considering only the Carney-Latham SBs with $P$~=~20\,-\,2,000~days and $K_1$~$>$~6~km~s$^{-1}$, where their survey is relatively complete (Fig.~\ref{Latham_fM}), the SB fraction of metal-poor halo stars ([m/H]~$<$~$-$1.0) is $\approx$\,1.9 times higher than metal-rich disk stars ([m/H]~$>$~$-$0.5).  Similarly, the observed SB companions to metal-poor giants ($-$3.5~$\lesssim$~[Fe/H]~$\lesssim$~$-$1.5) in the \citet{Carney2003} and \citet{Hansen2015,Hansen2016a} samples are concentrated toward $K_1$~$>$~7~km~s$^{-1}$ and $P$~=~35\,-\,3,000 days (Fig.~\ref{Carney_fM}), implying the bias-corrected close binary fraction of metal-poor solar-type dwarfs is $F_{\rm close}$~$\approx$~40\%\,-\,60\%.

\vspace*{0.2cm}

\noindent {\it APOGEE Radial Velocity Variables}. The APOGEE RV variability fraction of GK stars decreases by a factor of 4.0\,$\pm$\,0.5 across −0.9~$<$~[Fe/H]~$<$~0.5 at the 22$\sigma$ significance level (Fig.~\ref{DeltaRV}), consistent with the conclusions of \citet{Badenes2018}.  We measure the same trend independent of spectral type, surface gravity, and RV threshold, indicating both metal-poor and metal-rich binaries with $M_1$~$\approx$~0.6\,-\,1.5\,\Msun\ follow the same short-end tail of a log-normal period distribution.  After correcting the APOGEE RV variability survey of GK\,IV/V stars for incompleteness, the intrinsic close binary fraction decreases from $F_{\rm close}$~=~41\%\,$\pm$\,7\% at [Fe/H]~=~$−$0.8 to $F_{\rm close}$~=~11\%\,$\pm$\,2\% at [Fe/H]~=~+0.4 (Fig.~\ref{RV_binfrac}).  The median metallicities of close solar-type binaries are $\Delta$[Fe/H]~=~$-$0.13\,$\pm$\,0.03 dex lower than single stars (Fig.~\ref{cumRV}).  

\vspace*{0.2cm}

\noindent {\it Kepler Eclipsing Binaries}. For a large sample of {\it Kepler} solar-type dwarfs in which the metallicities have been measured photometrically to $\delta$[Fe/H]~$\approx$~0.3~dex precision, the observed EB fraction decreases by a factor of 3.4\,$\pm$\,0.5 across $-$0.9~$<$~[Fe/H]~$<$~0.3 at the 9$\sigma$ confidence level (Fig.~\ref{EBfrac}).  For a smaller subsample in which the metallicities have been measured spectroscopically to $\delta$[Fe/H]~$\approx$~0.1~dex precision, the observed EB fraction also decreases by a factor of $\approx$\,3.5 across the narrower interval $-$0.6~$<$~[Fe/H]~$<$~0.4 to 3$\sigma$ significance.  Metal-poor and metal-rich EBs both have the same period and eclipse depth distributions (Fig.~\ref{PvsF}), implying the period and mass-ratio distributions of close solar-type binaries are metallicity invariant. After accounting for various selection biases, the corrected solar-type close binary fraction decreases from  $F_{\rm close}$~=~52\%\,$\pm$\,14\% across $−$1.7~$<$~[Fe/H]~$<$~$-$1.1 to $F_{\rm close}$~=~13\%\,$\pm$\,3\% across 0.1~$<$~[Fe/H]~$<$~0.5 (Fig.~\ref{binfrac_EB}).

\vspace*{0.2cm}

\noindent {\it Combined Observational Constraints}. After correcting for incompleteness, all five samples of solar-type stars exhibit a quantitatively consistent anti-correlation: $F_{\rm close}$ = 53\%\,$\pm$\,12\%, 40\%\,$\pm$\,6\%, 24\%\,$\pm$\,4\% and 10\%\,$\pm$\,3\% at [Fe/H]~=~$-$3.0, $-$1.0, $-$0.2 (mean field metallicity), and +0.5, respectively (Fig.~\ref{allbin}). It is highly improbable that each of the different methods, with different biases, could conspire to produce consistent results. In contrast to close binaries, the wide binary fraction ($a$~$\gtrsim$~200~AU) of solar-type stars is relatively independent of metallicity.  The close binary fraction of $M_1$~$\approx$~10\,\Msun\ primaries is quite high ($F_{\rm close}$~=~70\%\,$\pm$\,11\%) and does not vary significantly with metallicity. As solar-type stars decrease in metallicity to [Fe/H]~$\lesssim$~$-$1.0, their close binary fraction ($F_{\rm close}$ $\approx$~50\%), overall binary fraction ($F_{\rm binary}$ $\approx$~90\%), triple/quadruple star fraction ($F_{\rm triple}$~+~$F_{\rm quadruple}$~$\approx$~35\%), and companion period distribution ($a_{\rm peak}$~$\approx$~10~AU) all approach that of early-B stars (Fig.~\ref{Pdist}).

\vspace*{0.2cm}

\noindent {\it Fragmentation Models}. Turbulent fragmentation of molecular cores on large spatial scales is relatively independent of metallicity, which is why the overall IMF and wide binary fraction are constant across $-$1.5~$\lesssim$~[Fe/H]~$<$~0.5.  Even at solar-metallicity, the disks of massive protostars are highly unstable and prone to fragmentation, explaining the high close binary fraction of massive stars.  Decreasing the metallicity of massive protostars can only marginally further increase the likelihood for disk fragmentation. For solar-type protostars with log(Z/\Zsun)~=~0.5, only the small fraction of disks that attain stochastic excursions to accretion rates $\dot{M}$~$\approx$~20$\langle \dot{M}_{\rm in} \rangle$ well above the mass-weighted average infall rates are capable of fragmentation at large radii $r_d$~$\approx$~200~AU.  With decreasing metallicity, (1)~the expected infall rates from hotter cores increase and (2)~the temperatures of the optically thick disks decrease, which both simultaneously drive the disk toward instability.  For solar-type protostars, the probability of disk fragmentation dramatically increases from log(Z/\Zsun)~=~+0.5 to $-$1.0, consistent with the observed increase in the close binary fraction. Metal-poor low-mass disks tend to fragment on smaller scales, possibly as small as $r_d$~=~10~AU, which is consistent with the observed shift in the peak of the overall solar-type binary period distribution.  

\vspace*{0.2cm}

\noindent {\it Implications for Binary Evolution}. Most solar-type stars with [Fe/H]~$<$~$-$1.0 will interact with a close binary companion, either through Roche lobe overflow or wind accretion. This has important consequences for binary evolution in old and metal-poor environments such as the galactic halo, bulge, thick disk, globular clusters, dwarf galaxies, and high-redshift universe.  Future studies must consider the effect of a close binary fraction versus metallicity anti-correlation on the inferred rates, properties, and progenitors of blue stragglers, barium stars, planetary nebulae, evolved giants, symbiotics, cataclysmic variables, novae, and Type Ia supernovae.  

\vspace*{0.2cm}

M.M. acknowledges financial support from NASA's Einstein Postdoctoral Fellowship program PF5-160139. K.M.K. acknowledges financial support from National Science Foundation under Grant No. AST-1410174 and NASA under Grant No. ATP-140078 and ATP-170070. We thank Andrei Tokovinin and Kevin Schlaufman for enlightening discussions that helped motivate our analysis.

\bibliographystyle{apj}                       
\bibliography{moe_biblio}

\end{document}